\documentstyle[12pt,twoside,titlepage,epsf]{article}  
\newcommand{\ncm}{\newcommand}
\renewcommand{\theequation}{\thesection.\arabic{equation}}
\newcommand{\sectiona}[1]{\setcounter{equation}{0}\section{#1}}
\newcommand{\sectie}{\def\thesection{\arabic{section}}}
\newcommand{\aanhangsel}{\def\thesection{\Alph{section}}
\setcounter{section}{0}}
\ncm{\oH}{\bar{H}}
\ncm{\us}{\quad\mbox{using}\quad}
\ncm{\ra}{\rightarrow}
\ncm{\ot}{\otimes}
\ncm{\DH}{D(H)}
\ncm{\DW}{D^{\omega}(H)}
\ncm{\TH}{T(\oH)}
\ncm{\ssc}{\displaystyle}
\ncm{\oq}{\eta} 
\ncm{\im}{\imath}
\ncm{\ba}{\begin{array}}
\ncm{\ea}{\end{array}}
\ncm{\ul}{\underline}
\ncm{\ol}{\overline}

\ncm{\str}{\rule{0cm}{3.5mm}}
\ncm{\om}{\omega}
\ncm{\ep}{\epsilon}
\newlength{\extraspace}
\newlength{\extraspaces}
\newcommand{\be}{\begin{equation}
\addtolength{\abovedisplayskip}{\extraspaces}
\addtolength{\belowdisplayskip}{\extraspaces}
\addtolength{\abovedisplayshortskip}{\extraspace}
\addtolength{\belowdisplayshortskip}{\extraspace}}
\newcommand{\ee}{\end{equation}}

\makeatletter

\def\eqnfourarray{\stepcounter{equation}%
  \def\@currentlabel{\p@equation\theequation}\global\@eqnswtrue \m@th
  \global\@eqcnt\z@ \tabskip\@centering
  \let\\\@eqncr \let\@@eqncr\@@eqnfourcr
  $$\everycr{}\halign to\displaywidth \bgroup
  \hskip\@centering $\displaystyle \tabskip\z@skip {##}$\@eqnsel &%
  \global\@eqcnt\@ne \hskip\tw@\arraycolsep \hfil ${##}$\hfil &%
  \global\@eqcnt\tw@ \hskip\tw@\arraycolsep $\displaystyle {##}$\hfil &%
  \global\@eqcnt\thr@@ \hskip\tw@\arraycolsep \hfil \hbox{##}%
    \tabskip\@centering &%
  \global\@eqcnt 4 \hbox\bgroup \hss ##\egroup
    \tabskip\z@skip \cr}

\def\endeqnfourarray{\@@eqncr \egroup
  \global\advance\c@equation\m@ne $$\global\@ignoretrue}

\def\@@eqnfourcr{\let\reserved@a\relax
  \ifcase\@eqcnt \def\reserved@a{& & & &}%
    \or \def\reserved@a{& & &}%
    \or \def\reserved@a{& &}%
    \or \def\reserved@a{&}%
    \else \let\reserved@a\@empty
      \@latex@error{Too many columns in eqnfourarray environment}\@ehc
  \fi
  \reserved@a \if@eqnsw\@eqnnum \stepcounter{equation}\fi
  \global\@eqnswtrue \global\@eqcnt \z@\cr}

\makeatother

\newcommand{\bea}{\begin{eqnarray}
\addtolength{\abovedisplayskip}{\extraspaces}
\addtolength{\belowdisplayskip}{\extraspaces}
\addtolength{\abovedisplayshortskip}{\extraspace}
\addtolength{\belowdisplayshortskip}{\extraspace}}
\newcommand{\eea}{\end{eqnarray}}
\newcommand{\beas}{\begin{eqnarray*}
\addtolength{\abovedisplayskip}{\extraspaces}
\addtolength{\belowdisplayskip}{\extraspaces}
\addtolength{\abovedisplayshortskip}{\extraspace}
\addtolength{\belowdisplayshortskip}{\extraspace}}
\newcommand{\eeas}{\end{eqnarray*}}
\ncm{\Z}{{\mbox{\bf Z}}}
\ncm{\al}{\alpha}
\ncm{\bt}{\beta}
\ncm{\gm}{\gamma}
\ncm{\dl}{\delta}
\ncm{\varep}{\varepsilon}
\ncm{\zt}{\zeta}
\ncm{\et}{\eta}
\ncm{\th}{\theta}
\ncm{\kp}{\kappa}
\ncm{\lm}{\lambda}
\ncm{\rh}{\rho}
\ncm{\hl}{\hline}
\ncm{\sg}{\sigma}
\ncm{\ta}{\tau}
\ncm{\ph}{\phi}
\ncm{\phv}{\varphi}
\ncm{\ch}{\chi}
\ncm{\ps}{\Phi}
\ncm{\nn}{\nonumber}
\title{ 
{\flushleft {\normalsize June 1996
\hfill PAR--LPTHE 96--17 and ITFA 96--16, hep-th/9606029}}\\[2.5cm]
SPONTANEOUSLY BROKEN \\
ABELIAN CHERN-SIMONS THEORIES \vspace{1.5cm} }
\author{ 
Mark de Wild Propitius\thanks{e-mail: mdwp@lpthe.jussieu.fr}\\
\normalsize{ {\em
Laboratoire de Physique Th\'eorique et Haute Energies}}\thanks{Laboratoire 
associ\'e No.\ 280 au CNRS}\\
\normalsize{{\em Universit\'e Pierre et Marie Curie - PARIS VI}} \\
\normalsize{{\em Universit\'e Denis Diderot - PARIS VII}} \\
\normalsize{{\em 4 place Jussieu, Boite 126, Tour 16, 1$^{er}$ \'etage}}\\
\normalsize{{\em F-75252 Paris CEDEX 05, France}}
\vspace{6cm}} 
\date{}
\begin{document}
\maketitle
\begin{abstract}
A detailed analysis of Chern-Simons (CS) theories in which a compact abelian 
direct product gauge group $U(1)^k$ is spontaneously broken down to a direct 
product of cyclic groups 
$H \simeq \Z_{N^{(1)}} \times  \cdots \times \Z_{N^{(k)}}$ is presented. 
The spectrum features global $H$ charges, vortices carrying 
magnetic flux labeled by the elements of $H$ and dyonic combinations.
Due to the Aharonov-Bohm effect these particles exhibit topological 
interactions. The remnant of the $U(1)^k$ CS term in the discrete ${ H}$ gauge 
theory describing the effective long distance physics of such a model is shown 
to be a 3-cocycle for ${H}$ summarizing the nontrivial topological 
interactions for the magnetic fluxes implied by the ${U(1)^k}$ CS term. 
It is noted that there are in general three types of 3-cocycles for a finite 
abelian gauge group ${H}$: one type describes topological interactions 
between vortices carrying flux w.r.t.\ the same cyclic group in the direct 
product $H$, another type gives rise to topological interactions among 
vortices carrying flux w.r.t.\ two different cyclic factors 
of $H$ and a third type leading to topological interactions between vortices 
carrying flux w.r.t.\ three different cyclic factors. Among other things, 
it is demonstrated that only the first two types can be obtained from a 
spontaneously broken $U(1)^k$ CS theory. The 3-cocycles that can not be 
reached in this way turn out to be 
the most interesting. They render the theory nonabelian and in general lead to 
dualities with planar theories with a nonabelian finite gauge group. In 
particular, the CS theory with finite gauge group 
$H \simeq \Z_2 \times \Z_2 \times \Z_2$ defined by such a 3-cocycle is shown 
to be dual to the planar discrete $D_4$ gauge theory with $D_4$ the dihedral 
group of order 8.
\end{abstract}
\newpage

\sectiona{Introduction}

A characteristic feature of 2+1 dimensional space time is the possibility
to endow a gauge theory with a Chern-Simons (CS) term. Ever since the 
pioneering work by Schonfeld and Deser, Jackiw and Templeton~\cite{schonfeld}
in the early eighties, these topological terms have had an impact in various 
seemingly unrelated branches of physics and mathematics. Notably, in a 
seminal paper~\cite{witten} Witten pointed out the significance of pure 
CS theories in the setting of knot invariants and in so doing revealed
a deep connection between pure CS theory and 1+1 dimensional rational 
conformal field theory. Earlier on, it was demonstrated by Hagen and 
Arovas, Schrieffer and Wilczek~\cite{aro} that sources coupled to 
abelian CS gauge fields in general behave as anyons~\cite{leinaas}, 
i.e.\ particles with fractional spin and quantum statistics intermediate 
between bosons and fermions. Anyons and CS theories gained further attention
after it was shown that an ideal gas of anyons is 
superconducting~\cite{anyonbook}. Moreover, it is by know well-established
that anyons are realized in nature as quasi-particles in fractional quantum 
Hall liquids~\cite{laughlin}. This remarkable observation due to Lauglin
and the aforementioned general results initiated a large body of work
in which fractional quantum Hall systems have served as a playground
for applications of 1+1 dimensional conformal field theory and 2+1 dimensional
CS theories (e.g.\ \cite{confhall} and references therein). Finally,  
CS theory also plays a role in 2+1 dimensional gravity~\cite{witgrav}.

In this paper, the main focus is on the implications of adding a CS term  
to planar gauge theories which are spontaneously broken down to a finite 
residual gauge group via the Higgs mechanism. That is, the models under 
consideration are governed by an action of the form
\bea                               \label{alg}
S &=& S_{\mbox{\scriptsize YMH} } + S_{\mbox{\scriptsize matter}} 
+ S_{\mbox{\scriptsize CS}} \, ,
\eea
where the Yang-Mills Higgs action $S_{\rm{\mbox{\scriptsize YMH}}}$  
gives rise to the spontaneous breakdown of some continuous compact gauge 
group $G$ to a finite subgroup ${ H}$ and $S_{\mbox{\scriptsize matter}}$ 
describes a conserved matter current  minimally coupled to the gauge fields. 
Finally, $S_{\mbox{\scriptsize CS}}$ denotes the CS action for the gauge 
gauge fields. 

The discrete $H$ gauge theories describing the long distance physics of 
the models~(\ref{alg}) without CS term $S_{\mbox{\scriptsize CS}}$ 
for the broken gauge group $G$ have been studied by various authors both 
in 2+1 and 3+1 dimensional space time and are by now completely understood. 
(For a recent detailed treatment and an up to date account of the literature 
on these models, the interested reader is referred to the lecture 
notes~\cite{banff}.)  The spectrum features topological defects which in 
2+1 dimensional space time appear as (particle-like) vortices carrying 
magnetic flux labeled by the elements of the finite gauge group $H$. In case 
$H$ is nonabelian, the vortices generally exhibit a nonabelian Aharonov-Bohm 
(AB) effect~\cite{ahabo}: upon braiding two vortices their fluxes affect each 
other through conjugation~\cite{bais}. Under the action of the residual 
global gauge group $H$, the fluxes also transform through
conjugation and the conclusion is that the different magnetic vortices
are labeled by the conjugacy classes of $H$. This is in a nutshell the physics
described by the Yang-Mills Higgs part $S_{\mbox{\scriptsize YMH} }$
of the action~(\ref{alg}). 
The matter fields, covariantly coupled to the gauge fields in the matter 
part $S_{\mbox{\scriptsize matter}}$ of the action, form multiplets which 
transform irreducibly under the broken gauge group ${ G}$. In the broken 
phase, these branch to irreducible representations of the residual gauge 
group ${ H}$. So, the matter fields introduce point charges in the broken  
phase labeled by the unitary irreducible representations (UIR's) $\Gamma$ of 
$H$. If such a charge encircles a magnetic flux $h \in H$, 
it also undergoes an AB effect~\cite{krawil,preskra,sam}. That is, it  returns 
transformed by the matrix $\Gamma(h)$ assigned to the group element $h$ in 
the representation $\Gamma$ of $H$. Since the gauge fields in these models 
are completely massive, the foregoing topological AB effects  
form the only long range interactions among the charges and vortices. 
Of course, the complete spectrum also features the dyons obtained 
by composing the vortices and charges. These are labeled by the conjugacy 
classes of $H$ paired with a nontrivial centralizer 
representation~\cite{spm}. Finally, as has been pointed out in 
reference~\cite{spm} as well, this spectrum of charges, vortices and dyons 
and the spin, braiding and fusion properties of these particles is fully 
described by the representation theory of the quasitriangular Hopf 
algebra $D(H)$ resulting~\cite{dpr} 
from Drinfeld's double construction~\cite{drin} applied to the abelian algebra 
${\cal F}(H)$ of functions on the finite group ${ H}$.

The presence of a CS term $S_{\mbox{\scriptsize CS}}$ for the broken gauge 
group $G$ in the action~(\ref{alg}) naturally has a bearing on the long 
distance physics. In~\cite{spm1,sm}, it was argued on general grounds that
the remnant of a CS term in the discrete $H$ gauge theory describing the 
long distance physics of the model is a 3-cocycle $\omega \in H^3(H,U(1))$ 
for the residual finite gauge group $H$, which governs the additional AB
interactions among the vortices implied by the original CS term 
$S_{\mbox{\scriptsize CS}}$ for the broken gauge group $G$. Accordingly,
the related algebraic structure now becomes the quasi-Hopf algebra $\DW$ 
being a natural deformation of $D(H)$ depending on this 3-cocycle 
$\omega$.  These general results were just explicitly illustrated  by the 
example of the abelian CS Higgs model in which the (compact) gauge group
$G \simeq U(1)$ is broken down to a cyclic subgroup $H \simeq \Z_N$.
In the present paper, this analysis is extended to the general 
case of spontaneously broken abelian  CS  theories~(\ref{alg}).
That is, we will concentrate on symmetry breaking schemes 
\bea                                        \label{scheme}
{ G} \; \simeq \; U(1)^k   &\longrightarrow&  { H} \, ,
\eea 
with  $U(1)^k$ being the direct product of $k$ compact $U(1)$ gauge 
groups and the finite subgroup $H$ a direct  product of $k$ cyclic groups  
$\Z_{N^{(i)}}$ of order $N^{(i)}$
\bea       \label{genab}
{ H} &\simeq& \Z_{N^{(1)}} \times \Z_{N^{(2)}} \times \cdots \times
\Z_{N^{(k)}} \, .
\eea
One of the main aims is to give a complete classification of these
broken abelian planar gauge theories.

In fact, unbroken CS theory with direct product gauge group $U(1)^k$ 
endowed with minimally coupled matter fields has received considerable 
attention recently (e.g.\ \cite{hagen, diamant} and references therein). 
One of the motivations to study these theories is that they find an 
application in multi-layered quantum Hall systems. To give a brief 
sketch of the main results, the CS terms for the gauge group $U(1)^k$ are 
known to fall into two types. On the one hand, there are CS terms (type~I) 
that describe self-couplings of the various $U(1)$ gauge fields. On the other 
hand, there are terms (type~II) that establish couplings between two different 
$U(1)$ gauge fields. To be concrete, the most general CS action for the 
gauge group $G \simeq U(1) \times U(1)$, for instance, is of the form
\bea 
S_{\mbox{\scriptsize CS}} &=& \int d\,^3x \; \left( \, 
\frac{\mu^{(12)}}{2} \epsilon^{\kappa\sigma\rho}
                              A^{(1)}_{\kappa} \partial_{\sigma} 
                              A^{(2)}_{\rho}
\: + \: 
\sum_{i=1}^2 \frac{\mu^{(i)}}{2} \epsilon^{\kappa\sigma\rho}
                              A^{(i)}_{\kappa} \partial_{\sigma} 
                              A^{(i)}_{\rho} \, \right) ,
\eea 
with $A_{\kappa}^{(1)}$ and $A_{\kappa}^{(2)}$ the two $U(1)$ gauge fields 
and $\epsilon^{\kappa\sigma\rho}$ the three dimensional anti-symmetric 
Levi-Civita tensor normalized such that $\epsilon^{012}=1$. The parameters 
$\mu^{(1)}$ and $\mu^{(2)}$ denote the topological masses characterizing the 
CS actions of type~I and  $\mu^{(12)}$ the topological mass characterizing 
the CS action of type~II. In the unbroken phase, these CS terms assign 
magnetic fluxes to the quantized matter charges $q^{(1)}$ and $q^{(2)}$ 
coupled to the two compact $U(1)$ gauge fields. Specifically, the type~I CS 
term for the gauge field $A_{\kappa}^{(i)}$ attaches a magnetic flux 
$\phi^{(i)}=-q^{(i)}/\mu^{(i)}$ to a matter charge 
$q^{(i)}=n^{(i)} e^{(i)}$ with $n^{(i)} \in \Z$ and $e^{(i)}$ the coupling 
constant for $A_{\kappa}^{(i)}$. As a consequence, there are nontrivial 
topological AB interactions among these charges. When a charge $q^{(i)}$ 
encircles a remote  charge $q^{(i)'}$ in a counterclockwise fashion, the wave 
function acquires~\cite{goldmac} the AB phase 
$\exp(-\im q^{(i)} q^{(i)'}/\mu^{(i)})$.  The CS term of type~II, in turn, 
attaches  fluxes which belong to one $U(1)$ gauge group to the matter charges 
of the other. That is, a charge $q^{(1)}$ induces a flux 
$\phi^{(2)}=-2q^{(1)}/\mu^{(12)}$ and a charge $q^{(2)}$ induces a flux 
$\phi^{(1)}=-2q^{(2)}/\mu^{(12)}$. Hence, the type~II CS term gives rise to 
topological interactions among matter charges of the two different $U(1)$ 
gauge groups~\cite{hagen}. A counterclockwise monodromy of a charge  
$q^{(1)}$ and a charge  $q^{(2)}$, for example, yields the AB phase 
$\exp(-2\im q^{(1)} q^{(2)}/\mu^{(12)})$.

The spontaneously broken versions~(\ref{scheme}) of these abelian CS 
theories, however, have not yet been fully explored. Among other things,
I will argue that in the broken case the $U(1)^k$ CS term
gives rise to nontrivial AB phases among the vortices labeled by the elements 
of the residual gauge group~(\ref{genab}). To be specific, the $k$ different 
vortex species carry quantized flux 
$\phi^{(i)} = \frac{2\pi a^{(i)}}{N^{(i)} e^{(i)}}$ with  
$a^{(i)} \in \Z$ and $N^{(i)}$ the order of the $i^{\rm th}$ cyclic group
of the product group~(\ref{genab}). A type~I CS term for the gauge field  
$A_{\kappa}^{(i)}$  then implies the AB phase 
$\exp(\im \mu^{(i)} \phi^{(i)} \phi^{(i)'})$ for a 
counterclockwise monodromy of  a vortex $\phi^{(i)}$ and a vortex 
$\phi^{(i)'}$. A CS term of type~II coupling the gauge fields 
$A_{\kappa}^{(i)}$ and $A_{\kappa}^{(j)}$, in turn, gives rise to the AB
phase $\exp(\im \mu^{(ij)} \phi^{(i)} \phi^{(j)})$ for the process in which 
a vortex $\phi^{(i)}$ circumnavigates a vortex $\phi^{(j)}$ in a 
counterclockwise fashion. In agreement with the general remarks in an earlier
paragraph, these additional AB phases among the vortices are shown to be the 
only distinction with the abelian discrete $H$ gauge theory describing the 
long distance physics in the absence of a CS action for the broken gauge group 
$U(1)^k$. That is, as was already pointed out for the simplest case
$U(1)\rightarrow \Z_N$ in~\cite{sam,spm1,sm}, the Higgs mechanism removes 
the fluxes attached the matter charges $q^{(i)}$ in the unbroken CS phase. 
Hence, contrary to the unbroken CS phase, there are {\em no} AB interactions 
among the matter charges in the CS Higgs phase. The canonical AB interactions 
$\exp(\im q^{(i)}\phi^{(i)})$ between the matter charges $q^{(i)}$ and 
the magnetic vortices $\phi^{(i)}$ persist though.

A key role in the analysis of this paper is played by the Dirac 
monopoles~\cite{dirac} that can be introduced in these compact 
$U(1)^k$ CS theories. There are $k$ different species corresponding to 
the $k$ different compact $U(1)$ gauge groups. It is 
known~\cite{diamant,hen,pisar} that a consistent incorporation of these 
monopoles requires the quantization of the topological masses characterizing 
the type~I and type~II CS terms. Moreover, it has been argued that in 
contrast to ordinary 2+1 dimensional compact QED~\cite{polyakov}, the 
presence of Dirac monopoles does {\em not} lead to confinement of the 
charges $q^{(i)}$ in the unbroken CS phase~\cite{pisar,affleck,klee}. 
Instead, the monopoles in these CS theories describe tunneling events leading 
to the creation or annihilation of charges $q^{(i)}$ with magnitude depending 
on the integral CS parameter. That is, the spectrum just features a finite 
number of stable charges depending on the integral CS 
parameter~\cite{sm,diamant,klee}. As usual, the presence of Dirac monopoles in 
the broken phase implies that the magnetic fluxes $a^{(i)}$ carried by the 
vortices are conserved modulo $N^{(i)}$, but the flux decay driven by the 
monopoles is now accompanied by the creation of matter charge where the 
species of the charge depends on the type of the CS term and the magnitude
is  again proportional to the integral CS parameter (see also~\cite{spm1,sm}).
Finally, it is shown that the quantization of topological mass implied 
by the presence of Dirac monopoles is precisely such that the $U(1)^k$ CS 
terms indeed boil down to a 3-cocycle for the residual finite gauge group
$H$ in the broken phase.


The organization of this paper is as follows.
In section~\ref{bgt}, I start by briefly recalling a result  
due to Dijkgraaf and Witten~\cite{diwi} stating that the different 
CS actions $S_{\mbox{\scriptsize CS}}$ for a compact gauge group $G$ are 
labeled by the elements of the cohomology group $H^4(BG, \Z)$ of the 
classifying space $BG$. A classification which for finite groups $H$ 
boils down to the cohomology group $H^3(H,U(1))$ of the group $H$ itself. 
In other words, the different CS theories for a finite gauge group $H$ 
correspond to the inequivalent 3-cocycles $\omega \in H^3(H,U(1))$. 
The new observation in this section is that the effective 
long distance physics of a CS theory in which the gauge group $G$ is broken 
down to a finite subgroup $H$ via 
the Higgs mechanism is described by a discrete $H$ CS theory 
defined by the 3-cocycle $\omega \in H^3(H,U(1))$
determined by the original CS action $S_{\mbox{\scriptsize CS}} \in 
H^4(BG,\Z)$ for the broken gauge group $G$ through the natural 
homomorphism $H^4(BG,\Z) \rightarrow H^3(H,U(1))$ induced by the 
inclusion $H \subset G$. 
Section~\ref{fincoh} subsequently 
contains a short introduction to the cohomology groups $H^n(H,U(1))$
of finite abelian groups $H$.
In particular, the explicit realization of the complete set 
of independent 3-cocycles $\omega \in H^3(H,U(1))$
for the abelian groups~(\ref{genab}) is presented there.
It turns out that these split up into three different types,
namely 3-cocycles (type~I) which give rise to nontrivial AB interactions 
among fluxes of the same cyclic gauge group in 
the direct product~(\ref{genab}), 
those (type~II) that describe interactions between 
fluxes corresponding to two different cyclic gauge groups 
and finally 3-cocycles (type~III) that lead to additional AB interactions 
between fluxes associated to three different cyclic gauge groups.
Section~\ref{multiple} then deals with the classification of 
CS actions for the compact gauge group $U(1)^k$. As mentioned before, 
these come in two types: 
CS actions (type~I) that describe self couplings of the different 
$U(1)$ gauge fields and CS action (type~II) establishing pairwise
couplings between different $U(1)$ gauge fields. 
The natural conclusion is  that the homomorphism 
$H^4(B(U(1)^k), \Z) \to H^3(H,U(1))$ induced by the spontaneous symmetry 
breakdown~(\ref{scheme}) is not onto.
That is, the only CS theories with finite abelian 
gauge group~(\ref{genab}) that may arise from a spontaneously broken 
$U(1)^k$ CS theory are those
corresponding to a 3-cocycle of type~I and/or type~II, while 3-cocycles of 
type~III do not occur.   

Section~\ref{symalg} is devoted to a discussion of 
the quasi--Hopf algebra $\DW$ related to an abelian discrete $H$ 
CS theory defined by the 3-cocycle $\omega \in H^3(H,U(1))$.
The emphasis is on the unified description this algebraic framework gives
of the spin, braid and fusion properties of the magnetic vortices, charges
and dyons constituting the spectrum of such a discrete $H$ CS theory.

In the next sections, the foregoing general considerations are illustrated 
by some representative examples. Specifically, section~\ref{typeI} deals 
with the abelian CS Higgs model in which the compact 
gauge group $G \simeq U(1)$ is broken down to the cyclic subgroup 
$H \simeq \Z_N$. First, the unbroken $U(1)$ phase of this model is briefly 
reviewed. In particular, it is recalled that a consistent 
implementation of Dirac monopoles requires the topological mass to be 
quantized as $\mu=\frac{pe^2}{\pi}$ with $p\in\Z$ and $e$ the 
coupling constant, which is in agreement with the fact that the 
different  CS actions for a compact gauge group $U(1)$
are classified by the integers: $H^4(BU(1), \Z) \simeq \Z$. 
Subsequently, the broken phase of the model is discussed. Among other things,
it is established that the long distance physics is indeed described by a 
$\Z_N$ CS theory with 3-cocycle $\omega \in H^3( \Z_N ,U(1))\simeq \Z_N$
fixed by the natural homomorphism $H^4(BU(1), \Z) \to H^3( \Z_N ,U(1))$.
In other words, the integral CS parameter $p$ becomes periodic 
in the broken phase with period $N$. 
Section~\ref{typeII} then contains a similar treatment of a 
CS theory of type~II with gauge group
$G \simeq U(1) \times U(1)$ spontaneously broken down 
to $H \simeq \Z_{N^{(1)}} \times \Z_{N^{(2)}}$. The effective 
long distance physics of this model is described by a 
$\Z_{N^{(1)}} \times \Z_{N^{(2)}}$ CS theory defined by a 3-cocycle of type~II.
The abelian discrete $H$ CS theories which do not occur as the remnant of a 
spontaneously broken $U(1)^k$ CS theory are actually the most interesting. 
These are the CS theories defined by  the aforementioned 3-cocycles of 
type~III.  The simplest example of such a theory, namely that with gauge group 
$H \simeq \Z_2 \times \Z_2 \times \Z_2$, is treated in full detail in 
section~\ref{typeIII}. It is pointed out that the incorporation of the 
corresponding 3-cocycle of type~III renders the theory nonabelian.
That is, the resulting type~III CS theory exhibits nonabelian phenomena 
like Alice fluxes, Cheshire charges, nonabelian Aharonov-Bohm sacttering and 
the multi-particle configurations generally satisfy nonabelian braid 
statistics. Probably the most striking result of this section is that 
this theory turns out to be dual to the ordinary $D_4$ planar 
gauge theory with $D_4$ the nonabelian dihedral group of order $8$. 
Moreover, it is argued  
that the $\Z_2 \times \Z_2 \times \Z_2$ CS theory defined by the product
of the 3-cocycle of type~III and either one of the three 3-cocycles of type~I
is dual to the ordinary $\bar{D}_2$ planar gauge theory with $\bar{D}_2$ the 
quaternion group being  the other nonabelian group of order $8$.

Finally, section~\ref{dijwit} presents some new results on the 
Dijkgraaf-Witten invariant for lens spaces based on the three different types 
of  3-cocycles for various finite abelian groups $H$, whereas some
concluding remarks and an outlook can be found in section~\ref{concl}.

In addition, there are three appendices. In appendix~\ref{gc}, I have 
collected the derivation of some identities in the theory 
of group cohomology used in the main text. In particular, it contains a 
derivation of the content of the cohomology group $H^3(H,U(1))$ for an 
arbitrary abelian finite group $H$ of the form~(\ref{genab}) and a derivation 
of the content of the cohomology group $H^4(B(U(1)^k), \Z)$.
Further, one of the novel observations in this paper is that 
rather than representations of the ordinary braid groups the 
multi-particle systems in abelian {\em discrete} $H$ CS theories 
realize representations of so-called truncated braid groups being 
factor groups of the ordinary braid groups. The precise definition of these 
truncated braid groups is given in appendix~\ref{trubra} along with  
useful identifications of some of them with well-known finite groups. 
Finally, the characteristic features of a planar gauge theory with 
finite nonabelian gauge group the dihedral group $D_4$ (being dual to 
the $\Z_2 \times \Z_2 \times \Z_2$ CS theory defined by a 3-cocycle of type~III
as argued in section~\ref{emduality}) are briefly 
discussed in appendix~\ref{dvier}.

In passing, the treatment of the examples 
in sections~\ref{typeI}, \ref{typeII}
and~\ref{typeIII} is more or less self contained. 
So, the more physically inclined reader who may not be so much interested in 
the rather mathematical classification side of the problem could well start 
with section~\ref{typeI} and occasionally go back to earlier sections 
for definitions and technicalities.

As for conventions, throughout this paper natural units in which 
$\hbar=c=1$ are employed. We will exclusively work in 2+1 dimensional 
Minkowsky space with signature $(+,-,-)$. Spatial coordinates are 
denoted by $x^1$ and $x^2$ and the time coordinate by $x^0 =t$. 
As usual, greek indices run from 0 to 2, while spatial components 
are labeled by latin indices$\in 1,2$. 
Unless stated otherwise, we will use Einstein's summation convention.

\sectiona{Group cohomology and symmetry breaking}
\label{bgt}

As has been argued by Dijkgraaf and Witten~\cite{diwi}, 
the  CS actions $S_{\mbox{\scriptsize CS}}$ for 
a compact gauge group $G$ are in one-to-one correspondence
with the elements of the cohomology 
group $H^4 (B{ G}, \Z)$ of the classifying space
$B{ G}$ with integer coefficients $\Z$.~\footnote{Let $EG$ be 
a contractible space with a free action of $G$. 
A classifying space $BG$ for $G$ is then given by dividing out 
the action of $G$ on $EG$. That is, $BG=EG/G$ (e.g.\ \cite{novikov}).}
In particular, this classification includes 
the case of finite gauge groups $H$. 
The isomorphism~\cite{milnor}
\bea                            \label{miln}
H^n(B{ H},{\mbox{\bf Z}}) &\simeq& H^n({ H},{\mbox{\bf Z}}) \, ,
\eea 
which only holds for finite groups ${ H}$, shows that the cohomology 
of the classifying space $BH$ is the same as that of the group $H$ itself.
In addition, we have the isomorphism 
\bea                  \label{clasi}         
H^{n} (H, \Z)  & \simeq &   H^{n-1} ({ H}, U(1))    
\qquad \forall  \, n>1 \, .
\eea
A derivation of this result, using the universal coefficients theorem, 
is contained in appendix~\ref{gc}.
Especially, we now arrive at the identification 
\bea          \label{clasor}                 
H^4 ({BH}, \Z) & \simeq & H^3 ({ H}, U(1)) \, ,
\eea 
which expresses the fact that the different CS theories 
for a finite gauge group $H$ are, in fact,  defined by the different 
elements $\omega \in H^3({ H}, U(1))$, i.e.\ algebraic 
3-cocycles $\omega$ taking values in $U(1)$. 
These 3-cocycles can be interpreted
as $\omega = \exp(\im S_{\mbox{\scriptsize CS}})$, 
where $S_{\mbox{\scriptsize CS}}$ denotes a
CS action for the finite gauge group $H$~\cite{diwi}.
With abuse of language, we will  usually call $\omega$ itself
a CS action for $H$.

Let $K$ be a subgroup of a compact group $G$.
The inclusion $K \subset G$ induces a 
natural homomorphism 
\bea    \label{restric} 
H^4 (B{ G}, \Z)  &\longrightarrow&   H^4 (BK, \Z) \, ,
\eea 
called the restriction (e.g.\ \cite{brown}). 
This homomorphism determines the fate 
of a given CS action $S_{\mbox{\scriptsize CS}} \in H^4 (B{ G}, \Z)$ 
when the gauge group $G$ is spontaneously broken 
down to $K$ via the Higgs mechanism.
That is, the mapping~(\ref{restric}) 
fixes the CS action~$\in H^4 (BK, \Z)$ for the residual
gauge group $K$ to which $S_{\mbox{\scriptsize CS}}$ reduces in the broken 
phase. In the following, we will only be concerned with CS theories
in which a continuous (compact) gauge  group $G$ is broken down 
to a finite subgroup $H$.  
The long distance physics of such a model 
is described by a discrete $H$ CS theory 
with 3-cocycle $\omega \in H^3 ({ H}, U(1))$ determined by 
the original CS action $S_{\mbox{\scriptsize CS}}$ 
for the broken gauge group $G$ through the natural homomorphism 
\bea                          
H^4 (B{ G}, \Z)  &\longrightarrow&   H^3 ({ H}, U(1)) \, , 
\label{homo}   
\eea 
being the composition of the restriction 
$H^4 (B{ G}, \Z)  \to  H^4 (BH, \Z)$ induced by the inclusion 
$H \subset G$, and the isomorphism~(\ref{clasor}).
As will become clear in the following sections, 
the 3-cocycle $\omega$ governs the additional 
AB phases among the magnetic fluxes (labeled by the elements $h \in H$) 
in the broken phase implied by the CS action $S_{\mbox{\scriptsize CS}}$.

The restrictions~(\ref{restric}) and~(\ref{homo}) for continuous 
subgroups $K \subset G$ and finite subgroups $H \subset G$, respectively,
are not necessarily onto. Hence, it is not 
guaranteed that all CS theories with continuous gauge group $K$ (or 
finite gauge group $H$)
can be obtained from spontaneously  broken CS theories
with gauge group $G$. 
Particularly, in section~\ref{multiple}, we will see that the natural 
homomorphism $H^4(B(U(1)^k), \Z) \to H^3(H,U(1))$ induced
by the symmetry breaking~(\ref{scheme}) is not onto.

\sectiona{Cohomology of finite abelian groups}
\label{fincoh}

In this section, I give a brief introduction to the 
cohomology groups $H^n({ H},U(1))$ of a finite abelian group $H$.
The plan is as follows. 
In subsection~\ref{reldef}, I recall  
the basic definitions and subsequently focus on the cocycle 
structure occurring in an abelian discrete $H$ CS theory. 
Finally, subsection~\ref{aa3c} contains the explicit realization
of all independent 3-cocycles $\omega \in H^3({ H},U(1))$
for an arbitrary abelian group ${ H}$.

\subsection{$H^n(H,U(1))$}
\label{reldef}

In the (multiplicative) 
algebraic description of the cohomology groups  $H^n(H,U(1))$ the 
$n$-cochains are represented as $U(1)$ valued functions 
\bea                \label{4}
c :  \; \underbrace{{ H} \times \cdots \times { H}}_{n\; 
\mbox{\scriptsize times}} 
&\longrightarrow& U(1) \, .
\eea
The set of all $n$-cochains forms the 
abelian group $C^n({ H},U(1)) := C^n$ 
with pointwise multiplication 
$
(c \cdot d) \, (A_1,\ldots,A_n) =   
c \, (A_1,\ldots,A_n)
\; d \, (A_1,\ldots,A_n),
$
where  the capitals $A_j$ (with $1 \leq j \leq n$) 
denote  elements of the finite group ${ H}$ and $c,d \in C^n$. 
The coboundary operator $\delta$ then establishes
a  mapping 
\beas \delta :   \;
 C^n & \longrightarrow & C^{n+1}  \\
           c & \longmapsto & \delta c \, , \nn
\eeas
given by 
\bea                     \label{coboundop}
\lefteqn{ \delta c \, (A_1, \ldots , A_{n+1}) \; := } \\
& & c \, (A_2,\ldots,A_{n+1})
\; c \, (A_1,\ldots  ,  A_n)^{(-1)^{n+1}} \; \prod_{i=1}^{n} 
c \, (A_1,\dots,A_i \cdot A_{i+1},\ldots,A_{n+1})^{(-1)^i}\, , \nn
\eea
which acts as  a derivation. That is, 
$\delta(c \cdot d) = \delta c \cdot \delta d$.
It can be checked explicitly that $\delta$ is indeed 
nilpotent: $\delta^2 =1$. 
The coboundary operator $\delta$ 
naturally defines two subgroups $Z^n$ and $B^n$
of $C^n$. Specifically, the subgroup $Z^n\subset C^n$
consists of $n$-cocycles  being the $n$-cochains $c$ in the kernel of $\delta$
\bea                                         
\label{ach}
\delta c &=& 1 \qquad \forall \,\, c \in Z^n \, ,
\eea
whereas the subgroup $B^n \subset Z^n \subset C^n$ 
contains the $n$-coboundaries or exact $n$-cocycles 
\bea                            \label{cobou}
c &=& \delta b  \qquad \forall \,\,c \in B^n \, .
\eea 
with $b$ some cochain $\in C^{n-1}$.  As usual, the cohomology group
$H^n({ H},U(1))$ is then defined as $
H^n({ H},U(1)) := Z^n/B^n$.
In other words, the elements of $H^n({ H},U(1))$ correspond 
to the different classes of 
$n$-cocycles~(\ref{ach}) with equivalence relation  
$c \sim c \delta b$.

The so-called slant product $i_A$ with arbitrary but fixed 
$A \in H$ is  a 
mapping in the opposite direction 
to the coboundary operator (e.g.\ \cite{spanier})
\beas
i_A : \; C^n & \longrightarrow & C^{n-1}\\
      c & \longmapsto & i_A c \, , \nn 
\eeas
defined as
\bea  \label{ig}
\lefteqn{ i_A c \, (A_1,\ldots,A_{n-1})  \; := } \\
& & c \, (A,A_1,\ldots  ,  
A_{n-1})^{(-1)^{n-1}}   \prod_{i=1}^{n-1} 
c \, (A_1,\ldots,A_i, A, A_{i+1},\ldots,A_{n-1})^{(-1)^{n-1+i}} \, .     
\nn 
\eea
It can be shown (e.g.\ \cite{spanier}) that 
the slant product satisfies the   relation 
$
\delta ( i_A  c ) =  i_A \, \delta c  
$
for all $n$-cochains $c$. Notably, if $c$ is a $n$-cocycle,
we immediately infer from this relation that  $i_A  c$ becomes 
a $(n-1)$-cocycle: 
$\delta (i_A  c) = i_A \, \delta c =  1$.
Hence, the slant product establishes a homomorphism 
$i_A: 
H^n({ H},U(1))
\rightarrow  H^{n-1}({ H},U(1))
$ for each $A \in H$.

Let us finally turn to the cocycle structure appearing  in 
an abelian discrete $H$ gauge theory with CS 
action $\omega \in H^3({ H},U(1))$. 
First of all, as indicated by~(\ref{coboundop}) and~(\ref{ach}), 
the 3-cocycle $\omega$ satisfies the relation
\bea
\label{pentagon}
\omega(A,B,C)\;\omega(A,B \cdot C,D)\;\omega(B,C,D) &=& 
\omega(A \cdot B,C,D)\;\omega(A,B,C \cdot D) \, ,   \qquad
\eea
for all $A,B,C \in H$.  To continue,
the slant product~(\ref{ig}) as applied to $\omega$
gives rise to a set of 2-cocycles $c_A \in H^2({ H},U(1))$
\bea     \label{c}
c_A (B,C) \; := \; i_A \omega(B,C) \; = \;
\frac{\omega(A,B,C)\;\omega(B,C,A)}{\omega(B,A,C)} \, ,
\eea
which are labeled by the different elements $A$ of $H$.
As will become clear in section~\ref{symalg}, 
these 2-cocycles enter the definition  of the projective dyon charge 
representations associated to the magnetic fluxes in this
abelian discrete $H$ CS gauge theory. 
To be specific, the different charges we can assign to a given 
abelian magnetic flux $A \in H$ to form dyons
are labeled by the inequivalent unitary irreducible projective 
representations $\alpha$ of $H$ defined as
\bea                \label{project} 
{\alpha}(B)\, \cdot \, {\alpha}(C) &=& c_A(B,C) \;  {\alpha}(B \cdot C) \, .
\eea
Here, the 2-cocycle relation satisfied by $c_A$ 
\bea
c_A (B,C) \; c_A (B \cdot C, D) &=& c_A (B,C\cdot D) \; c_A (C, D) \, ,
\label{tweeko} 
\eea
implies that the representations $\alpha$ are associative.
To conclude, as follows from~(\ref{coboundop}) and~(\ref{ach}),
the 1-cocycles obey the relation $c \, (B)\; c \, (C) = c \, (B \cdot C)$.
In other words, the different 1-cocycles being the elements of 
the cohomology group $H^1({ H},U(1))$ correspond to the inequivalent 
ordinary UIR's of the abelian group ${ H}$. These label the conceivable
{\em free} charges in a CS theory with finite abelian 
gauge group $H$.

\subsection{Chern-Simons actions for finite abelian groups}   
\label{aa3c}

An abstract group cohomological derivation 
(contained in appendix~\ref{gc}) reveals  
the following results for the first three cohomology 
groups of the finite abelian group $H$ being the direct product 
$\Z_N^k$ of $k$ cyclic groups $\Z_N$ of order $N$ 
\bea
H^1(\Z_N^k,U(1)) &\simeq& \Z_N^k   \label{conj1e}            \\
H^2(\Z_N^k,U(1)) &\simeq& \Z_N^{\frac{1}{2}k(k-1)} \label{conj2e}  \\
H^3(\Z_N^k,U(1)) 
&\simeq& \Z_N^{k+\frac{1}{2}k(k-1) +\frac{1}{3!}k(k-1)(k-2)} \, .   
\label{conj3e}
\eea
As we have seen in the previous subsection, the 
first result labels the inequivalent  UIR's of $\Z_N^k$, 
the second the different 2-cocycles entering the  
projective representations of $\Z_N^k$,
and the last the number of different 3-cocycles or 
CS actions for $\Z_N^k$.  
The derivation of the isomorphism~(\ref{conj3e}) in appendix~\ref{gc} 
pointed out that there are,
in fact, three dissimilar types of 3-cocycles.
The explicit realization of these 3-cocycles involves some 
notational conventions which I establish first.

Let $A,B$ and $C$ denote elements of $\Z_N^k$, i.e.\
\bea 
A &:=& (a^{(1)} , a^{(2)}, 
\ldots, a^{(k)}) \qquad \mbox{with 
$a^{(i)} \in \Z_N$   
for $i=1,\ldots, k$}\, ,
\eea
and similar decompositions for  $B$ and $C$. I adopt the additive 
presentation for the abelian group $\Z_N^k$, that is, 
the elements $a^{(i)}$ of $\Z_N$ take values in the range 
$ 0,\ldots, N-1$, and group multiplication 
is defined as
\bea
A \cdot B \; = \; [A+B] \; := \; ([a^{(1)}+b^{(1)}],  \ldots ,
[a^{(k)}+b^{(k)}]) \, .
\eea
Here, the rectangular brackets denote modulo $N$ 
calculus such that the sum always lies in the range $0, \ldots, N-1$.
With these conventions, the three  types of 3-cocycles
for the direct product group $\Z_N^k$ can then be presented as
\begin{eqnfourarray} 
\omega_{\rm I}^{(i)}(A,B,C) \!\!\!   &=& \!\!\!
\exp \left( \frac{2 \pi \im p^{(i)}_{\mbox{\scriptsize I}}}{N^2} 
a^{(i)}(b^{(i)} +c^{(i)} -[b^{(i)}+c^{(i)}]) \right)
 &  $\!\! 1 \leq i \leq  k$ \, \label{type1}    \\
\omega_{\mbox{\scriptsize II}}^{(ij)}(A,B,C) \!\!\!&=& \!\!\!
\exp \left( 
\frac{2 \pi \im p^{(ij)}_{\mbox{\scriptsize II}}}{N^2} 
a^{(i)}(b^{(j)} +c^{(j)} -
[b^{(j)}+c^{(j)}]) \right) &    $\!\! 1 \leq i < j \leq k$
\label{type2} \,  \\
\omega_{\mbox{\scriptsize III}}^{(ijl)}(A,B,C) \!\!\! &=& \!\!\!
\exp \left( \frac{2 \pi \im  
p^{(ijl)}_{\mbox{\scriptsize III}}}{N} 
a^{(i)}b^{(j)}c^{(l)} \right) &
 $\!\! 1 \leq i < j < l \leq k$ , \label{type3}
\end{eqnfourarray}
where the integral parameters 
$p_{\rm I}^{(i)}$, $p_{\mbox{\scriptsize II}}^{(ij)}$ and 
$p^{(ijl)}_{\mbox{\scriptsize III}}$  label the different elements of the 
cohomology group $H^3(\Z_N^k,U(1))$.
In accordance with~(\ref{conj3e}), the 3-cocycles are periodic functions
of these parameters with period $N$. 
For the 3-cocycles of type~III
this periodicity is obvious, while for the 3-cocycles 
of type~I and~II  it is immediate
after the observation that the factors  
$(b^{(i)} +c^{(i)} - [b^{(i)}+c^{(i)}])$, with $1\leq i \leq  k$,
either vanish or equal $N$.  It is also readily checked that the 
3-cocycles~(\ref{type1})--(\ref{type3}) indeed satisfy the 3-cocycle
relation~(\ref{pentagon}).

The $k$ different 3-cocycles~(\ref{type1}) of type~I  describe 
self-couplings, i.e.\ couplings between  the magnetic fluxes 
($a^{(i)}$,$b^{(i)}$ and $c^{(i)}$) associated to the same gauge 
group $\Z_N$ in the direct product $\Z_N^k$. 
In this counting procedure, it is, of course, 
understood that every 3-cocycle actually stands for a set of 
$N-1$ nontrivial 3-cocycles labeled by the periodic 
parameter $p_{\mbox{\scriptsize I}}^{(i)}$. 
The 3-cocycles~(\ref{type2}) of type~II, in turn,  establish pairwise 
couplings between the magnetic fluxes corresponding to
different gauge groups $\Z_N$  
in the direct product $\Z_N^k$.
Note that the 3-cocycles 
$\omega_{\mbox{\scriptsize II}}^{(ij)}$ and 
$\omega_{\mbox{\scriptsize II}}^{(ji)}$ are 
equivalent, since they just differ by a 3-coboundary~(\ref{cobou}).
In other words, there are only $\frac{1}{2}k(k-1)$ 
distinct 3-cocycles of type~II.
A similar argument holds for the 3-cocycles~(\ref{type3}) of type~III.
A permutation of the labels $i$, $j$ and $k$ in these 
3-cocycles yields an equivalent 3-cocycle. Hence, we end up with
$\frac{1}{3!}k(k-1)(k-2)$ different 3-cocycles of type~III, which
realize couplings between the fluxes associated to three 
distinct $\Z_N$ gauge groups in the direct product $\Z_N^k$.

We are now well prepared to discuss the 3-cocycle structure for
general abelian groups $H$ being direct 
products~(\ref{genab}) of cyclic groups possibly of 
different order. Let us assume that ${ H}$ consists of $k$ cyclic factors.
The abstract analysis in appendix~\ref{gc}  shows that depending 
on the divisibility of the orders of the different 
cyclic factors, there are again $k$ distinct 3-cocycles of type~I,
$\frac{1}{2}k(k-1)$ different 3-cocycles of type~II and 
$\frac{1}{3!}k(k-1)(k-2)$ different 3-cocycles of type~III.
It is easily verified that the associated generalization of the 3-cocycle 
realizations~(\ref{type1}),~(\ref{type2}) and~(\ref{type3})
becomes 
\bea
\omega_{\mbox{\scriptsize I}}^{(i)}(A,B,C)    &=& 
\exp \left( \frac{2 \pi \im p^{(i)}_{\mbox{\scriptsize I}}}{N^{(i)\;2}} \;
a^{(i)}(b^{(i)} +c^{(i)} -[b^{(i)}+c^{(i)}]) \right)  \label{type1do} \\
\omega_{\mbox{\scriptsize II}}^{(ij)}(A,B,C) &=&             
\exp \left( 
\frac{2 \pi \im p_{\mbox{\scriptsize II}}^{(ij)}}{N^{(i)}N^{(j)}}  \;
a^{(i)}(b^{(j)} +c^{(j)} - [b^{(j)}+c^{(j)}]) \right)  \label{type2do} \\
\omega_{\mbox{\scriptsize III}}^{(ijl)} (A,B,C) &=& \exp \left( \frac{2 \pi \im
p_{\mbox{\scriptsize III}}^{(ijl)}}{{\gcd}(N^{(i)}, N^{(j)},N^{(l)})} \;
a^{(i)}b^{(j)}c^{(l)} \right),            \label{type3do}
\eea
where $N^{(i)}$ (with $1\leq i \leq  k$) denotes the order 
of the $i^{\rm th}$
cyclic factor of the direct product group $H$. 
In accordance with the isomorphism~(\ref{conj1do}) of 
appendix~\ref{gc}, the 3-cocycles of type~III are cyclic in the 
integral parameter $p_{\mbox{\scriptsize III}}^{(ijl)}$ with period 
the greatest common divisor 
${\gcd}(N^{(i)}, N^{(j)},N^{(l)})$ of $N^{(i)}$, $N^{(j)}$ and 
$N^{(l)}$. The periodicity of the 3-cocycles of 
type~I coincides with the order $N^{(i)}$ of the associated cyclic factor of
$H$. Finally, the 3-cocycles of type~II  are  
periodic in the integral parameter $p_{\mbox{\scriptsize II}}^{(ij)}$ with
period the greatest common divisor  ${\gcd}(N^{(i)},N^{(j)})$ 
of $N^{(i)}$ and $N^{(j)}$.  This last periodicity  becomes clear upon
using the chinese remainder theorem 
\bea            \label{theorem}
\frac{\gcd(N^{(i)},N^{(j)})}{N^{(i)}N^{(j)}} &=& 
\frac{x}{N^{(i)}} + \frac{y}{N^{(j)}} \qquad\qquad \mbox{with $x,
y\in \Z$} \, ,
\eea
which indicates that~(\ref{type2do}) boils down to a 3-coboundary for
$p_{\mbox{\scriptsize II}}^{(ij)} = {\gcd}(N^{(i)},N^{(j)})$.

Let us finally focus on the 2-cocycles 
following from the three different types 
of 3-cocycles through the slant product~(\ref{c}). 
Upon substituting the expressions~(\ref{type1do}) 
and~(\ref{type2do}) in~(\ref{c}), we infer 
that the resulting 2-cocycles $c_A$ associated to the 3-cocycles of 
type~I and~II, respectively,  
correspond to the trivial element 
of the second cohomology group $H^2({ H}, U(1))$. To be precise,
these 2-cocycles are 2-coboundaries 
\be      \label{repphase}
c_A(B,C) \; = \; \delta \varepsilon_A (B,C) \; = \;
\frac{\varepsilon_A(B) \; \varepsilon_A(C)}{\varepsilon_A(B \cdot C)} \, ,
\ee
where the 1-cochains $\varepsilon_A$ of type~I and type~II read
\bea                                  \label{epi}
\varepsilon^{\mbox{\scriptsize I}}_{A}(B) &=& \exp 
\left( \frac{2 \pi \im p^{(i)}_{\mbox{\scriptsize I}}}
{N^{(i)\,2}} \; a^{(i)} b^{(i)} \right) \\
\varepsilon^{\mbox{\scriptsize II}}_{A}(B) &=& 
\exp \left(
\frac{2 \pi \im p^{(ij)}_{\mbox{\scriptsize II}}}{N^{(i)}N^{(j)}} 
\; a^{(i)} b^{(j)} \right). \label{epii}
\eea
Hence, the dyon charges in an abelian discrete $H$ gauge theory
endowed with a CS action of type~I and/or type~II  
correspond to trivial projective representations~(\ref{project}) of $H$
of the form $\alpha = \varepsilon_A \Gamma$, where $\Gamma$ 
denotes an ordinary UIR of $H$.
In contrast, the 2-cocycles $c_A$ obtained from the 
3-cocycles~(\ref{type3do}) of type~III
correspond to nontrivial elements
of the  cohomology group $H^2({ H}, U(1))$. 
The conclusion is that the dyon charges featuring in
an abelian discrete $H$ gauge theory
with a CS action of type~III are 
nontrivial (i.e.\ higher dimensional) 
projective representations of $H$.

\sectiona{Chern-Simons actions for $U(1)^k$ gauge theories}      
\label{multiple}

This section is concerned with the classification of the 
CS actions for the compact gauge group $U(1)^k$.
In addition, it is established which CS theories with finite abelian
gauge group $H$ may result from a spontaneous breakdown
of the corresponding $U(1)^k$ CS theories.

As mentioned in the introduction, the most general CS action for a planar  
$U(1)^k$ gauge theory is of the form~\cite{hagen}
\bea    \label{act}
S_{\mbox{\scriptsize CS}}&=& 
\int d\,^3x \;( {\cal L}_{\mbox{\scriptsize CSI}} + 
{\cal L}_{\mbox{\scriptsize CSII}}) 
\\
{\cal L}_{\mbox{\scriptsize CSI}} &=& 
\sum_{i=1}^k  \; \frac{\mu^{(i)}}{2} \epsilon^{\kappa\sigma\rho}
                              A^{(i)}_{\kappa} \partial_{\sigma} 
                              A^{(i)}_{\rho} \label{CSt1}  \\
{\cal L}_{\mbox{\scriptsize CSII}} &=&
\sum_{i<j=1}^k  \frac{\mu^{(ij)}}{2} \epsilon^{\kappa\sigma\rho}
                              A^{(i)}_{\kappa} \partial_{\sigma} 
                              A^{(j)}_{\rho} \, ,     \label{CSt2}
\eea
where $A^{(i)}_\kappa$ (with $i=1,\dots,k$)
denote the  various $U(1)$  gauge fields, $\mu^{(i)}$,
$\mu^{(ij)}$ the topological masses and $\epsilon^{\kappa\sigma\rho}$ 
the three dimensional anti-symmetric Levi-Civita tensor normalized
such that $\epsilon^{012}=1$. Hence,
there are  $k$ distinct CS terms~(\ref{CSt1}) describing 
self couplings of the  $U(1)$ gauge fields.
In analogy with the terminology developed in the previous section,
we will call these terms CS terms of type~I.
In addition, there are  $\frac{1}{2}k(k-1)$ 
distinct CS terms of type~II 
establishing  pairwise couplings between different $U(1)$ gauge fields. 
Note that by a partial integration a 
term labeled by $(ij)$ becomes a term $(ji)$. 
Therefore, these  terms are equivalent and should not be counted separately.
Also note that up to a total 
derivative the CS terms of type~I and type~II are 
indeed invariant under $U(1)^k$ gauge transformations
$ A^{(i)}_{\rho}  \rightarrow A^{(i)}_{\rho} - 
\partial_{\rho} \Omega^{(i)} $ with $i=1,\ldots,k$,
while the requirement of abelian gauge invariance immediately 
rules out `CS terms of type~III' $
\sum_{i<j<l=1}^k \frac{\mu^{(ijl)}}{2}
\epsilon^{\kappa\sigma\rho}   A^{(i)}_{\kappa} A^{(j)}_{\sigma}
                              A^{(l)}_{\rho} \, , $
which would establish a coupling  between three 
different $U(1)$ gauge fields.

Let us now assume that this abelian 
gauge theory is compact and features a family 
of Dirac monopoles~\cite{dirac}
for each compact $U(1)$ gauge group. That is, 
the complete spectrum of Dirac monopoles consists of 
the magnetic charges 
$g^{(i)} = \frac{2\pi m^{(i)}}{e^{(i)}}$ with $m^{(i)} \in \Z$, 
$1 \leq i \leq k$ and $e^{(i)}$  
the fundamental charge associated with the  compact
$U(1)$ gauge  group being the $i^{\rm th}$ factor 
in the direct product $U(1)^k$. 
In this 2+1 dimensional Minkowsky setting, 
these monopoles are, of course,  
instantons tunneling between states with flux difference
$\Delta \phi^{(i)}= \frac{2\pi m^{(i)}}{e^{(i)}}$. 
A consistent implementation of these monopoles/instantons requires  
that the topological masses in~(\ref{CSt1}) and~(\ref{CSt2}) are 
quantized as
\bea                                  
\mu^{(i)} &=& \frac{p^{(i)}_{\mbox{\scriptsize I}} e^{(i)}e^{(i)}}{\pi} 
\;\qquad \qquad 
\mbox{with $p^{(i)}_{\mbox{\scriptsize I}} \in {\mbox{\bf Z}}$}       
\label{quantmui}   \\
\mu^{(ij)} &=&\frac{p^{(ij)}_{\mbox{\scriptsize II}} e^{(i)}e^{(j)}}{\pi}
\qquad \qquad
\mbox{with $p^{(ij)}_{\mbox{\scriptsize II}} \in {\mbox{\bf Z}}$} \, .
\label{quantmuij}  
\eea
This will be shown  in sections~\ref{rev} and~\ref{revii},
where we will discuss these models in further detail.
The integral CS parameters 
$p^{(i)}_{\mbox{\scriptsize I}}$ 
and $p^{(ij)}_{\mbox{\scriptsize II}}$ now  label the different 
elements of the cohomology group 
\bea                    \label{u1k}
H^4(B(U(1)^k), \Z) &\simeq& \Z^{ k + \frac{1}{2}k(k-1)} \, ,
\eea
where a derivation of the isomorphism~(\ref{u1k}) 
is contained in appendix~\ref{gc}.

We now have all the ingredients to make explicit the 
homomorphism~(\ref{homo}) accompanying the spontaneous symmetry breakdown 
of the gauge group $U(1)^k$ to the finite abelian group 
$H \simeq \Z_{N^{(1)}} \times \cdots \times
\Z_{N^{(k)}}$.
In terms of the integral CS parameters 
in~(\ref{quantmui}) and~(\ref{quantmuij}), it takes the form
\bea
H^4(B(U(1)^k), \Z) 
&\longrightarrow & 
H^3({ H}, U(1)) 
\label{homo1en2} \\
p^{(i)}_{\mbox{\scriptsize I}} & 
\longmapsto & p^{(i)}_{\mbox{\scriptsize I}} \qquad \; \bmod  N^{(i)} 
\label{homoI} \\
p^{(ij)}_{\mbox{\scriptsize II}} & \longmapsto & 
p^{(ij)}_{\mbox{\scriptsize II}} \qquad \bmod  
\gcd(N^{(i)}, N^{(j)}) \, ,     \label{homoII}
\eea
where the periodic parameters being the images of this mapping label the 
different 3-cocycles~(\ref{type1do}) and~(\ref{type2do}) of type~I and 
type~II. The conclusion is that
the long distance physics of a spontaneously broken 
$U(1)^k$ CS theory of type~I/II is described by 
a CS theory of type~I/II with the residual finite 
abelian gauge group $H$.
We will illustrate this result with two representative examples 
in sections~\ref{typeI} and~\ref{typeII}. 
As a last obvious remark, from~(\ref{homo1en2}) we also learn that
abelian discrete $H$ gauge theories with 
a CS action of type~III can not be obtained from a spontaneously 
broken $U(1)^k$ CS theory.

\sectiona{Quasi-quantum doubles}
\label{symalg}

There are deep connections between two dimensional rational conformal field 
theory, three dimensional topological field theory and quantum groups or 
Hopf algebras, e.g.\ \cite{witten,alvarez1} and references therein. 
Planar discrete $H$ gauge theories, being examples of 
three dimensional topological field theories, naturally fit in this 
general scheme. In~\cite{spm}, see also reference~\cite{banff}, the Hopf 
algebra related to the discrete $H$ gauge theory  describing the long 
distance physics of the spontaneously broken model~(\ref{alg}) without 
CS term has been identified as the quasitriangular Hopf algebra  $D(H)$   
being the result~\cite{dpr} of applying Drinfeld's quantum double 
construction~\cite{drin} to the abelian algebra ${\cal F}(H)$ 
of functions on the finite group ${ H}$. 
Following reference~\cite{banff}, we will simply refer to the 
Hopf algebra $D(H)$ as the quantum double.
To proceed, according to the discussion
of section~\ref{bgt}, in the presence of a nontrivial 
CS term $S_{\mbox{\scriptsize CS}} \in H^4 (BG, \Z)$ 
for the broken gauge group $G$ in the action~(\ref{alg}), 
the long distance physics of the model is described by a discrete $H$ CS 
theory with 3-cocycle $\omega \in H^3 ({ H}, U(1))$ determined by 
the natural homomorphism~(\ref{homo}). 
As has been pointed in~\cite{spm1}, see also the references~\cite{sm,thesis},
the related Hopf algebra now becomes the so-called quasi-quantum double
$\DW$ being a natural deformation of $D(H)$ depending on the 
3-cocycle $\omega \in H^3 ({ H}, U(1))$. 

To put the results outlined in the previous paragraph in historical 
perspective, the quantum double $D({ H})$ and the corresponding 
quasi-quantum doubles $\DW$ were first proposed by Dijkgraaf, Pasquier 
and Roche~\cite{dpr}. They identified these as the Hopf algebras associated 
with certain two dimensional holomorphic orbifolds of rational conformal field 
theories~\cite{dvvv} and the related three dimensional topological 
field theories with finite gauge group $H$ introduced by Dijkgraaf 
and Witten~\cite{diwi}. One of the essentially new observations in the 
references~\cite{banff,spm,spm1,sm,thesis} in this respect 
was that such a topological field 
theory finds a natural realization as the residual discrete $H$ (CS) gauge 
theory describing the long range physics of (CS) gauge theories~(\ref{alg}) 
in which some continuous gauge group $G$ is spontaneously 
broken down to a finite subgroup $H$.

In this section, I recall the basic features of the quasi-quantum double 
$D^\omega ({ H})$ for  abelian finite groups $H$ and subsequently 
elaborate on the unified description this algebraic framework gives 
of the spin, braid and fusion properties of the particles in the spectrum 
of a discrete $H$ gauge theory with CS action $\omega \in H^3 ({ H}, U(1))$.
For a general study of quasi-Hopf algebras, the interested reader
is referred to the original papers by Drinfeld~\cite{drin} 
and the excellent book by Shnider and Sternberg~\cite{shnider}.

\subsection{$\DW$ for abelian $H$}

The quasi-quantum double $D^\omega ({ H})$ 
for an abelian finite group $H$ is spanned by the basis 
elements~\footnote{In this paper, I cling to the notation set 
in the discussion of the quantum double $D(H)$ in reference~\cite{banff}.} 
$\{ {\mbox{P}}_A \, B \}_{A,B\in { H}} $
representing a global symmetry transformation $B \in H$ followed
by the operator ${\mbox{P}}_A$ projecting 
out the magnetic flux $A\in H$.
The deformation of the quantum double $D({ H})$ into the 
{\em quasi}-quantum double $D^\omega ({ H})$ amounts to
relaxing the coassociativity condition for the comultiplication. 
That is, the comultiplication
$\Delta$ for  $D^\omega ({ H})$ now satisfies the
{\em quasi}-coassociativity condition~\cite{dpr}
\bea \label{quasicoas}
({\mbox{id}} \ot \Delta) \, \Delta( \, {\mbox{P}}_A \, B \, ) &=& 
\varphi\cdot (\Delta \ot {\mbox{id}}) \, \Delta( \, {\mbox{P}}_A B \, ) 
\cdot\varphi^{-1} \, ,
\eea
with the invertible associator $\varphi \in D^\omega ({ H})^{\ot 3}$ 
defined in terms of the 3-cocycle $\omega$ for $H$ as
\bea                  
\varphi &:=& \sum_{A,B,C}\,\omega^{-1}(A,B,C) \;
{\mbox{P}}_A \otimes {\mbox{P}}_B \otimes {\mbox{P}}_{C} \, .  \label{isom}  
\eea
The multiplication and  comultiplication are  deformed accordingly
\bea
{\mbox{P}}_A \, B \cdot {\mbox{P}}_D \, C &=& 
\delta_{A,D} \;\; {\mbox{P}}_A  \, B \cdot C 
\;\; c_A(B,C) \label{algebra}       \\
 \Delta(\,{\mbox{P}}_A \, B \,) &=& 
\sum_{C\cdot D=A} \; {\mbox{P}}_C \, B \ot {\mbox{P}}_D \, B       \;\;
c_B(C,D) \, , \label{coalgebra}
\eea
where $c$ denotes the 2-cocycle obtained from $\omega$  through  the 
slant product~(\ref{c}) and $\delta_{A,B}$ the Kronecker delta function
for the group elements of $H$. 
The 2-cocycle relation~(\ref{tweeko}) satisfied by $c$ implies that  the 
multiplication~(\ref{algebra}) is associative and, in addition, 
that the comultiplication~(\ref{coalgebra}) is indeed
quasi-coassociative~(\ref{quasicoas}).
By repeated use of the 3-cocycle relation~(\ref{pentagon})
for $\omega$, one also easily verifies the relation
\bea
c_A (C,D) \; c_B (C,D) \; c_C (A,B) \; c_D (A,B)  &=& 
c_{A \cdot B} (C,D) \;  c_{C \cdot D} (A,B) \, ,     \label{alamos}
\eea
which indicates that the comultiplication~(\ref{coalgebra}) defines 
an algebra morphism from  $D^{\omega}({ H})$ to 
$D^{\omega}({ H})^{\ot 2}$.

As mentioned before, the particles in the associated 
discrete $H$ gauge theory with CS action $\omega$
are labeled by a magnetic flux $A \in H$ paired with a projective  
UIR $\alpha$ of $H$ defined as~(\ref{project}).
Thus the spectrum can be presented as
\bea         \label{repo}
( \, A, {\alpha} \,) \, ,
\eea
where $A$ runs over the different elements of $H$ 
and $\alpha$ over the range of inequivalent projective 
UIR's~(\ref{project}) of $H$ associated with the 2-cocycle 
$c_A$ given in~(\ref{c}).
The spectrum~(\ref{repo}) constitutes the complete set 
of inequivalent irreducible representations of the quasi-quantum 
double $D^{\omega}({ H})$.   The internal Hilbert 
space $V_{\alpha}^A$ assigned to a given  particle 
$( \, A, {\alpha} \,)$  is spanned by the states
\bea \label{intqstat}
 \{|\, A,\,^{\alpha} v_j\rangle\}_{j=1,\ldots,d_\alpha} \, , 
\eea
with $^{\alpha}\!v_j$ a basis vector and $d_\alpha$ the dimension of 
the representation space associated with $\alpha$.
The irreducible representation $\Pi^A_{\alpha}$ of $D^{\omega}({ H})$
carried by $V_{\alpha}^A$ is then given by~\cite{dpr} 
\bea \label{13}                                              
\Pi^A_{\alpha}(\, {\mbox{P}}_B \, C \,) \; |\,A ,\,^{\alpha} v_j \rangle &=&
\delta_{A,B}\;\, |\,A,\,\alpha(C)_{ij}\,^{\alpha} v_i \rangle \, .
\eea
So, the global symmetry transformations $C\in H$ affect
the projective dyon charge $\alpha$ and 
leave the abelian magnetic flux $A$ invariant. 
The projection operator ${\mbox{P}}_B$
subsequently projects out the flux $B \in H$.
Note that although the dyon charges $\alpha$ are projective 
representations of ${ H}$,
the action~(\ref{13}) defines an ordinary representation 
of the quasi-quantum double: 
$\Pi^A_{\alpha}(\, {\mbox{P}}_B \, C \,) \cdot 
\Pi^A_{\alpha}(\, {\mbox{P}}_D \, E\,) \; = \; 
\Pi^A_{\alpha}( \, {\mbox{P}}_B \, C \cdot {\mbox{P}}_D \, E \,)$.

As follows from the discussion in section~\ref{aa3c},
we may now  distinguish two cases.
Depending on the actual 3-cocycle $\omega$ at hand, 
the 2-cocycle $c_A$ obtained from the slant product~(\ref{c})
is either trivial or nontrivial.  
When $c_A$ is trivial, it can be written as the coboundary~(\ref{repphase}) 
of a 1-cochain or phase factor $\varepsilon_A$. 
This situation occurs for the 2-cocycles $c_A$
related to the 3-cocycles~(\ref{type1do}) of type~I, the 
3-cocycles~(\ref{type2do}) of type~II and products thereof.
From the relations~(\ref{project}) and~(\ref{repphase}),
we obtain that the inequivalent (trivial) 
projective dyon charge representations for this case are of the form 
\bea                   
{\alpha}(C) &=&             \varepsilon_A(C) \;  \;
 \Gamma^{n^{(1)} \cdots \;  n^{(k)}} (C) \, ,    \label{rei}      
\eea
where $\Gamma^{n^{(1)}\cdots \; n^{(k)}}$ denotes an ordinary 
(1-dimensional) UIR of  ${ H}$
\bea              \label{hrepz}
\Gamma^{n^{(1)} \cdots \; n^{(k)}} (C) 
&=&   \exp \left(\sum_{l=1}^k \frac{2 \pi \im}{N^{(l)}} \, n^{(l)} c^{(l)}
\right) \, .
\eea 
For a 3-cocycle of type~I, the epsilon factor appearing
in the  dyon charge representation~(\ref{rei}) 
is given by~(\ref{epi}), while a 3-cocycle of type~II leads to the 
factor~(\ref{epii}). If we are dealing with a 3-cocycle $\omega$ being 
a product of various 3-cocycles of type~I and~II, then the total
epsilon factor naturally becomes the product of the epsilon factors 
related to the 3-cocycles of type~I and~II constituting the total 
3-cocycle~$\omega$.
The 2-cocycles $c_A$ associated to the 3-cocycles~(\ref{type3do}) 
of type~III, in contrast, are nontrivial. 
As a consequence, the dyon charges correspond to  
nontrivial  higher dimensional irreducible projective 
representations of ${ H}$ when the total 3-cocycle $\omega$ 
contains a factor of type~III.

\begin{figure}[htb]    \epsfxsize=8.6cm
\centerline{\epsffile{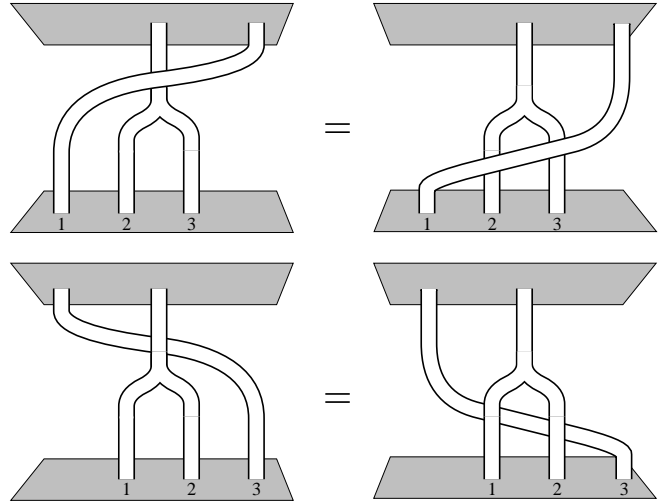}}
\caption{\sl  Compatibility of fusion
and braiding as expressed by the quasitriangularity conditions.
It  makes no difference whether a third particle  braids with two 
particles separately or with the composite that arises after fusing these two 
particles. The ribbons represent the trajectories of the particles.}
\label{qu1zon}
\end{figure}

There is a spin assigned to the particles~(\ref{repo}). 
In a counterclockwise rotation 
over an angle of $2 \pi$, the dyon charge $\alpha$ 
of the particle $(A,\alpha)$
is transported around the flux $A$ and as a result of the AB effect
picks up a global transformation $\alpha(A)$ by this flux.~\footnote{ 
Of course, a small separation between the dyon charge $\alpha$
and the flux $A$ is required for this interpretation.}  
The element of $\DW$ that implements this effect 
is the central element $\sum_B \; {\mbox{P}}_B \, B$.
It signals the flux of a given quantum state~(\ref{intqstat}) and implements 
this flux on the dyon charge:
\bea \label{spin13}                                              
\Pi^A_{\alpha}(\, \sum_B \; {\mbox{P}}_B \, B \,) \; 
|\,A ,\,^{\alpha} v_j \rangle &=&
 |\,A,\,\alpha(A)_{ij}\,^{\alpha} v_i \rangle \, .
\eea   
Upon using~(\ref{project}) and subsequently~(\ref{c}), we infer that 
the matrix $\alpha(A)$ commutes with all other matrices appearing in 
the projective UIR $\alpha$ of ${ H}$
\bea
\alpha(A) \, \cdot \, \alpha(B) 
\;= \; \frac{c_A(A,B)}{c_A(B,A)} \;  \alpha(B) \, \cdot \, \alpha(A)
\; = \; \alpha(B) \, \cdot \, \alpha(A)    \qquad \forall B \in { H}.
\eea 
From Schur's lemma, we then conclude that $\alpha (A)$ is proportional 
to the unit matrix in this irreducible projective  representation of $H$
\bea                           \label{anp}
\alpha(A) &=& e^{2 \pi \im s_{(A,\alpha)}} \; {\mbox{\bf 1}}_\alpha \, ,
\eea
where $s_{(A,\alpha)}$ denotes the spin carried by the particle 
$( \, A, {\alpha} \,)$ and ${\mbox{\bf 1}}_\alpha$ the unit matrix. 
Relation~(\ref{anp}), in particular, 
reveals the physical relevance of the epsilon factors
entering the definition~(\ref{rei}) of the dyon charges 
in the presence of CS actions of type~I and/or type~II.
Under a counterclockwise rotation over an angle of $2\pi$,
they give rise to an additional spin factor $\varepsilon_A(A)$ 
in the internal quantum state describing a particle carrying the magnetic 
flux $A$. To keep track of the writhing of the trajectories  
of the particles and the associated nontrivial spin factors~(\ref{anp}), 
the particle trajectories are depicted by ribbons instead of lines in the 
following. See, for instance, figure~\ref{qu1zon}.

The action~(\ref{13}) of the quasi-quantum double $D^\omega(H)$ 
is extended to two-particle states by means of the 
comultiplication~(\ref{coalgebra}).
Specifically, the tensor product 
representation $(\Pi^A_{\alpha} \ot \Pi^B_{\beta}, V_{\alpha}^A \ot 
V_{\beta}^B)$ of $D^\omega(H)$ associated to
a system consisting of the two particles $(\,A, \alpha \,)$ 
and $(\,B, \beta \,)$ is defined by the action
$\Pi^A_{\alpha} \ot \Pi^B_{\beta}( \Delta (\, {\mbox{P}}_A \, B \, ))$. 
The tensor product representation of the quasi-quantum double related to
a system of three particles $(\,A, \alpha \,)$, $(\,B, \beta \,)$ and 
$(\,C, \gamma \,)$ may now be defined
either through $(\Delta \ot {\mbox{id}} ) \, \Delta$
or through $({\mbox{id}} \ot \Delta) \, \Delta$.
Let $(V_{\alpha}^A \ot V_{\beta}^B) \ot V_{\gamma}^C$ denote 
the representation space corresponding to 
$(\Delta \ot {\mbox{id}} ) \, \Delta$
and $V_{\alpha}^A \ot (V_{\beta}^B \ot V_{\gamma}^C)$ the one 
corresponding to $({\mbox{id}} \ot \Delta) \, \Delta$. 
The quasi-coassociativity condition~(\ref{quasicoas}) indicates that
these representations are equivalent. To be precise, their equivalence 
is established by the nontrivial isomorphism or intertwiner
\bea                              \label{fi}     
\Phi : \;  (V_{\alpha}^A \ot V_{\beta}^B) \ot V_{\gamma}^C 
&\longrightarrow&  V_{\alpha}^A \ot (V_{\beta}^B \ot V_{\gamma}^C) \, ,
\eea
with  $   
\Phi   :=  \Pi_{\alpha}^A \! \ot \! \Pi_{\beta}^B \! \ot 
\Pi_{\gamma}^C \,(\varphi)  =   \omega^{-1} (A,B,C)$,
where I used relation~(\ref{isom}) in the last equality sign.
Finally, the 3-cocycle relation~(\ref{pentagon}) 
implies  consistency in rearranging the brackets,
i.e.\ commutativity of the following pentagonal diagram~\footnote{Here, 
we momentarily use the abbreviation $V_A:= V_{\alpha}^A$.}
\[
\ba{ccccc}
\! ((V_A \ot V_B) \ot V_C)  \ot  V_D   \! \! \!
&\stackrel{\scriptscriptstyle{\Phi \ot {\mbox{\tiny \bf 1}}} }{\rightarrow} 
& \! \! \!
(V_A \ot (V_B \ot V_C)) \ot V_D
\!  \! \! &
\stackrel{\scriptscriptstyle{
({\mbox{\tiny id}} \ot \Delta \ot {\mbox{\tiny id}})(\Phi)}}{\longrightarrow} 
& \! \!  \! \!
V_A \ot ((V_B \ot V_C) \ot V_D)
\\
           & & & &            \\
\downarrow {\scriptstyle (\Delta \ot  {\mbox{\scriptsize id}} 
\ot {\mbox{\scriptsize id}})(\Phi)}
& & & & 
\downarrow {\scriptstyle {\mbox{\scriptsize \bf 1}} \ot \Phi}           \\
           & & & &            \\
\! (V_A \ot V_B) \ot (V_C \ot V_D) 
& &\stackrel{({\mbox{\scriptsize id}} \ot 
{\mbox{\scriptsize id}} \ot \Delta)(\Phi)}{\longrightarrow} & &
\! \! \!  V_A \ot (V_B \ot (V_C \ot V_D)) \, .
\ea
\]

The braid operation is implemented by 
the so-called universal $R$-matrix being the element 
$R  =   \sum_{C,D} \; {\mbox{P}}_C \ot {\mbox{P}}_D \, C $
of $D^\omega(H)^{\ot 2}$ 
which acts on a given two-particle state as a global symmetry transformation
on the second particle by the flux of the first particle.
The physical braid operator ${\cal R}$ establishing  
a counterclockwise interchange of two particles $(\, A, \alpha \,)$
and $(\, B, \beta \,)$ is then 
defined as the action of this $R$-matrix followed by a permutation 
$\sigma$ of the two particles, i.e.\  
\bea \label{jordicr}
{\cal R}_{\alpha\beta}^{AB} &:=& 
\sigma\circ(\Pi_{\alpha}^A\otimes\Pi_{\beta}^B)( \, R \, ) \, .
\eea  
So, on the two-particle 
internal Hilbert space $V_{\alpha}^A \ot V_{\beta}^B$, 
the braid operator ${\cal R}$ acts as
\bea                 \label{braidaction}
{\cal R} \;| \, A,\, ^{\alpha} v_j\rangle
|\, B,\,^{\beta} v_l\rangle &=& 
|\,B,\,{\beta}(A)_{ml} \, ^{\beta} v_m\rangle 
|\,A,\, ^{\alpha} v_j\rangle \, .
\eea   
From~(\ref{rei}) and~(\ref{braidaction}), we then learn  
that the particles in an abelian 
discrete $H$ gauge theories endowed with a CS action of type~I and/or type~II  
obey abelian braid statistics. That is, the effect of braiding two particles
in these theories is just an AB phase in the (scalar) internal wave function, 
where on top of the conventional AB phase 
$\Gamma^{n^{(1)} \cdots \; n^{(k)}} (A)$ for a global $H$ charge
$n^{(1)} \cdots \; n^{(k)}$
and a magnetic flux $B$ the epsilon 
factors~(\ref{epi}) and~(\ref{epii}) represent  
additional AB phases generated between the magnetic fluxes.
This picture changes drastically in the presence of a CS action 
of type~III.  In that case, the expression~(\ref{braidaction})
indicates that the higher dimensional internal charge 
of a particle $( \, B,\beta \,)$ picks up an AB {\em matrix} $\beta(A)$ upon 
encircling another  remote
particle $(\, A, \alpha \,)$ in a counterclockwise fashion. In the same
process, the particle $( \, A,\alpha \,)$ picks up the AB 
{\em matrix} $\alpha(B)$. Thus, the introduction of a CS action
of type~III in an abelian discrete gauge theory
leads to {\em non}abelian phenomena.  In particular, 
the multi-particle configurations in such a theory generally  
realize nonabelian braid statistics.

Relation~(\ref{c}) implies that the comultiplication~(\ref{coalgebra}),
the associator~(\ref{fi}) and the braid operator~(\ref{jordicr})
satisfy the so-called quasitriangularity conditions:
\bea
{\cal R}\: \Delta(\, {\mbox{P}}_A\, B\,)           \label{grcom}
&=& \Delta(\, {\mbox{P}}_A\, B\,) \:{\cal R}  \\
({\mbox{id}} \ot \Delta)({\cal R}) &=& 
\Phi^{-1} \; {\cal R}_2 \; \Phi \; {\cal R}_1 \;
\Phi^{-1}
\label{tria2}       \\
(\Delta \ot {\mbox{id}})({\cal R}) &=& \Phi \; {\cal R}_1 \; \Phi^{-1} \; 
{\cal R}_{2} \; \Phi \, .
\label{tria1}
\eea
Here, the braid operator ${\cal R}_1$ 
acts as ${\cal R} \ot {\mbox{\bf 1}}$ on the three particle
internal Hilbert space $(V_{\alpha}^A \ot V_{\beta}^B) \ot V_{\gamma}^C$ and 
${\cal R}_2$ as  ${\mbox{\bf 1}} \ot {\cal R}$ on 
$V_{\alpha}^A \ot (V_{\beta}^B \ot V_{\gamma}^C)$.
The condition~(\ref{grcom}) obviously states that the action of  
$\DW$ on a two-particle internal Hilbert space 
commutes with the braid operation. The conditions~(\ref{tria2}) 
and~(\ref{tria1}), in turn, indicate that the following hexagonal 
diagrams commute
\beas
\ba{ccccc} 
V^A_\alpha \ot (V^B_\beta \ot V^C_\gamma)
& 
\stackrel{\Phi^{-1}}{\rightarrow}
& 
(V^A_\alpha \ot V^B_\beta) \ot V^C_\gamma   
& 
\stackrel{{\cal R}_1}{\rightarrow}
&
(V^B_\beta \ot V^A_\alpha ) \ot V^C_\gamma  \\
\downarrow {\scriptstyle ({\mbox{\scriptsize id}} 
\ot \Delta) ({\cal R})} & & & & \downarrow 
{\scriptstyle \Phi } \\
(V^B_\beta \ot V^C_\gamma) \ot V^A_\alpha & 
\stackrel{\Phi^{-1}}{\leftarrow}
& 
V^B_\beta \ot (V^C_\gamma \ot V^A_\alpha)  & 
\stackrel{{\cal R}_2}{\leftarrow}
& V^B_\beta \ot (V^A_\alpha  \ot V^C_\gamma)       \\
& & & & \\
(V^A_\alpha  \ot V^B_\beta) \ot V^C_\gamma       
& 
\stackrel{\Phi}{\rightarrow}   
& 
V^A_\alpha \ot (V^B_\beta \ot V^C_\gamma)            
& 
\stackrel{{\cal R}_2}{\rightarrow}    
&
V^A_\alpha \ot (V^C_\gamma \ot V^B_\beta)     \\
\downarrow {\scriptstyle (\Delta \ot {\mbox{\scriptsize id}})({\cal R})} 
& & & & 
\downarrow {\scriptstyle \Phi^{-1}}  \\
V^C_\gamma \ot (V^A_\alpha  \ot V^B_\beta)         & 
\stackrel{\Phi}{\leftarrow}
& 
(V^C_\gamma \ot V^A_\alpha ) \ot V^B_\beta      & 
\stackrel{{\cal R}_1}{\leftarrow}
&   (V^A_\alpha  \ot V^C_\gamma) \ot V^B_\beta \, .
\ea
\eeas    
In other words, these conditions express the compatibility of braiding 
and fusion as depicted in figure~\ref{qu1zon}.

Due to the finite order of the braid operator, 
multi-particle systems in  planar discrete gauge theories 
without CS action realize representations of factor 
groups~\footnote{The definition of these so-called truncated braid 
groups can be found in appendix~\ref{trubra}.} of 
the well-known braid groups~\cite{banff,thesis}.
This property persists if one adds a CS action 
$\omega \in H^3({ H}, U(1))$ to an abelian 
discrete $H$ gauge theory (or a nonabelian one for that matter).
However, from the quasitriangularity conditions~(\ref{grcom})--(\ref{tria1}), 
we  infer that instead of the ordinary Yang-Baxter equation,
the braid operators now satisfy the quasi-Yang-Baxter equation
\bea       \label{qujaba}
{\cal R}_1 \; \Phi^{-1} \, {\cal R}_2 \, \Phi \; {\cal R}_1 &=&
\Phi^{-1} \, {\cal R}_2 \, \Phi \; {\cal R}_1 \; 
\Phi^{-1} \, {\cal R}_2 \, \Phi \, .
\eea  
Hence, the truncated braid group representations 
realized by the multi-particle systems in abelian 
discrete CS gauge theories in principle 
involve the associator~(\ref{isom}), which 
takes care of the rearrangement of brackets.
Let $(((V_{\alpha_1}^{A_1} \ot V_{\alpha_2}^{A_2})\ot \cdots 
\ot V_{\alpha_{n-1}}^{A_{n-1}}) \ot V_{\alpha_n}^{A_n})$ denote 
an internal Hilbert space for a system of $n$ particles.
Thus, all left brackets occur at the beginning. 
Depending on whether we are dealing with a system of identical particles,
distinguishable particles, or a system consisting of different subsystems
of identical particles, the associated $n$-particle 
internal Hilbert space 
$(((V_{\alpha_1}^{A_1} \ot V_{\alpha_2}^{A_2})\ot \cdots 
\ot V_{\alpha_{n-1}}^{A_{n-1}}) \ot V_{\alpha_n}^{A_n})$
then carries an unitary representation of an ordinary truncated braid group, 
a colored truncated braid group or a partially colored truncated braid 
group on $n$ strands respectively. This representation is defined by the 
formal assignment~\cite{altsc1} 
\bea                                           \label{brareco}
\tau_i & \longmapsto& \Phi_i^{-1}  \; {\cal R}_i \;  \Phi_i   \, ,
\eea
with  $1 \leq i \leq n-1$ and 
\bea 
{\cal R}_i  &:= &
{\mbox{\bf 1}}^{\ot (i-1)} \ot {\cal R} \ot {\mbox{\bf 1}}^{\ot (n-i-1)}   \\
\Phi_i & :=& 
\left(\bigotimes_{i=1}^n \Pi_{\alpha_i}^{A_i}\right) 
\left( \Delta_L^{i-2} (\varphi) \ot 1^{\ot (n-i-1)} \right) \, .    
\label{intermi}
\eea
Here, $\varphi$ is the associator~(\ref{isom}), whereas the object 
$\Delta_L$  stands for  the mapping 
\beas
\Delta_L (\, {\mbox{P}}_{C_1} \, D_1  \ot {\mbox{P}}_{C_2} \, D_2
\ot \cdots \ot {\mbox{P}}_{C_m} \, D_m\,) 
& :=& \Delta(\, {\mbox{P}}_{C_1} \, D_1 \, ) \ot {\mbox{P}}_{C_2} \, D_2 \ot 
\cdots \ot {\mbox{P}}_{C_m} \, D_m\, ,
\eeas 
from $D^{\omega}({ H})^{\ot m}$ to $D^{\omega}({ H})^{\ot (m+1)}$ and
$\Delta_L^k$ for the associated mapping from 
$D^{\omega}({ H})^{\ot m}$ to $D^{\omega}({ H})^{\ot (m+k)}$ being the
result of applying $\Delta_L$ $k$ times.
The isomorphism~(\ref{intermi}) 
now  parenthesizes the adjacent internal Hilbert spaces
$V_{\alpha_i}^{A_i}$  and $V_{\alpha_{i+1}}^{A_{i+1}}$ 
and  ${\cal R}_i$ acts as~(\ref{braidaction}) 
on this pair of internal Hilbert spaces.   
At this point, it is important to note that 
the 3-cocycles of type~I and type~II, 
displayed in~(\ref{type1do}) and~(\ref{type2do}), are 
symmetric in the two last entries, i.e.\ $\omega(A,B,C) = \omega(A,C,B)$. 
This implies that the isomorphism $\Phi_i$  
commutes with the braid operation ${\cal R}_i$ for these 3-cocycles. 
A similar observation appears for the 3-cocycles of type~III 
given in~(\ref{type3do}). To start with, 
$\Phi_i$ obviously commutes with ${\cal R}_i$, iff the exchanged particles
carry the same fluxes, that is, $A_i=A_{i+1}$. Since the 3-cocycles 
of type~III are not symmetric in their last two entries, 
this no longer holds when the particles carry different fluxes
$A_i \neq A_{i+1}$. In this case, however, only the monodromy operation
${\cal R}_i^2$ is relevant, which clearly  commutes with the isomorphism
$\Phi_i$.
The conclusion is that the isomorphism $\Phi_i$ drops out 
of the formal definition~(\ref{brareco}) of the 
truncated braid group representations in 
CS theories with an abelian finite gauge group $H$.
It should be stressed, though, 
that this simplification only occurs for abelian 
gauge groups $H$. In CS theories with a nonabelian 
finite gauge group, in which the fluxes exhibit flux 
metamorphosis~\cite{bais}, the isomorphism $\Phi_i$ has to be taken 
into account~\cite{thesis}.

All in all, the internal Hilbert space of a multi-particle system 
in an abelian discrete gauge theory with CS action 
$\omega \in H^3({ H}, U(1))$ carries a representation of 
he quasi-quantum double $\DW$ and some truncated braid group.
Both representations are in general reducible.
It is now easily verified that relation~(\ref{grcom}) 
extends to an internal Hilbert space describing
an arbitrary number of particles and states that the action of the 
quasi-quantum double commutes with the action of the related 
truncated braid group. Hence, the multi-particle internal Hilbert space 
in these theories, in fact, decomposes into irreducible subspaces 
under the action of the direct product of $\DW$ and the related truncated
braid group. I will discuss this decomposition and the relation with 
the spins assigned to the particles in further detail in the 
following subsection. 

As a last remark, it can be shown~\cite{dpr} that the deformation 
of the quantum double $D(H)$ into the quasi-quantum double
$\DW$ just depends on the cohomology class of  $\omega$ in $H^3(H,U(1)$.
That is, the quasi-quantum double $D^{\omega\delta \beta}({ H})$ with 
$\delta \beta$ a 3-coboundary is isomorphic to $D^{\omega}({ H})$, which
is consistent with the fact (see section~\ref{aa3c}) that these 
3-cocycles define equivalent CS theories.

\subsection{Fusion, spin and braid statistics}
\label{fuspbr}

Let $(\Pi^A_{\alpha}, V_\alpha^A)$ and
$(\Pi^B_{\beta}, V_\beta^B)$  be  two irreducible 
representations of $\DW$ defined in~(\ref{13}).
The tensor product representation 
$(\Pi^A_{\alpha} \ot \Pi^B_{\beta},V_\alpha^A \ot V_\beta^B)$ constructed
by means of the comultiplication~(\ref{coalgebra}) 
in general decomposes into a direct sum of irreducible representations
\bea               \label{piet}
\Pi^A_{\al}\otimes\Pi^B_{\beta}& = & \bigoplus_{C , \gamma}
N^{AB\gamma}_{\alpha\beta C} \; \Pi^C_{\gamma} \, ,
\eea
with $N^{AB\gamma}_{\alpha\beta C}$  the multiplicity of the irreducible 
representation $(\Pi^C_{\gamma}, V^C_{\gamma})$ given by~\cite{dpr}
\bea          \label{Ncoef}
N^{AB\gamma}_{\alpha\beta C} &=& \frac{1}{|{ H}|} \, \sum_{D,E}
            \mbox{tr}  \left( \Pi^A_{\alpha} \ot \Pi^B_{\beta}
                      (\Delta (\, {\mbox{P}}_E \, D \, )) \right)  \;
            \mbox{tr} \left(  \Pi^C_{\gamma} (\,{\mbox{P}}_E \,D \,) 
\right)^* \\     &=& \delta_{C,A \cdot B} \; \frac{1}{|{ H}|} \,
\sum_D \mbox{tr} \left( \alpha(D) \right) \; 
       \mbox{tr} \left( \beta(D)  \right) \; 
\mbox{tr} \left( \gamma(D) \right)^* \;   c_D(A,B) \, .      \nn
\eea
Here, $\mbox{tr}$ stands for taking the trace, $|{ H}|$ for the order of 
the abelian group ${ H}$, $*$ for complex conjugation and  $c_D(A,B)$ 
for the 2-cocycle~(\ref{c}).
The so-called fusion rule~(\ref{piet}) determines which particles 
$(\,C,\gamma \, )$ can be formed in the composition 
of  two given particles $(\, A,\alpha \,)$ and  $(\, B,\beta\,)$,
or if read backwards, gives the decay channels of the particle
$(\,C ,\gamma\,)$. The Kronecker delta in~(\ref{Ncoef}) then 
indicates that the various composites $(\,C,\gamma \, )$ 
which  may result  from fusing the particles $(\, A, \alpha \, )$ 
and $(\, B, \beta \, )$ carry the flux $C=A \cdot B$, 
whereas the rest of the formula determines the 
composition rules for the dyon charges $\alpha $ and $\beta$.

The fusion algebra spanned by the elements
$\Pi^A_{\alpha}$ with multiplication rule~(\ref{piet}) is 
commutative and associative and can therefore be diagonalized.
The matrix  implementing this diagonalization is the so-called
modular $S$ matrix~\cite{ver0} 
\bea                                
S^{AB}_{\alpha\beta} &=& \frac{1}{|{ H}|} \, \mbox{tr} \; {\cal 
R}^{-2 \; AB}_{\; \; \; \; \; \alpha\beta} \; = \;\frac{1}{|{ H}|} \,
\mbox{tr} \left(\alpha(B) \right)^* \; \mbox{tr} \left(\beta (A)\right)^* \, . 
\label{fusion}  
\eea
This matrix contains all information concerning 
the fusion algebra~(\ref{piet}).
In particular, the multiplicities~(\ref{Ncoef}) can be expressed in 
terms of the modular $S$ matrix by means of Verlinde's formula~\cite{ver0}
\bea      \label{verlindez}
N^{AB\gamma}_{\alpha\beta C}&=&\sum_{D,\delta}\frac{
S^{AD}_{\alpha\delta}S^{BD}_{\beta\delta}
(S^{*})^{CD}_{\gamma\delta}}{S^{eD}_{0\delta}} \, .
\eea 
Whereas the modular $S$ matrix is determined through the monodromy operator
following from~(\ref{braidaction}), the modular matrix $T$ contains the spin 
factors~(\ref{anp}) assigned to the particles in the spectrum of 
an abelian discrete $H$ CS theory 
\bea
T^{AB}_{\alpha\beta} &=& 
\delta_{\alpha,\beta} \, \delta^{A,B} \; 
\exp(2\pi \im s_{(A,\alpha)})
 \; = \;  \delta_{\alpha,\beta} \, \delta^{A,B} 
\frac{1}{d_\alpha} \, \mbox{tr} \left( \alpha(A) \right) \, ,     \label{modut}
\eea
with $d_\alpha$ the dimension of the projective dyon charge representation 
$\alpha$. The  matrices~(\ref{fusion}) and~(\ref{modut}) now 
realize an unitary representation of the discrete modular group $SL(2,\Z)$
with the following  relations~\cite{dvvv}
\begin{eqnfourarray}
{\cal C} &=&(ST)^3 \; = \; S^2\, ,      &         \label{charconj} \\
S^* &=& {\cal C} S \; = \;S^{-1}\, , & $\qquad S^t \;=\; S \, , $      
  \label{sun}      \\
T^* &=&T^{-1} \, , & $\qquad T^t \;=\; T \, ,$                   \label{tun}
\end{eqnfourarray}  
The  relations~(\ref{sun}) and~(\ref{tun}) express the fact that 
the matrices~(\ref{fusion}) and~(\ref{modut}) 
are symmetric and unitary. To proceed, the matrix ${\cal C}$ defined 
in~(\ref{charconj}) represents the charge conjugation operator which
assigns an unique anti-partner 
${\cal C} \, (\, A ,\alpha \,)=(\, \bar{A},\bar{\alpha} \,)$
to every particle  $(\, A,\alpha \,)$ in the spectrum
such that the vacuum channel occurs in the fusion rule~(\ref{piet}) 
for the particle/anti-particle pairs.
Also note that the complete set of relations~(\ref{sun})--(\ref{tun}) 
indicate that the charge conjugation matrix ${\cal C}$ commutes with 
the modular matrix $T$, which implies
that a given particle carries the same spin as its anti-partner.

We are now well prepared to address the issue of braid statistics 
and the fate of the spin-statistics connection in these 
2+1 dimensional models.
Let me emphasize from the outset that much of what follows has 
been established elsewhere in a more general setting.
See for instance~\cite{alvarez1,moseco} and the references therein 
for the 1+1 dimensional conformal field theory point of view 
and~\cite{witten,frohma} for the related 2+1 dimensional space time 
perspective.

\begin{figure}[htb]    \epsfxsize=7cm
\centerline{\epsffile{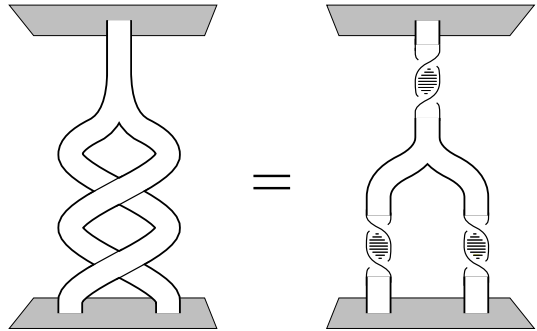}}
\caption{\sl Generalized spin-statistics connection. 
The displayed ribbon diagrams are homotopic as can 
be checked with a pair of pants.
In other words,
a monodromy of two particles in a given fusion channel followed  
by fusion of the pair  can be continuously deformed into the process
describing a rotation over an angle of $-2\pi$ of 
the two particles seperately followed by fusion of the pair and 
a final rotation over an angle of $2\pi$ of the composite. }
\label{kanaal}
\end{figure}

We first focus on a system consisting of two distinguishable 
particles $(\, A ,\alpha \,)$ and  $(\, B ,\beta\,)$. The associated 
two particle internal Hilbert space $V_\alpha^A \ot V_\beta^B$ carries 
a representation of the cyclic truncated colored braid group 
$P(2,m)\simeq \Z_{m/2}$ (defined in appendix~\ref{trubra}) with 
$m/2 \in \Z$ the order of the monodromy matrix ${\cal R}^2$ depending on 
the nature of the two particles. The aforementioned
representation decomposes into a direct sum of one dimensional irreducible 
subspaces, each being labeled by the associated eigenvalue of the monodromy 
matrix ${\cal R}^2$. 
As follows immediately from relation~(\ref{grcom}), the monodromy operation 
commutes with the action of the quasi-quantum double. This implies that 
the decomposition~(\ref{piet}) simultaneously diagonalizes the monodromy 
matrix. That is, the two particle 
total flux/charge eigenstates spanning a given 
fusion channel $V_\gamma^C$ all carry the same monodromy eigenvalue,
which in addition can be shown to satisfy 
the generalized spin-statistics connection~\cite{dpr}
\bea             \label{gekspist}
K^{ABC}_{\alpha\beta\gamma} \;
{\cal R}^2 &=& 
e^{2\pi \im(s_{(C,\gamma)}-s_{(A,\alpha)}-s_{(B,\beta)})} \;\;
K^{ABC}_{\alpha\beta\gamma} \, .
\eea
Here,  $K^{ABC}_{\alpha\beta\gamma}$ stands for the projection on  
the irreducible component $V_\gamma^C$  of  $V_\alpha^A \ot V_\beta^B$.  
So, the monodromy operation on a two particle state in a given
fusion channel is the same as a clockwise rotation over an angle of $2\pi$ 
of the two particles separately accompanied by a counterclockwise
rotation over an angle of $2\pi$ of the single particle state emerging 
after fusion. This is consistent with the fact that these two processes can be 
continuously deformed into each other, which is easily verified with the 
associated ribbon diagrams depicted in figure~\ref{kanaal}.
The discussion can now be summarized by the statement that 
the total internal Hilbert space $V_\alpha^A \ot V_\beta^B$ decomposes 
into the following  direct sum of irreducible representations of the direct 
product  $\DW \times P(2,m)$ 
\bea
 \bigoplus_{C,\gamma} N^{AB\gamma}_{\alpha\beta C} \; 
(\Pi^C_{\gamma}, \Lambda_{C-A-B} ) \, ,
\eea 
where $\Lambda_{C-A-B}$ denotes the one dimensional 
irreducible representation of $P(2,m)$ in which the  
generator $\gamma_{12}$ of $P(2,m)$  acts as~(\ref{gekspist}).

The discussion for a system 
of two identical particles $(\, A ,\alpha \,)$
is similar. The total internal Hilbert space 
$V_\alpha^A \ot V_\alpha^A$ now decomposes
into one dimensional irreducible subspaces under the action 
of the cyclic truncated braid group $B(2,m) \simeq \Z_m$. In the conventions 
set in appendix~\ref{trubra}, $m$ denotes the order 
of the braid operator ${\cal R}$, which depends on the system
under consideration. 
By the same argument as before, the two particle total flux/charge 
eigenstates spanning 
a given fusion channel $V_\gamma^C$ all 
carry  the same one dimensional representation
of $B(2,m)$.  The quantum statistics phase assigned to this channel
now satisfies the square root version of the generalized spin-statistics 
connection~(\ref{gekspist})
\bea             \label{gespietst}
K^{AAC}_{\alpha\alpha\gamma} \;
{\cal R} &=& \epsilon \;
e^{ \pi \im(s_{(C,\gamma)}-2 s_{(A,\alpha)})} \; \;
K^{AAC}_{\alpha\alpha\gamma} \, ,
\eea
with $\epsilon$ a sign depending on whether the 
fusion channel $V_\gamma^C$ appears in a symmetric or an
anti-symmetric fashion~\cite{alvarez1}. 
Thus, the internal space Hilbert space for a system of two 
identical particles $(\, A ,\alpha \,)$ breaks up into the following
irreducible representations of the direct product $\DW \times B(2,m)$
\bea                           \label{nil!}
 \bigoplus_{C,\gamma} N^{AA\gamma}_{\alpha\alpha C} \; 
(\Pi^C_{\gamma}, \Lambda_{C-2A} ) \, ,
\eea 
with $\Lambda_{C-2A}$ the one dimensional representation 
of the truncated braid group $B(2,m)$ defined in~(\ref{gespietst}).

\begin{figure}[htb]    \epsfxsize=9cm
\centerline{\epsffile{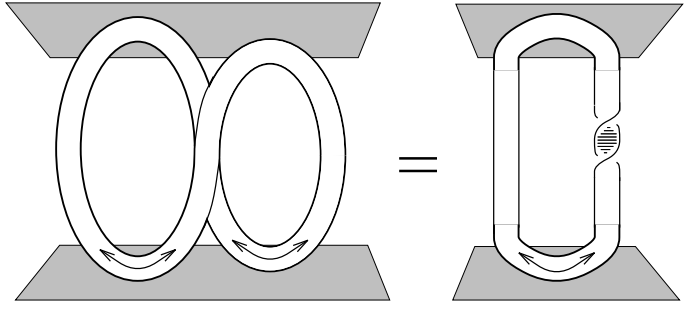}}
\caption{\sl Canonical spin-statistics connection. The particle 
trajectories describing a counterclockwise interchange of two particles in
separate particle/anti-particle pairs (the 8 laying on its back)
can be continuously deformed into a single pair in which the particle 
undergoes a counterclockwise rotation over an angle of $2\pi$ around its 
own centre (the 0 with a twisted leg).}
\label{spinstafig}
\end{figure}

In fact, the generalized spin-statistics connection~(\ref{gespietst})
incorporates the so-called canonical one. 
This can be seen using a topological proof of the canonical spin-statistics 
connection orginally due to Finkelstein and Rubinstein~\cite{fink}. 
Finkelstein and Rubinstein restricted themselves to skyrmions in 3+1 
dimensions, but their argument naturally extends to particles in 2+1 
dimensional space time. (See also reference~\cite{balach} 
and~\cite{frohma} for an algebraic approach.)
The crucial ingredient in their topological proof of the 
canonical spin-statistics connection for a given model is the existence 
of an anti-particle for every particle in the spectrum such that the pair 
can annihilate into the vacuum after fusion. Given this, one may then 
consider the process depicted at the l.h.s.\ of the equality
sign in figure~\ref{spinstafig}. It describes the creation of 
two separate identical particle/anti-particle pairs from the vacuum,
a subsequent counterclockwise exchange of the particles of the 
two pairs and a final annihilation of the two pairs.
It is readily checked that the closed ribbon associated with the 
process just explained  can be continuously deformed into the ribbon at the 
r.h.s.\ of figure~\ref{spinstafig}  corresponding to a counterclockwise 
rotation of the particle over an angle of
$2\pi$ around its own centre. In other words, the effect of interchanging
two identical particles in a consistent 
quantum description should be the same as the effect 
of rotating one particle over an angle of $2\pi$ around its centre. 
The effect of this rotation in the wave function is the spin factor
$\exp (2\pi \im s)$ with $s$ the spin of the particle 
(which may take any real value in 2+1 dimensional space time). 
Therefore, the result of exchanging  the two identical particles
necessarily boils down to a quantum statistical phase 
factor $\exp (\im \Theta)$ in the wave function being
the same as the spin factor
\bea                          \label{spistath}
\exp (\im \Theta) &=& \exp (2\pi \im s) \, .
\eea
This relation is known as the canonical spin-statistics connection. 
Actually, a further consistency condition can be inferred from this 
ribbon argument. The writhing in the particle trajectory can be 
continuously deformed into a writhing with the same orientation in 
the anti-particle trajectory. Hence, the 
anti-particle necessarily carries the same spin and statistics 
as the particle.

The basic assertions for the foregoing topological proof of the 
canonical spin-statistics connection are satisfied in the abelian 
discrete $H$ CS theories under consideration. 
That is to say, for every particle $(\, A ,\alpha \,)$ in the spectrum 
there exists an anti-particle $(\, {\bar{A}} ,\bar{\alpha} \,)$ such that 
under the proper composition the pair acquires the quantum numbers of the 
vacuum and may decay. Moreover,  as indicated by the fact that the 
charge conjugation operator ${\cal C}$ commutes with the modular matrix $T$,
every particle carries the same spin as its
anti-partner.
It should be noted now that the ribbon argument in figure~\ref{spinstafig}
actually {\em only} applies to states in which
the particles that propagate along the exchanged ribbons are 
in strictly  identical internal states. Otherwise the ribbons can not be
closed. Indeed, we find that the action~(\ref{braidaction}) of the 
braid operator on two particles in identical internal flux/charge eigenstates
\bea                 \label{braidactid}
{\cal R} \; |\, A ,\,^{\alpha}\!v_j \rangle 
|\,A,\,^{\alpha}\!v_j \rangle
&=& 
|\, A,\,\alpha\,(\,^A\! h_1\,)_{mj} \,
^{\alpha}\!v_m \rangle
|\, A,\,^{\alpha}\!v_j \rangle \, ,
\eea  
boils down to the diagonal matrix~(\ref{anp}) and therefore to the 
same spin factor~(\ref{spist}) 
\bea                             \label{spist}
\exp(\im \Theta_{(A,\alpha)}) &=& \exp(2\pi \im s_{(A,\alpha)}) \, ,
\eea
for all  $j$.
The conclusion is that the canonical spin-statistics connection holds 
in the fusion channels spanned by linear combinations 
of the states~(\ref{braidactid}) in which the particles are in strictly 
identical internal flux/charge eigenstates. 
The quantum statistics phase~(\ref{gespietst}) assigned to 
these channels reduces to the spin factor in~(\ref{spist}). Thus the 
effect of a counterclockwise interchange 
of the two particles in the states in these channels is the same 
as the effect of rotating one of the particles over an angle of $2\pi$.
To conclude, the closed ribbon proof does not apply to the other channels
and we are left with the more involved connection~(\ref{gespietst}) following 
from the open ribbon argument displayed in figure~\ref{kanaal}.

Higher dimensional irreducible truncated braid group 
representations are conceivable for systems consisting of more than 
two particles in abelian discrete $H$ gauge theories with a type~III CS 
action~(\ref{type3do}). 
The occurrence of such representations simply means that the generators of 
the associated truncated braid group can not be diagonalized simultaneously. 
What happens in this situation is that under the full set of braid 
operations, the system jumps between isotypical fusion channels, i.e.\ 
fusion channels of the same type or `color'. 
Let us make this statement more precise. To keep the discussion 
general, we do not specify the nature of the 
particles in the system. Depending  on whether the system consists of 
identical particles, distinguishable particles or some `mixture', 
we are dealing with a truncated braid group, a colored truncated 
braid group or a partially colored truncated braid group respectively.
The decomposition of the internal Hilbert for a system of more then two 
particles into a direct sum of irreducible subspaces (or fusion channels) 
under the action of the quasi-quantum double $\DW$ simply follow from 
the fusion rules~(\ref{piet}) and the fact that the fusion algebra 
is associative.  
Given that the action of the associated truncated 
braid group commutes with that of the quasi-quantum double, we 
are left with two possibilities.
On the one hand, there will in general be some 
fusion channels being  separately invariant under
the action of the associated truncated braid group.
As in the two particle systems discussed before, 
the total flux/charge eigenstates spanning such a 
fusion channel, say $V_\gamma^C$, carry the same one 
dimensional irreducible representation 
$\Lambda_{\mbox{\scriptsize abelian}}$ of the
related truncated braid group. That is, these states realize abelian 
braid statistics with the same quantum statistics or monodromy phase.
So, the fusion channel $V_\gamma^C$ carries the irreducible representation 
$(\Pi_\gamma^C, \Lambda_{\mbox{\scriptsize abelian}})$ of the direct product 
of the quasi-quantum double and the related truncated braid group.
On the other hand, it is also feasible 
that states carrying the {\em same} total flux and charge 
in {\em different} (isotypical) fusion channels 
are mixed under the action of 
the related truncated braid group.  In that case, we are dealing with 
a higher dimensional irreducible representation of the 
truncated braid group or nonabelian braid statistics. 
Note that nonabelian braid statistics is conceivable, 
if and only if some fusion channel, say $V_\delta^D$, occurs more 
then once  in the decomposition of the Hilbert space
under the action of $\DW$. 
Only then there are some orthogonal states with 
the same total flux and charge available to span an higher dimensional 
irreducible representation of the associated truncated braid group.
The number $n$ of fusion channels $V_\delta^D$ 
related by the action of the braid operators
now constitutes the dimension of the irreducible representation 
$\Lambda_{\mbox{\scriptsize nonabelian}}$ of the braid group and the 
multiplicity of this representation 
is the dimension $d$ of the fusion channel $V_\delta^D$. 
To conclude, the direct sum of these $n$ fusion 
channels $V_\delta^D$ then carries an $n \cdot d$ dimensional 
irreducible representation 
$(\Pi_\delta^D, \Lambda_{\mbox{\scriptsize nonabelian}})$
of the direct product of $\DW$ and the associated 
truncated braid group.

\sectiona{$U(1)$ Chern-Simons theory}
\label{typeI}

We turn to the simplest example of a spontaneously broken
CS gauge theory, namely the planar abelian Higgs model equipped with a CS 
term~(\ref{CSt1}) for the gauge fields.
So, the action of the model under consideration reads
\bea        \label{action}
S &=& \int d \, ^3x \;
({\cal L}_{\mbox{\scriptsize YMH}} + 
{\cal L}_{\mbox{\scriptsize matter}} + {\cal L}_{\mbox{\scriptsize CSI}}) \\
{\cal L}_{\mbox{\scriptsize YMH}} &=& 
-\frac{1}{4}F^{\kappa\nu} F_{\kappa\nu} 
  +({\cal D}^\kappa \Phi)^*{\cal D}_\kappa \Phi - V(|\Phi|)  
\label{higgspa} \\
{\cal L}_{\mbox{\scriptsize matter}} &=& -j^{\kappa}A_{\kappa} \label{maco} \\
{\cal L}_{\mbox{\scriptsize CSI}} &=&  
\frac{\mu}{2} \epsilon^{\kappa\nu\tau} A_{\kappa} \partial_{\nu}
A_{\tau} \, ,     \label{CStyp1}
\eea
where the Higgs field $\Phi$ is assumed to carry the global $U(1)$ 
charge $Ne$. In our conventions, this means that 
the covariant derivative takes the form 
${\cal D}_\kappa \Phi = (\partial_\kappa +\im Ne A_\kappa) \Phi$.
Further, the Higgs potential 
\bea                         \label{pot}
V(|\Phi|) &=& \frac{\lambda}{4}(|\Phi|^2-v^2)^2  \qquad\qquad
\mbox{with $\lambda, v > 0 \, ,$}
\eea 
endows the Higgs field $\Phi$ with a nonvanishing
vacuum expectation value $|\langle \Phi \rangle|=v$.
So, the compact $U(1)$ gauge symmetry is spontaneously    
broken to the finite cyclic subgroup $\Z_N$ at the energy scale 
$M_H=v\sqrt{2\lambda}$. 
Finally, the matter charges $q$ introduced by the current $j^\kappa$
in~(\ref{maco}) are assumed to be 
multiples of the fundamental charge unit $e$. That is, 
$q=ne$ with $n\in \Z$.

In fact, with the incorporation of the topological CS term~(\ref{CStyp1}),
the complete phase diagram for a compact planar $U(1)$
gauge theory endowed with matter exhibits the following structure.
Depending on the parameters in our model~(\ref{action}) 
and the presence of Dirac monopoles/instantons, 
we can distinguish the phases:
\begin{itemize}
\item $\mu=v=0$ $\Rightarrow$ Coulomb phase.
   The spectrum consists of the quantized matter charges $q=ne$ exhibitting
   Coulomb 
   interactions, where the Coulomb potential depends logarithmically
   on the distances between the charges in this two spatial 
   dimensional setting.
\item $\mu=v=0$ with Dirac monopoles $\Rightarrow$ confining phase.
   As has been shown by Polyakov~\cite{polyakov}, 
   the contribution of monopoles/instantons to the 
   partition function leads to linear confinement 
   of the quantized charges $q$.
\item $v \neq 0, \,  \mu = 0$ $\Rightarrow$ $\Z_N$ 
   Higgs phase, e.g.\ \cite{banff,spm} and references therein.
   The spectrum  consists of screened matter charges $q=ne$, magnetic 
   fluxes quantized as $\phi= \frac{2\pi a}{Ne}$ with $a \in \Z$
   and dyonic combinations. 
   The long range interactions are topological AB 
   interactions: in the process of  
   circumnavigating a flux $\phi$  counterclockwise with 
   a matter charge $q$, for instance, the wave function of the system 
   picks up the AB phase $\exp (\im q \phi)$. Under these
   remaining long range interactions, the   charges and fluxes become $\Z_N$ 
   quantum numbers. Further, in the presence of Dirac monopoles/instantons, 
   magnetic flux $a$ is  conserved modulo $N$.
\item $v=0, \, \mu \neq 0 $ $\Rightarrow$  CS 
   electrodynamics~\cite{schonfeld}. The gauge fields 
   carry the topological mass $|\mu|$. 
   The charges $q=ne$ constituting the spectrum are screened by induced 
   magnetic fluxes $\phi= -q/\mu$. 
   The long range interactions between the matter charges
   are AB interactions with coupling constant $\sim 1/\mu$,
   i.e.\ a counterclockwise monodromy involving a charge $q$ and a 
   charge $q'$ gives rise~\cite{goldmac} to the AB phase 
   $\exp (-\im qq'/\mu)$. 
   It has been argued~\cite{pisar,affleck,klee} that the presence 
   of Dirac monopoles does {\em not} lead to confinement 
   of the matter charges in this massive CS phase. 
   A consistent implementation of Dirac monopoles 
   requires that the topological mass is quantized~\cite{pisar} as
   $\frac{pe^2}{\pi}$ with $p \in \Z$. The Dirac monopoles
   then describe tunneling events between particles with charge difference 
   $\Delta q = 2pe$ with $p$ the integral CS parameter.
   Thus, the spectrum only contains a total number of $2p-1$ 
   distinct stable charges in this case.
\item $v \neq 0, \, \mu \neq 0$ $\Rightarrow$ $\Z_N$ CS 
   Higgs phase~\cite{sam,spm1,sm,thesis}.
   Again, the spectrum features screened matter charges $q=ne$, magnetic
   fluxes quantized as $\phi= \frac{2\pi a}{Ne}$ with $a \in \Z$
   and dyonic combinations. In this phase, 
   we have the conventional long range AB 
   interaction $\exp (\im q \phi)$ between charges and fluxes, and,
   in addition, AB interactions $\exp (\im \mu \phi \phi')$
   between the fluxes themselves~\cite{sam,spm1}. 
   Under these interactions, the charges 
   then obviously remain  $\Z_N$ quantum numbers, whereas a compactification
   of the magnetic flux quantum numbers only occurs for fractional values
   of the topological mass $\mu$~\cite{spm1}. In particular, the aforementioned
   quantization of the topological mass required in the presence of Dirac 
   monopoles renders the magnetic fluxes to be $\Z_N$ quantum numbers.
   The flux tunneling $\Delta a = -N$ induced by the minimal 
   Dirac monopole is now accompanied by a charge jump $\Delta n = 2p$, with
   $p$ the integral CS parameter. Finally, as implied by the 
   homomorphism~(\ref{homoI}) for this case, the CS parameter
   becomes periodic in this broken phase, that is, there are just
   $N-1$ distinct $\Z_N$ CS Higgs phases in which both charges
   and fluxes are $\Z_N$ quantum numbers~\cite{spm1,sm}.
\end{itemize}

In this section,  we just focus on the phases summarized 
in the last two items. The discussion is organized as follows.
Subsection~\ref{ub} contains a brief exposition of CS electrodynamics 
featuring Dirac monopoles. In subsection~\ref{bp}, we then turn to  
the CS Higgs screening mechanism for the electromagnetic fields generated by  
the matter charges and the magnetic vortices in the broken phase and 
establish the above mentioned long range AB interactions between these 
particles. To conclude, a detailed discussion of the discrete $\Z_N$ CS 
gauge theory describing the long distance physics in the broken phase is 
presented in subsection~\ref{rev}.

\subsection{Dirac monopoles and topological mass quantization}
\label{ub}

For future use and reference, I begin by briefly reviewing the basic 
features of CS electrodynamics, i.e.\ we set the symmetry breaking scale 
in our model~(\ref{action}) to zero for the moment ($v=0$) and 
take $\mu \neq 0$. 
Varying the action~(\ref{action}) w.r.t.\ the vector 
potential $A_{\kappa}$ then yields the field equations
\bea                      \label{fieldequation}
\partial_{\nu} F^{\nu\kappa} + 
\mu\epsilon^{\kappa\nu\tau} \partial_{\nu}A_{\tau} &=& 
j^\kappa+j^\kappa_H \, , 
\eea   
where 
\bea       \label{Higgscur}
j^{\kappa}_H &=& \im Ne(\Phi^*{\cal D}^{\kappa}\Phi-
({\cal D}^{\kappa}\Phi)^*\Phi) \, ,
\eea 
denotes the Higgs current and $j^{\kappa}$ the matter current in~(\ref{maco}).
These field equations indicate that the gauge fields are massive.
To be precise, this model features a single component 
photon carrying the topological mass $|\mu|$~\cite{schonfeld}.
So, the electromagnetic fields generated by the currents 
in~(\ref{fieldequation}) are screened: they fall 
off exponentially with mass $|\mu|$. Hence, at distances 
$\gg 1/|\mu|$ 
the Maxwell term in~(\ref{fieldequation}) can be neglected which 
immediately reveals how the screening mechanism operating in CS 
electrodynamics works. 
The currents $j^{\kappa}$ and $j^{\kappa}_H$  induce 
magnetic flux currents 
$-\frac{1}{2}\epsilon^{\kappa\nu\tau} \partial_{\nu} A_{\tau}$
exactly screening the electromagnetic fields 
generated by $j^{\kappa}$ and $j^{\kappa}_H$. 
Specifically, from Gauss' law
\be                      \label{gauss}
Q=q +q_H+\mu \phi=0 \, ,
\ee  
with $Q=\int\! d\, ^2x\, \nabla \!\cdot\! {\mbox{\bf E}} =0$, 
$q=\int\! d\, ^2x \, j^0 $, $q_H=\int\! d\,^2x \, j^0_H $ 
and $\phi = \int \! d\,^2x \,\epsilon^{ij}\partial_i A^j$,
we learn  that the CS  screening  mechanism 
attaches fluxes $\phi=-q/\mu$ and 
$\phi_H=-q_H/\mu$ of characteristic size $1/|\mu|$ to the
point charges $q$ and $q_H$ respectively~\cite{schonfeld}.

The remaining long range interactions between these screened  
charges $q$ are the topological  AB  interactions~\cite{ahabo} 
implied by the matter coupling~(\ref{maco}) and the CS 
coupling~(\ref{CStyp1}). These can be 
summarized~\cite{goldmac} as~\footnote{In this paper, I adopt 
the conventions set in reference~\cite{banff} and~\cite{thesis}. 
Accordingly, the quantum state 
$|q \rangle := |q, \mbox{\bf x}\rangle$ describes a single particle 
carrying charge $q$ located at some position $\mbox{\bf x}$ in the plane.
Further, I use a gauge in which the nontrivial parallel transport 
in the gauge fields around the fluxes $\phi$ carried by the particles 
takes place in thin strips or Dirac strings starting at the locations 
of the particles and going off to spatial infinity in the direction of 
the positive vertical axis. Also, in constructing multi-particle wave 
function, the particle located most left in the plane by convention 
appears most left in the tensor product and so on. Finally, the 
topological interactions are absorbed in the boundary conditions of 
the (multi-) particle wave functions, i.e.\ I work with multi-valued 
wave functions propagating with completely free Lagrangians.} 
\bea           \label{ons}
{\cal R}^2 \; |q\rangle|q'\rangle &=&
e^{-\im \frac{q q'}{\mu}}|q\rangle|q'\rangle \\
{\cal R} \; |q\rangle |q\rangle &=& e^{-\im \frac{q q}{2\mu}} \, 
|q\rangle|q\rangle \, .     \label{onsan}
\eea
So, the particles in this theory realize abelian braid statistics.
Particularly, relation~(\ref{onsan}) indicates that 
identical particle configurations in general  
exhibit anyon statistics~\cite{leinaas} 
with quantum statistics phase 
$\exp (\im \Theta_q)= \exp(-\im \frac{q q}{2\mu})$
depending on the square of the charge $q$ of the particles and 
the inverse of the topological mass $\mu$.
Further, the assertions for the topological proof of the canonical 
spin-statistics connection~(\ref{spistath}) are obviously satisfied
in this model. Hence, a rotation of a charge $q$ over an angle of $2\pi$
gives rise to the spin factor $\exp(2\pi \im s_q) = 
\exp (\im \Theta_q)= \exp(-\im \frac{q q}{2\mu})$.

\begin{figure}[htb] 
\begin{center}
\begin{picture}(125,125)(-60,-60)
\put(-60,0){\line(1,0){120}}
\put(0,-60){\line(0,1){120}}
\thinlines
\multiput(-60,0)(10,0){13}{\line(0,1){2}}
\multiput(0,-60)(0,10){13}{\line(1,0){2}}
\multiput(-30,60)(20,-40){4}{\circle*{2.0}}
\multiput(-20,40)(20,-40){3}{\circle*{2.0}}
\put(-20,40){\circle{3.6}}
\put(-20,40){\vector(1,-2){20}} 
\put(5,50){\vector(0,1){5}}
\put(43,-3){\vector(1,0){5}}
\put(8,55){\small$\phi[\frac{\pi}{2e}]$}
\put(50,-6){\small$q[e]$}
\end{picture}
\vspace{0.5cm}
\caption{\sl Spectrum of unbroken $U(1)$ CS theory.
We depict the flux $\phi$ 
versus the global $U(1)$  charge $q$.
The CS parameter $\mu$ is set to its minimal nontrivial value 
$\mu=\frac{e^2}{\pi}$, i.e.\ $p=1$. 
The arrow represents the effect of a {\em charged} Dirac 
monopole/instanton, which shows that there is just one stable 
(fermionic) particle in this theory.}
\label{u1}
\end{center}  
\end{figure}
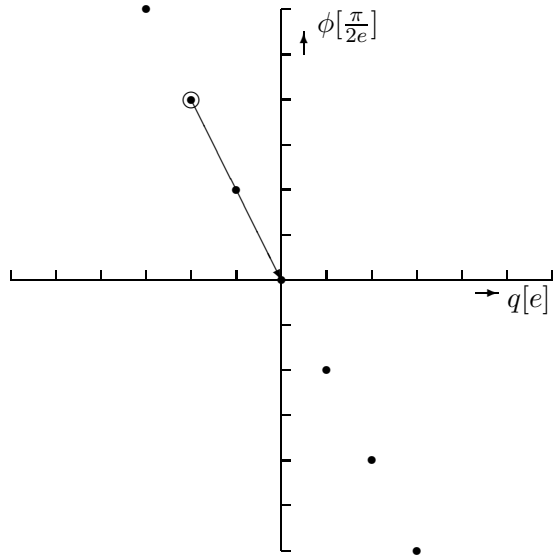

Let us now assume that  this compact $U(1)$ CS gauge theory features singular 
Dirac monopoles~\cite{dirac} carrying magnetic charges quantized 
as $g=\frac{2\pi m}{e}$ with $m \in \Z$. 
In this 2+1 dimensional model,
these monopoles become instantons corresponding to tunneling events 
between states with flux difference and to obey Gauss' law~(\ref{gauss}) 
also charge difference~\cite{hen,pisar,affleck,klee}. To be explicit, 
the minimal Dirac monopole induces the tunneling 
\bea                     \label{dcharge}  
\mbox{instanton:} & &
\left\{ \ba{lcl}    
\Delta \phi &=& -\frac{2\pi}{e} \\
\Delta q &=& \mu \frac{2\pi}{e} \, .  
\ea
\right.
\eea
A consistent implementation of these Dirac monopoles requires the  
quantization of the matter charges $q$ in multiples of $e$ {\em and} 
as a direct consequence quantization of the topological mass $\mu$.  
Dirac's argument~\cite{dirac} works also in the presence of a CS term. 
In this case, the argument goes as follows. The tunneling 
event~(\ref{dcharge}) corresponding to the  minimal Dirac monopole should 
be invisible to the monodromies~(\ref{ons}) with the charges $q$ present 
in our model. In other words, the AB phase 
$\exp (-\im \frac{ q \Delta q}{\mu})=\exp ( -\im \frac{2\pi q}{e})$
should be trivial, which implies the charge quantization $q=ne$ with 
$n \in \Z$. Furthermore, the tunneling event~(\ref{dcharge}) 
should respect this quantization rule for $q$.
So, the charge jump has to be a multiple of $e$: 
$\Delta q = \mu \frac{2\pi}{e}=pe$ with $p \in \Z$, 
which leads to the quantization $\mu = \frac{pe^2}{2\pi}$.
There is, however, a further restriction on the values of 
the topological mass $\mu$. So far, we have only considered the 
monodromies in this theory, but the particles connected by Dirac monopoles
should as a matter of course also have 
the same quantum statistics phase~(\ref{onsan})
or equivalently the same spin factor. In particular, the spin 
factor for the charge $\Delta q$ connected to the vacuum $q=0$ should be 
trivial: $\exp (-\im \frac{(\Delta q)^2}{2\mu})
=\exp (-\im \frac{\mu}{2} (\frac{2\pi}{e})^2)=1$. The conclusion then 
becomes that in the presence of Dirac monopoles the topological mass is 
necessarily quantized as~\footnote{The observation that a consistent 
implementation of Dirac monopoles implies the quantization of the 
topological mass $\mu$ was first made by Henneaux and Teitelboim~\cite{hen}. 
However, they only used the 
monodromy part of the above argument and did not implement the demand 
that the particles connected by Dirac monopoles should give rise the same
spin factor. As a consequence, they arrived at the aforementioned erroneous 
finer quantization $\mu=\frac{pe^2}{2\pi}$. Subsequently, Pisarski 
derived the correct quantization~(\ref{mu}) by considering 
gauge transformations in the background of a Dirac monopole~\cite{pisar}.}
\bea                 \label{mu}
\mu &=& \frac{pe^2}{\pi}   \qquad \qquad \mbox{with $p \in \Z$} \, ,
\eea
which is the result alluded to in~(\ref{quantmui}).
To conclude the argument, 
substituting relation~(\ref{mu}) into expression~(\ref{dcharge}) 
reveals that the presence of Dirac monopoles/instantons imply that 
the quantized charges $q=ne$ are conserved modulo $2pe$ with $p$ being 
the integral CS parameter. Thus, the spectrum of this unbroken $U(1)$ CS 
theory just consists of $2p-1$ stable charges $q=ne$ screened by the induced 
magnetic fluxes $\phi=-q/\mu$. (See also~\cite{sm,klee,moorseib} and 
references therein.) We have depicted this spectrum for $p=1$ in 
figure~\ref{u1}.

Alternatively, we may embed a $U(1)$ CS theory in a nonabelian compact CS 
gauge theory. In that case, the whole or part of the foregoing spectrum of
singular Dirac monopoles turns into regular 't Hooft-Polyakov 
monopoles (e.g.\ \cite{banff,later,thooftpolya} and references therein). 
As will be illustrated next with two representative examples, 
the correct quantization of the topological mass is automatic. 
This is as it should, since the regular monopoles/instantons, in any case,
can not be left out in such a theory. Let us first consider a 
CS theory in which the gauge group $SU(2)$ is spontaneously broken down 
to $U(1)$~\cite{spm1,sm}. 
It is well-known (e.g.\ Deser et al.\ in reference~\cite{schonfeld}) that 
in order to end up with a gauge invariant quantum theory, 
the topological mass for a $SU(2)$ CS theory is necessarily 
quantized as $\mu_{SU(2)} = \frac{k e^2}{4 \pi}$ with $k \in \Z$. 
At first sight, this finer quantization seems to be in conflict 
with~(\ref{mu}). This is not the case though. 
The point is that the effective low energy $U(1)$ CS theory of the 
foregoing model features $U(1)$ matter charges quantized 
as $q=\frac{ne}{2}$ with $n \in \Z$ and regular 't Hooft-Polyakov 
monopoles/instantons~\cite{banff,later,thooftpolya} 
with magnetic charge quantized as $g=\frac{4\pi m}{e}$ with $m \in \Z$.  
In other words, upon redefining $e \mapsto \frac{e}{2}$ the spectrum of 
matter and (Dirac) magnetic charges {\em and} the 
quantization~(\ref{mu}) of the topological mass for the compact $U(1)$ 
CS theory discussed in the previous paragraph coincides with that for 
the foregoing broken theory. In short, the finer quantization   
of the topological mass as compared to~(\ref{mu}) is perfectly consistent  
with the larger quantization of magnetic charge in this broken theory. 
For an $SO(3)$ CS gauge theory, in turn, a rather abstract  
argument~\cite{moorseib} (see also~\cite{diwi}) showed 
that the integer CS parameter must be divisible 
by four in units in which any integer is allowed for $SU(2)$. So,  
$\mu_{SO(3)} = \frac{p e^2}{\pi}$ with $p \in \Z$.
This is, in fact, precisely the quantization required
for a consistent implementation of the $\Z_2$ Dirac monopoles~\cite{later} 
(carrying magnetic charge $g=\frac{2\pi m_s}{e}$ with $m_s \in 0,1$) 
that can be introduced in a $SO(3)$ CS theory. If this theory is 
subsequently broken down to $U(1)$, we obtain regular 't Hooft-Polyakov 
monopoles~\cite{thooftpolya} carrying magnetic charge quantized as 
$g=\frac{4\pi m_{r}}{e}$ with $m_r \in \Z$. Hence, with the incorporation
of the aforementioned Dirac monopoles, the complete monopole spectrum
in this broken theory consists of the magnetic charges 
$g=\frac{2\pi m}{e}$ with $m \in \Z$.
As for the matter part, in the presence of the $\Z_2$ Dirac monopoles
in the original $SO(3)$ theory, matter fields carrying faithful
(i.e.\ half integral spin) representations of the universal covering group
$SU(2)$ of $SO(3)$ are ruled out.
Thus, only $U(1)$ matter charges $q$ carrying integral multiples of $e$ are 
conceivable in the foregoing broken theory.
All in all, we then arrive at the same spectrum of matter 
and magnetic charges as the compact $U(1)$ CS theory of the previous 
paragraph, while the correct quantization of the topological 
mass~(\ref{mu}) required for a consistent implemention of this spectrum
of singular and regular monopoles is again automatic.     
To conclude, from the foregoing discussion it is also immediate that
the natural homomorpisms~(\ref{restric}) accompanying the spontaneous
breakdown of these theories, i.e.\ the restrictions 
$H^4(BSU(2), \Z) \rightarrow H^4(BU(1), \Z)$ and 
$H^4(BSO(3), \Z) \rightarrow H^4(BU(1), \Z)$ induced by the inclusions
$U(1) \subset SU(2)$ and $U(1) \subset SO(3)$ respectively, are not just
onto, but even ono-to-one.

\subsection{Dynamics of the Chern-Simons Higgs medium}
\label{bp}

We continue with an analysis of  the Higgs 
phase of the model~(\ref{action}). So, from now on  $v \neq 0$. 
To keep the discussion general, however, the topological mass $\mu$  may 
take any real value in this subsection. The incorporation of Dirac 
monopoles in this phase, which requires the quantization~(\ref{mu}) 
of $\mu$, will be discussed in the next subsection.

Let me first recall some of the basic dynamical features of  this model.
To start with, the complex Higgs field $\Phi(x) = \rho(x)\exp (\im \sigma(x))$ 
describes two physical degrees of freedom: the charged Goldstone boson field
$\sigma(x)$ and the physical field $\rho(x)-v$ with mass 
$M_H=v \sqrt{2\lambda}$ corresponding to the charged neutral Higgs particles.
The  Higgs mass $M_H$ sets the characteristic energy scale of this model.
At energies larger then $M_H$, the massive Higgs particles can 
be excited. At energies smaller then $M_H$ on the other hand,
the massive Higgs particles are frozen in. 
For simplicity we will restrict ourselves to 
the latter low energy regime. In this case,
the Higgs field is completely condensed, i.e.\ it acquires ground 
state values everywhere: 
$\Phi(x)  \mapsto \langle \Phi(x) \rangle = v \exp (\im \sigma(x))$.
The condensation of the Higgs field implies
that the Yang-Mills Higgs part~(\ref{higgspa}) of the action reduces to 
\bea                               \label{efhi}
{\cal L}_{\mbox{\scriptsize YMH}} & 
\longmapsto & -\frac{1}{4}F^{\kappa\nu} F_{\kappa\nu} 
+\frac{M_A^2}{2} \tilde A^{\kappa}\tilde A_{\kappa} \, , 
\eea
with $\tilde{A}_{\kappa} := A_{\kappa} + \frac{1}{Ne}\partial_{\kappa}
\sigma $ and $M_A  :=  Ne v\sqrt{2}.$
Hence, in the low energy regime, 
our model  is governed by the effective action 
obtained from substituting~(\ref{efhi}) in~(\ref{action}).
The effective field equations which follow from varying the resulting 
effective action w.r.t.\  $A_\kappa$ and the Goldstone boson $\sigma$,
respectively, read    
\bea                      \label{fequus}
\partial_\nu F^{\nu\kappa} + 
\mu\epsilon^{\kappa\nu\tau} \partial_{\nu}A_{\tau} &=& 
j^\kappa+j^\kappa_{\mbox{\scriptsize scr}}                         \\
\partial_\kappa  j^\kappa_{\mbox{\scriptsize scr}} &=& 0 \, .
\label{fequusco}
\eea
The important difference with the field equations~(\ref{fieldequation}) 
for the unbroken case is that the 
Higgs current~(\ref{Higgscur}) has become the screening
current~\cite{sam} 
\bea                                \label{scrcur}
j^\kappa_{\mbox{\scriptsize scr}} &=& - M_A^2 \tilde{A}^\kappa \, .
\eea
From the equations~(\ref{fequus}) and~(\ref{fequusco}), 
it is then readily inferred that the two 
polarizations $+$ and $-$ of the photon field $\tilde{A}_\kappa$
in the CS Higgs medium carry the masses~\cite{pisa}
\bea
M_{\pm} &=& \sqrt{M_A^2 + \frac{1}{2}\mu^2 \pm \frac{1}{2}\mu^2\sqrt{
\frac{4 M_A^2}{\mu^2}+1}} \, ,
\label{mass}
\eea
which differ by the topological mass~$|\mu|$. As an aside, by setting 
$\mu = 0$ in~(\ref{mass}), we  restore the 
fact that in the  ordinary Higgs phase 
both polarizations of the photon carry the same mass 
$M_+ = M_-=M_A$.
Taking the limit $v \rightarrow 0$,
on the other hand, yields $M_+=|\mu|$ and $M_-=0$. The $-$ component then
ceases to be a physical degree of freedom~\cite{pisa} and we  
recover the fact that  unbroken CS electrodynamics features
a single component photon with mass $|\mu|$.

There are now two dually charged types of sources for electromagnetic 
fields in this CS Higgs medium: 
the quantized point charges $q=ne$ 
introduced by the matter current $j^\kappa$ and   
magnetic vortices~\cite{abri} corresponding to holes 
in the CS Higgs medium 
of characteristic size $\sim 1/M_H$
carrying quantized magnetic flux.
Specifically, inside the core of a vortex (i.e.\ $r< 1/M_H$ with $r$
the distance to the centre of the vortex), the Higgs field
vanishes ($\Phi(r=0)=0$), while outside the core 
($r > 1/M_H$) the Higgs field makes a noncontractible winding 
in the vacuum manifold: 
$\Phi(r > 1/M_H, \theta) = 
v \exp\left(\im \sigma(r > 1/M_H, \theta)\right)$. Here, $\theta$
denotes the polar angle defined w.r.t.\ the centre of the vortex
and  $\sigma$ the (multi-valued) Goldstone boson
\bea
\sigma(r > 1/M_H, \theta+ 2\pi) - 
\sigma(r > 1/M_H, \theta) &=& 2\pi a \, ,   \label{mvsi}
\eea  
with $a \in \Z$ to keep the Higgs field $\Phi$ itself single valued.
Demanding the vortex solution to be of minimal (finite) energy  
also implies that the covariant derivative of the Higgs field
vanishes away from the core:
\bea \label{phova}
{\cal D}_i \Phi(r > 1/M_H, \theta)=0 &\Rightarrow& 
\tilde{A}_i(r > 1/M_H, \theta)=0 \, . 
\eea 
So, the holonomy in the Goldstone boson field is accompanied by a 
holonomy in the gauge fields and the magnetic flux $\phi$ 
trapped inside the core of the vortex is quantized as 
\bea         \label{quflux}
\phi \; = \; 
\oint dl^i A^i(r > 1/M_H) \;= \; 
\frac{1}{Ne} \oint  dl^i\partial_i \sigma(r > 1/M_H)  \; = \; 
\frac{2\pi a}{Ne} \;\;\;\; \mbox{with $a \in \Z \, . $}
\eea   

Both the matter charges $q=ne$ and the magnetic vortices 
$\phi=\frac{2\pi a}{Ne}$ enter the field 
equations~(\ref{fequus}) describing the physics outside the cores 
of the vortices. The matter charges enter by means of the matter current 
$j^\kappa$ and the vortices through the magnetic flux current 
$-\frac{1}{2}\epsilon^{\kappa\nu\tau} \partial_{\nu}A_{\tau}$.
From these equations, we learn that 
both the matter current and the flux current generate electromagnetic 
fields, which are screened at large distances by an 
induced current $j^\kappa_{\mbox{\scriptsize scr}}$ in the 
CS Higgs medium~\cite{sam}.
This becomes clear from Gauss' law  for this case
\bea                    \label{higgsgauss}
Q=q +  q_{\mbox{\scriptsize scr}}  +\mu \phi  =0 \, ,
\eea
with $ 
q_{\mbox{\scriptsize scr}}  =  
\int d \, ^2x \, j^0_{\mbox{\scriptsize scr}} = - \int d \, ^2x \, 
M_A^2 \tilde{A}^0 $,
which indicates that both 
the  matter charges $q$ and the  magnetic vortices $\phi$
are surrounded by localized screening charge densities 
$j^0_{\mbox{\scriptsize scr}}$.
At large distances, the contribution to the long range Coulomb fields 
of the induced screening charges 
\bea               \label{scherm}     \ba{rcl}
q=ne        &\Rightarrow& q_{\mbox{\scriptsize scr}} =-q    \\
\phi=\frac{2\pi a}{Ne}   &\Rightarrow& q_{\mbox{\scriptsize scr}} = 
-\mu \phi \, ,
\ea
\eea 
then completely cancel those of the matter charges $q$ and the fluxes 
$\phi$ respectively.
Here, it is of course understood that the screening charge density 
$j^0_{\mbox{\scriptsize scr}}$ accompanying a magnetic vortex 
is localized in a ring outside the core, 
since inside the core the Higgs field vanishes and the 
CS Higgs medium is destroyed. 
Let me also stress that just as in the ordinary Higgs medium~\cite{banff}
(i.e.\ no CS term) the matter charges $q$ are screened by 
charges $q_{\mbox{\scriptsize scr}}=-q$ provided by the Higgs condensate  
in this CS Higgs medium 
and {\em not} by attaching fluxes to them as in the case of 
unbroken CS electrodynamics. This is already  
apparent from the simple fact that the irrational `screening' fluxes 
$\phi=-q/\mu$ would render the Higgs condensate multi-valued.

An important observation~\cite{sam} concerning the induced screening 
charges in~(\ref{scherm}) is that they do not exhibit the long range 
AB effect~\cite{ahabo} in the process of taking them around a remote 
vortex. (See also~\cite{banff}.) The point is that the screening charge 
$q_{\mbox{\scriptsize scr}}$ (attached either to a matter charge $q$ or 
a vortex $\phi$) not only  couples to the holonomy in the gauge connection 
$A_\kappa$ around a remote vortex, but also to the holonomy~(\ref{mvsi}) 
in the Goldstone boson field. This is immediate from the effective low 
energy action following from substituting~(\ref{efhi}) in~(\ref{action}).
Let $j_{\mbox{\scriptsize scr}}^\kappa$ be the screening 
current~(\ref{scrcur}) associated with some screening charge
$q_{\mbox{\scriptsize scr}}$. The second term at the l.h.s.\ 
of~(\ref{efhi}) couples this current to the massive photon field
$\tilde{A}_\kappa$ around the remote vortex:  
$j_{\mbox{\scriptsize scr}}^\kappa \tilde{A}_\kappa$.
As we have seen in~(\ref{phova}), outside the core of the vortex,
the holonomies in the gauge fields and the Goldstone boson are
related such that $\tilde{A}_\kappa$ strictly vanishes. 
Consequently, as long as the screening charge stays well away from the core 
of the vortex, the interaction term 
$j_{\mbox{\scriptsize scr}}^\kappa \tilde{A}_\kappa$
vanishes and therefore does not generate an AB phase in the process
of taking a screening charge around a remote vortex. 
This proves our claim. An immediate conclusion is that the screening 
charges $q_{\mbox{\scriptsize scr}}=-q$ attached to the matter charges 
$q$ screen the Coulomb interactions between the matter charges, but not
their AB interactions with the magnetic vortices. 
That is, a counterclockwise monodromy of a screened charge $q$ and a 
remote screened magnetic flux $\phi$ leads to the conventional AB phase 
\bea          \label{klam}
{\cal R}^2 \; |q \rangle |\phi  \rangle &=& e^{\im q \phi } \;
|q \rangle |\phi  \rangle \, ,  
\eea 
as implied by the coupling~(\ref{maco}). Relation~(\ref{klam}) 
summarizes all the long range interactions for the mater charges. 
So, in contrast with unbroken CS electrodynamics, 
there are no long range AB interactions between 
the matter charges themselves in this broken phase.
Instead, we now obtain nontrivial AB interactions among the 
screened magnetic fluxes
\bea        \label{fluxAB}
{\cal R}^2 \; |\phi \rangle |\phi ' \rangle &=& e^{\im \mu \phi \phi '} \;
|\phi \rangle |\phi ' \rangle   \\
{\cal R} \; |\phi \rangle |\phi  \rangle &=& 
e^{\im \frac{\mu}{2} \phi \phi} \;
|\phi \rangle |\phi \rangle \, ,     \label{qs}
\eea
entirely due to the CS coupling~(\ref{CStyp1}). 
From~(\ref{qs}), we conclude that depending on their flux 
and the topological mass, identical magnetic vortices 
realize anyon statistics.

In retrospect, the basic characteristics of the CS Higgs 
screening mechanism uncovered in~\cite{sam} and outlined above find their 
confirmation in results established in earlier  studies of the 
static magnetic vortex solutions of the abelian CS Higgs model.
In fact, the study of these so-called CS vortices was started 
by Paul and Khare~\cite{pakh}, who noted that they correspond to finite 
energy solutions carrying both magnetic flux and electric charge. 
Subsequently, various authors have obtained both analytical and numerical
results on these static vortex solutions. 
See for example~\cite{vega}--\cite{kleep} and the references therein.
Here, I just collect the main results.
In general, one takes the following ansatz for a static vortex solution 
of the field equations corresponding to~(\ref{action})
\bea
\Phi(r,\theta)  =  \rho(r) \exp \left(\im \sigma(\theta) \right) \, , 
\qquad A_0(r,\theta) =  A_0(r) \, , \qquad               
A_i (r,\theta)  =  -A(r) \partial_i \sigma(\theta) \, ,   
\label{higans} 
\eea
where $r$ and $\theta$ again denote the polar coordinates 
and $\sigma$ the multi-valued Goldstone boson 
\bea
\sigma(\theta+ 2\pi) - \sigma(\theta) &=& 2\pi a \, ,   \label{mvs}
\eea  
with $a \in \Z$ to render the Higgs field $\Phi$ itself single valued.
Regularity of the solution imposes the following boundary conditions as 
$r \rightarrow 0$ 
\bea
\ba{lll}
\rho \rightarrow 0 \, , \qquad  &  
A_0 \rightarrow {\mbox{constant}} \, , \qquad &  
A \rightarrow 0 \, ,
\ea
\eea  
whereas, for finite energy, 
the asymptotical behavior for $r \rightarrow \infty$  becomes 
\bea                                         \label{asy}
\ba{lll}
\rho \rightarrow v \, , \qquad   &  A_0 \rightarrow 0 \, , \qquad &  
A \rightarrow \frac{1}{Ne} \, .
\ea
\eea 
From~(\ref{asy}),~(\ref{higans}) and~(\ref{mvs}) it then follows that this 
solution corresponds to  the quantized magnetic flux~(\ref{quflux}).

Since the two polarizations of the photon carry 
the distinct masses~(\ref{mass}),
it seems, at first sight, 
that there are two different  vortex solutions 
corresponding to a long range exponential decay of 
the electromagnetic fields either with mass $M_-$ or with mass $M_+$. 
However, a careful analysis~\cite{ino} (see also~\cite{boya}) of the 
differential equations following from the field equations with this ansatz
shows that the $M_+$ solution does not exist for 
finite $r$. Hence,  we are left with the $M_-$ solution.
To proceed, it turns out that 
the modulus $\rho$ of the Higgs field $\Phi$ in~(\ref{higans})
grows monotonically from zero (at $r=0$) to its 
asymptotic ground state value~(\ref{asy}) at $r=1/M_H$, where the profile 
of this growth does not change much in the full range of the 
parameters, e.g.\ \cite{boya}.

An important issue is, of course, 
whether vortices will actually form or not, i.e.\ 
whether the superconductor we are describing is type~II or I
respectively. In this context, the competition between  
the penetration depth $1/M_-$ of the electromagnetic fields
and the coresize $1/M_H$ becomes important.
In ordinary superconductors ($\mu=0$), 
an evaluation of the free energy yields that we are dealing 
with a type~II superconductor if $M_H/M_A=\sqrt{\lambda}/Ne \geq 1$ 
and a type~I superconductor otherwise~\cite{gennes}.
Since $M_-$ is smaller then $M_A$, it is expected that in the presence 
of a CS term the type~II region is extended. A perturbative  
analysis for small $\mu$ shows that this is indeed the
case~\cite{jacobs}.
In the following, we will always 
assume that our parameters are adjusted  
such that we are in the type~II region.

\begin{figure}[htb]    \epsfxsize=8.7cm
\centerline{\epsffile{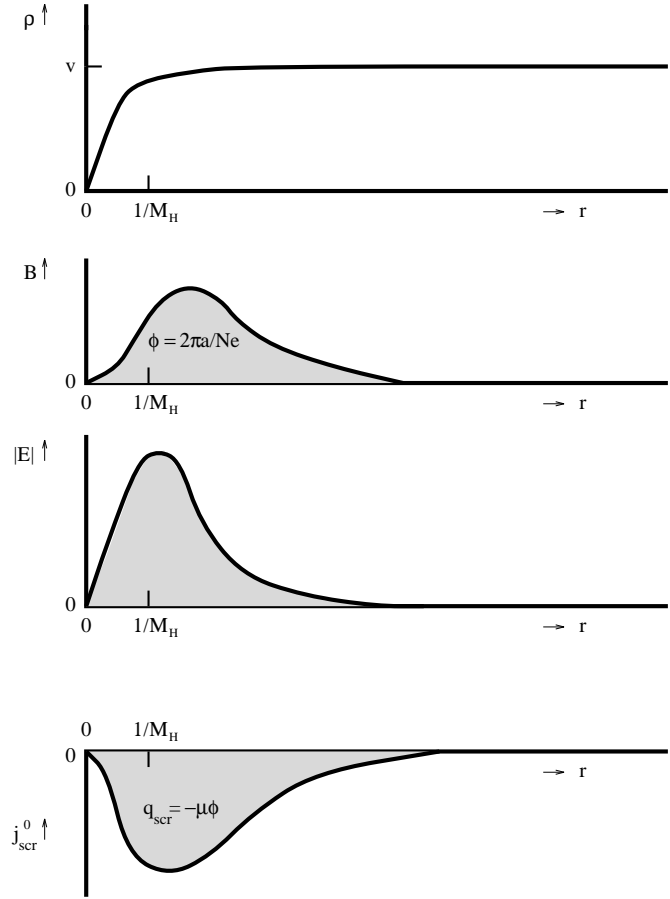}}
\caption{\sl Qualitative behavior of the vortex solution carrying 
the quantized magnetic flux $\phi=\frac{2\pi a}{Ne}$ 
in the CS limit.  
We have depicted the modulus of 
the Higgs field $\rho$, the magnetic
field $B$, the electric field $|{\mbox{\bf E}}|$ 
and the screening charge density $j^0_{\mbox{\protect\scriptsize scr}}=
-2(Ne)^2 \rho^2 A^0$,  respectively,
versus the radius $r$. 
The electromagnetic fields and the screening charge density vanish at $r=0$,
reach there maximal value 
outside the core at $r=1/M_H$ and subsequently drop off exponentially 
with mass $M_-$ at larger distances.}
\label{qCShigg}
\end{figure}

Let us now briefly recall the structure of the electromagnetic 
fields of the vortex solution in the full range of parameters. 
To start with, 
the distribution of the magnetic field $B=\partial_1 A^2-\partial_2 A^1$ 
strongly depends on the topological mass $\mu$. 
For $\mu=0$,  we  are dealing with 
the Abrikosov-Nielsen-Olesen vortex~\cite{abri}.
In that case, the magnetic field reaches its maximal value 
at the center ($r=0$) of the vortex and 
drops off exponentially with mass $M_A$ at distances $r > 1/M_H$.
For $\mu \neq 0$, the magnetic field 
then decays exponentially with mass $M_-$ at distances $r > 1/M_H$.
Moreover, as $|\mu|$ increases from zero, 
the magnetic field at the origin $r=0$ diminishes 
until it completely vanishes in the 
so-called CS limit: $e,|\mu| \rightarrow\infty$, 
with fixed ratio $e^2/\mu$~\cite{boya}. 
(Note that in case the topological mass $\mu$ is quantized as~(\ref{mu}), 
this limit simply means $e \rightarrow \infty$ leaving 
the CS parameter $p$ fixed.) 
Hence, in the CS limit, which amounts to 
neglecting the Maxwell term in~(\ref{higgspa}), the magnetic 
field is localized in a ring-shaped region around 
the core at $r=1/M_H$ as depicted
in figure~\ref{qCShigg} \cite{hong}--\cite{boya}.
Further, as indicated by the zeroth component of the 
field equation~(\ref{fieldequation}), a magnetic field distribution 
$B$ generates an electric field distribution ${\mbox{\bf E}}$ 
iff $\mu \neq 0$.  
These electric fields are localized in a ring shaped region around 
the core at $1/M_H$ for all values of $\mu \neq 0$. 
Specifically, 
they vanish at $r=0$ and fall off exponentially with mass $M_-$ at distances 
$r>1/M_H$.
We have seen in~(\ref{scherm}) how these electric fields 
induced by the magnetic field of the vortex 
are screened by the CS Higgs medium occurring at $r>1/M_H$. 
A screening charge density $j_{\mbox{\scriptsize scr}}^0$
develops in the neighborhood of the core of the vortex,
which falls off exponentially with mass $M_-$ at $r>1/M_H$. 
In this static case, the screening 
charge density boils down to $j^0_{\mbox{\scriptsize scr}} = -  M_A^2 A^0$, 
i.e.\ the Goldstone boson does not contribute. 
The analytical and numerical evaluations in for 
example~\cite{hong}--\cite{boya}
show that the distribution of $A_0$ is indeed of 
the shape described above.

The spin that can be calculated for this classical 
CS vortex solution takes the 
value  
\bea           \label{spinajax}
s \; = \; \int d\,^2 x \, \epsilon^{ij} x^i \, T^{0j} \;  = 
\; \frac{\mu \phi^2}{4\pi} \, ,
\eea
where $T^{0j}$ denotes the energy momentum tensor~\cite{hong}--\cite{boya}.
Note that this spin value is consistent with the quantum statistics phase
$\exp(\im \Theta)=\exp(\im \mu \phi^2/2)$ in~(\ref{qs}). 
That is, these vortices satisfy the canonical spin-statistics
connection~(\ref{spistath}).
This is actually a good point to resolve some inaccuracies in the 
literature. It  is often stated (e.g.\ \cite{hong,boya})
that it is the fact that the CS vortices carry 
the charge~(\ref{scherm}) which leads to nontrivial
AB interactions among these vortices. 
As we have argued, however, the screening 
charges $q_{\mbox{\scriptsize scr}}$ do not couple to the AB 
interactions~\cite{sam} and the  phases in~(\ref{fluxAB}) and~(\ref{qs}) 
are entirely due to the CS term~(\ref{CStyp1}).
In fact, erroneously assuming that the screening charges 
accompanying the vortices do couple to the AB 
interactions leads to the quantum statistics phase 
$\exp(-\im \mu \phi^2/2)$, which is inconsistent with 
the spin~(\ref{spinajax}) carried by these vortices. 
In this respect, we  remark  that 
the correct quantum statistics phase~(\ref{qs}) for the vortices 
has also been derived in the dual formulation of this model~\cite{kleep}.

To my knowledge, the nature of the static point charge solutions 
$j=(q \delta({\mbox{\bf x}}), 0,0)$ of the field equations~(\ref{fequus}) 
have not been studied in the literature so far. 
An interesting question in this context is with which mass~(\ref{mass})
the electromagnetic fields fall off around these  matter charges. 
It is tempting to
conjecture that this exponential decay corresponds to the mass $M_+$. 
The appealing overall picture would then become that the magnetic 
vortices $\phi$ excite the $-$~polarization of the massive photon in 
the CS Higgs medium, whereas the $+$~polarization is excited around the 
matter charges $q$.

\subsection{ $\Z_N$ Chern-Simons  theory}
\label{rev}

I turn to the inclusion of Dirac monopoles in the 
$\Z_N$ CS Higgs phase discussed in the previous subsection.
In other words, $\mu$ is quantized as~(\ref{mu}) in the 
following. Among other things, it will be argued that with 
this particular quantization the $\Z_N$ CS theory describing 
the long distance physics in this Higgs phase corresponds to the 
3-cocycle $\omega_I$ determined by the homomorphism~(\ref{homoI}) 
for this case~\cite{spm1,sm}.

\begin{figure}[tbh] 
\begin{center}
\begin{picture}(85,80)(-15,-15)
\put(-5,-5){\dashbox(40,40)[t]{}}
\put(-15,0){\line(1,0){60}}
\put(0,-15){\line(0,1){60}}
\thinlines
\multiput(-10,-10)(0,10){6}{\multiput(0,0)(10,0){5}{\circle*{2.0}}}
\multiput(0,-10)(0,10){6}{\multiput(0,0)(20,0){3}{\circle*{2.0}}}
\put(0,40){\circle{3.6}}
\put(0,40){\vector(1,-2){20}} 
\put(-3,42){\vector(0,1){5}}
\put(43,-3){\vector(1,0){5}}
\put(-10,50){\small$\phi[\frac{\pi}{2e}]$}
\put(50,-6){\small$q[e]$}
\put(18,-24){(a)}
\end{picture}
\hspace{2cm}
\begin{picture}(85,80)(-15,-15)
\put(-5,-5){\dashbox(40,40)[t]{}}
\put(-15,0){\line(1,0){60}}
\put(0,-15){\line(0,1){60}}
\thinlines
\multiput(-15,-10)(5,10){6}{\multiput(0,0)(10,0){5}{\circle*{2.0}}}
\multiput(-5,-10)(5,10){6}{\multiput(0,0)(20,0){2}{\circle*{2.0}}}
\multiput(-15,10)(5,10){4}{\multiput(0,0)(20,0){1}{\circle*{2.0}}}
\multiput(35,-10)(5,10){4}{\multiput(0,0)(20,0){1}{\circle*{2.0}}}
\put(-3,42){\vector(0,1){5}}
\put(43,-3){\vector(1,0){5}}
\put(20,40){\circle{3.6}}
\put(20,40){\vector(0,-1){40}}
\put(-10,50){\small$\phi[\frac{\pi}{2e}]$}
\put(50,-6){\small$-q_{\mbox{\scriptsize scr}}[e]$}
\put(18,-24){(b)}
\end{picture}
\end{center} 
\vspace{0.5cm}
\caption{\sl The spectrum of a $\Z_4$ CS Higgs phase compactifies
to the particles inside the dashed box.
We depict the flux $\phi$ versus the matter charge $q$
and the screening charge  $-q_{\mbox{\protect\scriptsize scr}}=q+\mu\phi$ 
respectively. 
The CS parameter $\mu$ is set to its minimal nontrivial value 
$\mu=\frac{e^2}{\pi}$, that is,  $p=1$. 
The arrows visualize the tunneling event induced by a minimal Dirac monopole.}
\label{z41}
\end{figure}
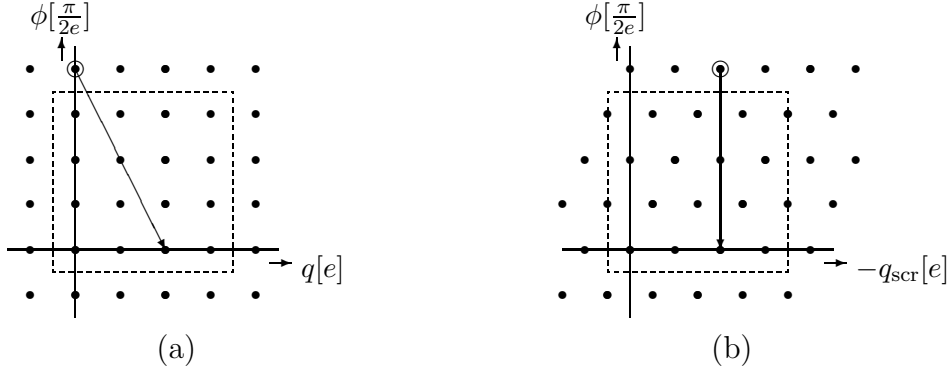

As we have seen in the previous subsection, the Higgs mechanism causes the 
identification of charge and flux occurring in unbroken CS 
electrodynamics to disappear.
That is, the  spectrum of the $\Z_N$ CS Higgs phase consists
of the quantized matter charges $q=ne$, the quantized magnetic fluxes 
$\phi=\frac{2\pi a}{Ne}$ and dyonic combinations of the two. 
We will label these particles as $(a,n)$ with  $a,n \in \Z$.
Upon implementing the quantization~(\ref{mass}) 
of the topological mass $\mu$, the AB 
interactions~(\ref{klam}), (\ref{fluxAB}) and~(\ref{qs}) 
can  be cast in the form
\bea                      \label{monoI}
{\cal R}^2\;|a,n\rangle |a',n'\rangle &=&
e^{\frac{2 \pi \im}{N} (n a' +n'a
+\frac{2p}{N}aa')}
\;|a,n\rangle |a',n'\rangle \\    \label{bradenI}
{\cal R} \; |a,n\rangle |a,n\rangle &=&
e^{\frac{2 \pi \im}{N} (n a +\frac{p}{N}aa)}
\;|a,n\rangle |a,n\rangle \\ \label{sfI}
T \;  |a,n\rangle &=& e^{\frac{2 \pi \im}{N}(na + \frac{p}{N}aa)} 
\; |a,n \rangle \, , 
\eea 
where $p$ denotes the integral CS parameter. Expression~(\ref{sfI}) 
contains the spin factors assigned to the particles. 
Under these topological interactions, the charge label $n$ obviously 
becomes a $\Z_N$ quantum number, i.e.\ at large distances 
we are only able to distinguish the charges $n$ modulo $N$.
Furthermore, in the presence of the Dirac monopoles/instantons~(\ref{dcharge}) 
magnetic flux $a$ is conserved modulo $N$. However, the 
flux decay events are now accompanied by charge creation~\cite{spm,sm}.
To be specific, in terms of the integral 
flux and charge quantum numbers $a$ and $n$,
the tunneling event induced by the minimal Dirac monopole
can be recapitulated as 
\bea                            \label{instanz}
\mbox{instanton:}  & &      \left\{ \ba{lcl}    
a &\mapsto& a -N  \\
n &\mapsto& n+2p \, .  
\ea
\right.          
\eea
I have depicted this effect of a Dirac monopole in the spectrum 
of a $\Z_4$ CS Higgs phase in figure~\ref{z41}.
Recall from section~\ref{ub} that  
the quantization~(\ref{mu}) of the topological mass  was such that 
the particles connected by monopoles were invisible to the 
monodromies~(\ref{ons}) and carried the same spin
in the unbroken phase.
This feature naturally  persists in this broken phase.
It is readily checked that the particles connected by 
the monopole~(\ref{instanz}) can not be distinguished 
by the AB interactions~(\ref{monoI}) and give rise to the same 
spin factor~(\ref{sfI}). 
As a result, the spectrum of this broken phase can be presented as
\bea                              \label{compsp}
(a,n)  \qquad \qquad \qquad \mbox{with} \qquad a,n \in 0,1, \ldots, N-1 \, ,
\eea 
where it is understood that the modulo $N$ calculus for the magnetic 
fluxes $a$ involves the charge jump~(\ref{instanz}).

Let us now explicitly verify that we are indeed dealing with a 
$\Z_N$ gauge theory with CS action~(\ref{type1}), i.e.
\bea     \label{seedorf}
\omega_{\mbox{\scriptsize I}} (a,b,c)    &=& 
\exp \left( \frac{2 \pi \im p}{N^2} \,
a(b +c -[b+c]) \right) \, ,
\eea 
where the rectangular brackets denote modulo $N$
calculus such that the sum always lies in the range $ 0,1, \ldots, N-1$.
First of all, the different particles~(\ref{compsp}) constitute
the compactified spectrum on which the quasi-quantum double 
$D^{\omega_I}(\Z_N)$ acts. The additional AB interactions 
among the fluxes are then absorbed in the definition of the 
dyon charges.~\footnote{In fact, the more accurate statement 
at this point~\cite{spm1} is that the fluxes $\phi$ enter 
the Noether charge $\tilde{Q}$ which generates the residual $\Z_N$ symmetry
in the presence of a CS term.  That is, 
$\tilde{Q}=q +\frac{\mu}{2} \phi$, with $q$ the usual 
contribution of a matter charge.}
To be specific, the dyon charge~(\ref{rei}) corresponding  
to the flux $a$ is given by  ${\alpha}(b) = \varepsilon_a(b) \;
 \Gamma^{n} (b)$ 
with $\varepsilon_a(b)$ defined in~(\ref{epi}) as
\bea        \label{vareen}
\varepsilon_a(b) &=& \exp \left( \frac{2 \pi \im p}{N^2} ab \right) ,
\eea 
and $\Gamma^{n} (b) = \exp \left( \frac{2 \pi \im }{N} nb \right)$
an UIR of $\Z_N$.
The action of the braid operator~(\ref{braidaction}) now gives rise to the 
AB phases presented in~(\ref{monoI}) and~(\ref{bradenI}),
whereas the action of the central element~(\ref{spin13}) yields the 
spin factor~(\ref{sfI}).   Furthermore, the fusion rules for 
$D^{\omega_I}(\Z_N)$  following from~(\ref{Ncoef}) 
\bea   \label{CSfusion}
(a,n) \times (a',n') &=& 
\left( [a+a'],[n+n'+\frac{2p}{N}(a+a'-[a+a'])] \right) \, ,
\eea  
express the tunneling properties of the Dirac monopoles.    
Specifically, if the sum of the fluxes $a+a'$ exceeds $N-1$, the composite 
carries unstable flux and tunnels back to the range~(\ref{compsp}) 
by means of the charged monopole~(\ref{instanz}).
Note that the charge jump induced by the monopole 
for CS parameter $p\neq 0$  implies that the fusion 
algebra now equals $\Z_{kN} \times \Z_{N/k}$~\cite{dvvv}.
Here, we defined $k := N/{\gcd}(p,N)$  for odd $N$ and 
$k := N/{\gcd}(2p,N)$ for even $N$, where  $\gcd$ stands for the 
greatest common divisor.
In particular, for odd $N$ and $p=1$, 
the complete spectrum is generated by the single magnetic flux $a=1$.
Finally, the charge conjugation operator 
${\cal C}=S^2$ following from~(\ref{fusion})   takes the form
\bea
{\cal C} \; (a,n) &=& \left(
[-a], [-n + \frac{2p}{N}(-a-[-a])]\right) \, .
\eea 
So, as usual, under the  action of the charge conjugation
operator the fluxes $a$ and charges $n$ reverse sign. Subsequently,
the `twisted' modulo $N$ calculus for the fluxes~(\ref{instanz})  
and the ordinary modulo $N$ calculus for the charges are applied 
to return to the range~(\ref{compsp}). Also, note 
that the particles and anti-particles in this theory
naturally carry the same spin, i.e.\  
the action~(\ref{sfI}) of the modular $T$ matrix indeed commutes with 
${\cal C}$.

Having established that the U(1) CS term~(\ref{CStyp1}) gives rise to 
the 3-cocycle~(\ref{seedorf}) in the residual $\Z_N$ gauge theory in the 
Higgs phase, I now turn to  the periodicity $N$ of the CS parameter $p$ 
indicated by the homomorphism~(\ref{homoI}). This periodicity can be made 
explicit as follows. From the braid properties~(\ref{monoI}), the spin 
factors~(\ref{sfI}) and the fusion rules~(\ref{CSfusion}), we infer that 
setting  the CS parameter to $p=N$ amounts to an automorphism 
$(a,n) \mapsto (a,[ n+2a])$ of the spectrum~(\ref{compsp}) for $p=0$. 
In other words, for $p=N$ the theory describes the same topological 
interactions between the particles as for $p=0$. We just have relabeled 
the dyons.

\begin{figure}[tbh]    \epsfysize=7.9cm
\centerline{\epsffile{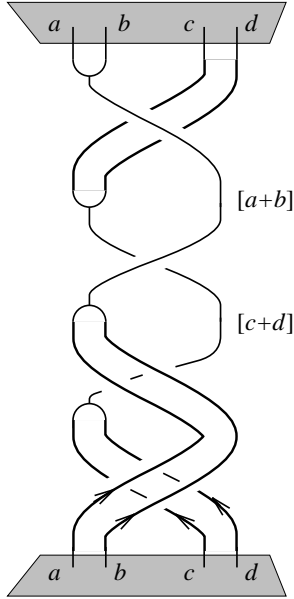}}
\caption{\sl  The 3-cocycle condition states that the topological
action $\exp(\im S_{\mbox{\protect\scriptsize CSI}})$ for this process is
trivial. 
The vertices in which the fluxes are fused correspond to a  minimal  
Dirac instanton iff the total flux of the composite is larger then $N-1$.}
\label{dreip}
\end{figure}    
 
Let me close this section by identifying the process corresponding to the 
CS action~(\ref{seedorf}). A comparison of the 
expressions~(\ref{seedorf}) and~(\ref{vareen}) yields
\bea                     \label{acmilan}
\omega_{\mbox{\scriptsize I}} ( a,b,c) &=& \varepsilon_b(a) \; 
\varepsilon_c(a) 
\; \varepsilon^{-1}_{[b+c]} (a) \, ,
\eea 
from which we immediately conclude 
\bea    \label{intermilan}
\omega_{\mbox{\scriptsize I}} ( a,b,c)
&=&
\exp (\im S_{\mbox{\scriptsize CSI}}) 
\left\{\vcenter{\epsfxsize=1,5cm\epsfbox{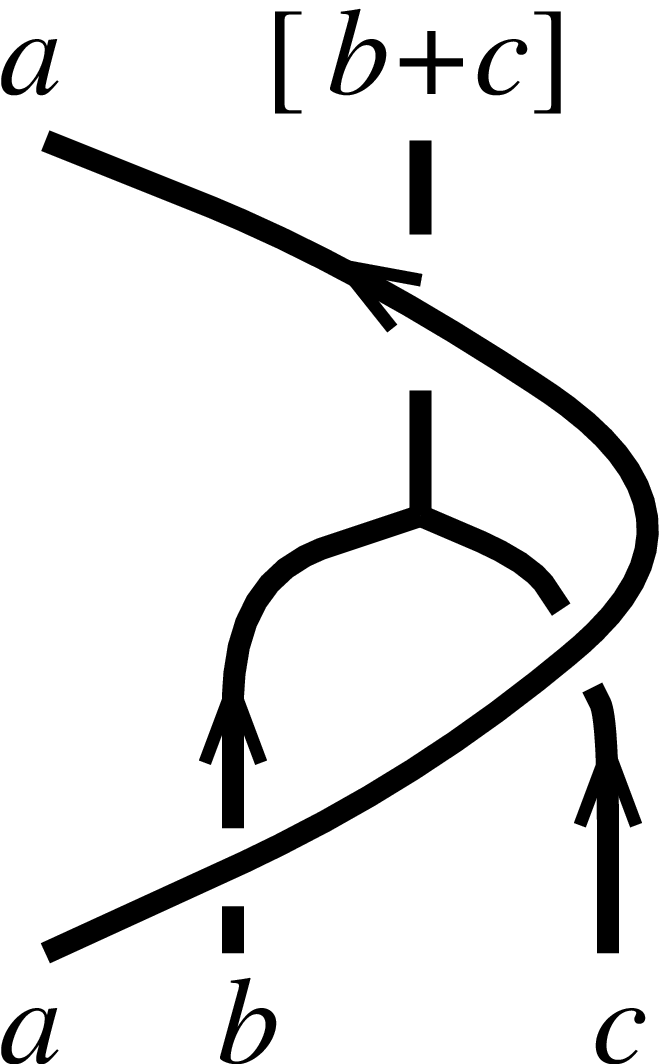}}\right\} \,.
\eea
Here, the fluxes $a$,$b$ and $c$ are again assumed to take values in 
the range $0,1,\ldots, N-1$. For convenience, the trajectories of the 
fluxes are depicted as lines rather than 
ribbons.~\footnote{To avoid confusion,
there is no writhing of particle trajectories involved in the 
following argument.}   
The vertex corresponding to fusion of  
the fluxes $b$ and $c$ then describes the tunneling event~(\ref{instanz}) 
induced by the minimal Dirac monopole iff the total flux $b+c$ of the 
composite exceeds $N-1$. Of course, the total AB phase for the process 
depicted in~(\ref{intermilan}), which also involves the matter 
coupling~(\ref{maco}), is trivial as witnessed by the 
fact that the quasitriangularity condition~(\ref{tria2}) is satisfied.
The contribution~(\ref{intermilan}) of the CS term~(\ref{CStyp1}) to this 
total AB phase, however, is nontrivial iff the vertex corresponds to a 
monopole. It only generates AB phases between magnetic fluxes and therefore
only notices the flux tunneling at the vertex and not the charge creation. 
Specifically, in the first  braiding of the process~(\ref{intermilan}), 
the CS coupling generates the AB phase $\varepsilon_b(a)$, 
in the second  $\varepsilon_c(a)$ and 
in the last $\varepsilon^{-1}_{[b+c]}(a)$. 
Hence, the total CS action for this process 
indeed becomes~(\ref{acmilan}).
With the prescription~(\ref{intermilan}), 
factorization of the topological action, 
the so-called skein relation
\bea     \label{skein}
\exp (\im S_{\mbox{\scriptsize CSI}})
\left\{\vcenter{\epsfxsize=1,4cm\epsfbox{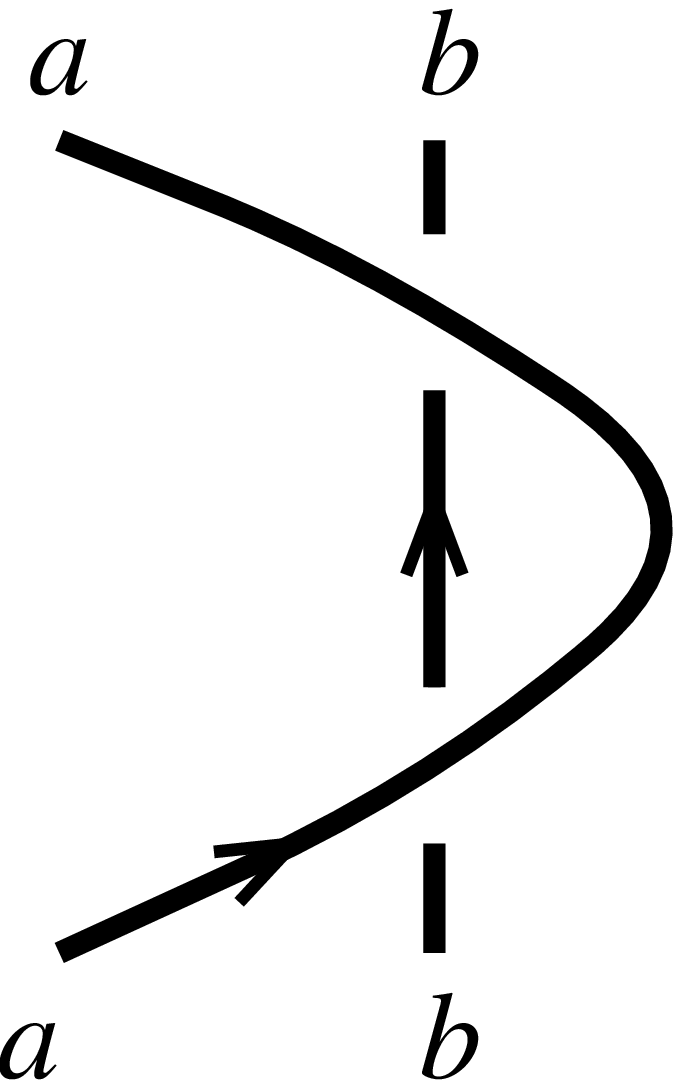}}\right\} \; = \;
\exp (\im S_{\mbox{\scriptsize CSI}})
\left\{\vcenter{\epsfxsize=1cm\epsfbox{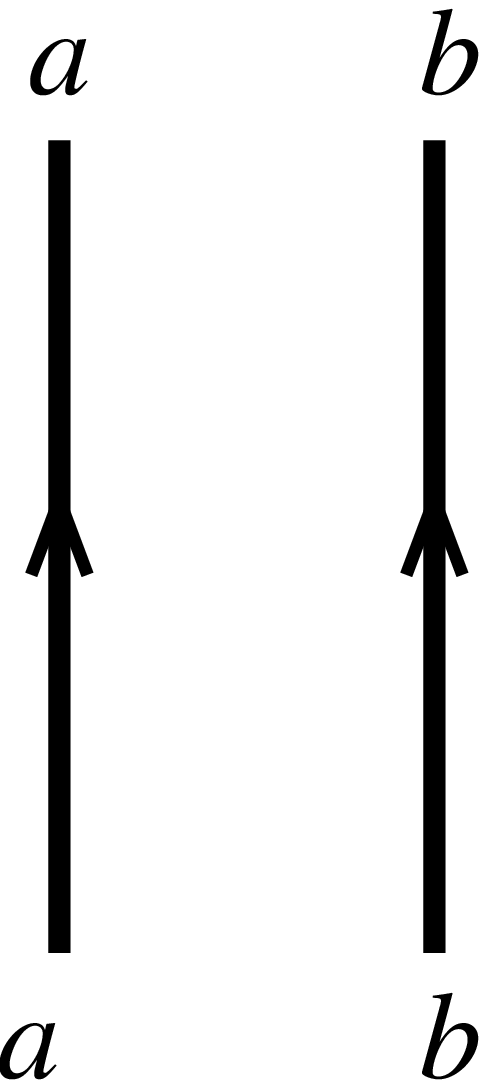}}\right\} \; = \;1 \, ,
\eea   
and the obvious relation
\bea
\omega_{\mbox{\scriptsize I}}^{-1} ( a,b,c)
&=&
\exp (\im S_{\mbox{\scriptsize CSI}}) 
\left\{\vcenter{\epsfxsize=1,4cm\epsfbox{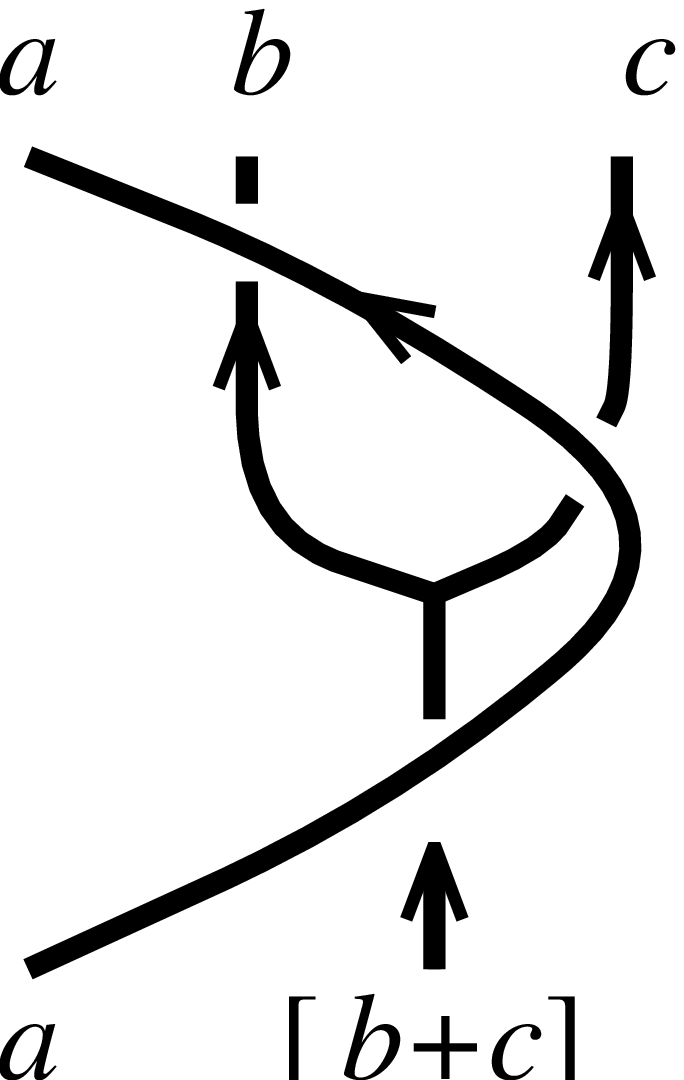}}\right\}\, ,
\eea
it is then  readily verified that  the 3-cocycle condition
\bea           \label{repetee}
\omega_{\mbox{\scriptsize I}} ( a,b,c) \;\,  
\omega_{\mbox{\scriptsize I}} ( a,[b+c], d) \;\,
\omega_{\mbox{\scriptsize I}}^{-1}  ( a,b,[c+d]) \;\,    
\omega_{\mbox{\scriptsize I}}^{-1}  ( [a+b],c,d ) \;\,    
\omega_{\mbox{\scriptsize I}} ( b,c,d ) \; = \; 1  \, ,  
\eea
boils down to the statement that the topological 
action $\exp(\im S_{\mbox{\scriptsize CSI}})$ for the process depicted 
in figure~\ref{dreip} is trivial. In fact, this condition
can now be interpreted as the requirement that the particles connected
by Dirac monopoles should give rise to the same spin factor, 
which, in turn, imposes the quantization~(\ref{mu}) of the topological mass.
To that end, I first note that iff the total flux of either one of the 
particle pairs in figure~\ref{dreip} does not exceed $N-1$, i.e.\ 
$a+b <N-1$ and/or $c+d < N-1$,  
the 3-cocycle condition~(\ref{repetee}) is trivially satisfied,
as follows from the skein relation~(\ref{skein}).
When both pairs carry flux larger then $N-1$, all vertices 
in figure~\ref{dreip} correspond to Dirac monopoles~(\ref{instanz}), 
transferring fluxes $N$ into the charges $2p$ and vice versa.
The requirement that the action $\exp(\im S_{\mbox{\scriptsize CSI}})$ for 
this process 
is trivial now becomes nonempty. Let us, for example, consider the case
$a+b=N$ and $c+d = N$. Each pair may 
then be viewed as a single  particle carrying either unstable flux $N$ 
or charge $2p$ depending on the vertex it has crossed.
The total CS  action $\exp(\im S_{\mbox{\scriptsize CSI}})$ for 
this case
then reduces to the quantum statistical parameter (or spin factor)
$\varepsilon_N(N)= \exp(2\pi \im p)$ generated  in the first
braiding.  Note that this AB phase is {\em not} cancelled 
by the one implied by the matter coupling~(\ref{maco}) for this process.
To be specific,  this AB phase becomes
$\exp(\im S_{\mbox{\scriptsize matter}})=\exp(-4\pi \im p)$ corresponding 
to  the 
second braiding in figure~\ref{dreip} where the charge $2p$ is exchanged
with the flux $N$ in a clockwise fashion. The last two braidings do not 
contribute. Upon demanding the 
total topological action 
$\exp(\im S_{\mbox{\scriptsize CSI}}+\im S_{\mbox{\scriptsize matter}})= 
\exp(-2\pi \im p)$
to be trivial, we finally rederive the fact that the CS parameter 
$p$ has to be  integral.
To conclude, the 3-cocycle condition~(\ref{repetee}) is necessary and 
sufficient for a consistent implementation of Dirac monopoles
in a $\Z_N$ CS gauge theory.

\sectiona{Type II $U(1) \times U(1)$ Chern-Simons theory} 
\label{typeII}

The type~II CS terms~(\ref{CSt2})   
establish pairwise couplings between 
the different $U(1)$ gauge fields $A^{(i)}_\kappa$ of a direct product 
gauge group $U(1)^k$.
In this section, I discuss the simplest example~\footnote{The 
generalization of the following analysis to $k>2$ is straightforward.}
of such a CS theory of type~II, namely
that with gauge group $U(1) \times U(1)$ spontaneously broken down to 
the product of two cyclic groups 
$\Z_{N^{(1)}} \times \Z_{N^{(2)}}$.
Specifically, the spontaneously broken planar 
$U(1) \times U(1)$ CS theory to be studied here is of the form  
\bea        \label{action12}
S &=&  \int d\,^3x \;  ({\cal L}_{\mbox{\scriptsize YMH}} + 
{\cal L}_{\mbox{\scriptsize matter}}
+ {\cal L}_{\mbox{\scriptsize CSII}})  \\
{\cal L}_{\mbox{\scriptsize YMH}} &=& 
\sum_{i=1}^2\{-\frac{1}{4}F^{(i)\kappa\nu} F^{(i)}_{\kappa\nu} + 
({\cal D}^\kappa \Phi^{(i)})^*{\cal D}_\kappa \Phi^{(i)} - V(|\Phi^{(i)}|)\} \\
{\cal L}_{\mbox{\scriptsize matter}} &=& 
-\sum_{i=1}^2 j^{(i)\kappa}A^{(i)}_{\kappa}  
\label{j12mat} \\
{\cal L}_{\mbox{\scriptsize CSII}} &=& \frac{\mu}{2} \epsilon^{\kappa\nu\tau} 
A^{(1)}_{\kappa} \partial_{\nu}A^{(2)}_{\tau} \, ,    \label{CSac12}
\eea 
where $A^{(1)}_\kappa$ and $A^{(2)}_\kappa$  denote 
the two different $U(1)$ gauge fields.
I assume that these gauge symmetries are realized with quantized charges,
i.e.\ the $U(1)$ gauge groups are compact.
To keep the discussion general, however,
different fundamental charges for the two different
compact $U(1)$ gauge  groups are allowed. The fundamental charge
associated to the gauge field $A^{(1)}_\kappa$ is denoted by $e^{(1)}$, 
while $e^{(2)}$ denotes the fundamental charge for $A^{(2)}_\kappa$. 
The  two Higgs fields $\Phi^{(1)}$ and $\Phi^{(2)}$, respectively,
are then assumed to carry charge $N^{(1)}e^{(1)}$ 
and $N^{(2)}e^{(2)}$, i.e.\ ${\cal D}_\kappa \Phi^{(i)}= 
(\partial_{\kappa}+\im N^{(i)}e^{(i)} A_{\kappa}^{(i)})\Phi^{(i)}$.
The charges introduced by the matter currents
$j^{(1)}$ and $j^{(2)}$ in~(\ref{j12mat}), in turn,  are quantized
as $q^{(1)}= n^{(1)} e^{(1)}$ and $q^{(2)}= n^{(2)} e^{(2)}$, respectively, 
with $n^{(1)},n^{(2)} \in \Z$. 
For convenience, both Higgs fields are endowed
with the same (nonvanishing) vacuum expectation value $v$
\bea                         \label{poti}
V(|\Phi^{(i)}|) &=& \frac{\lambda}{4}(|\Phi^{(i)}|^2-v^2)^2  \qquad\qquad
 \lambda, v > 0 \qquad \mbox{and $i=1,2 \,.$}
\eea 
Hence, both compact $U(1)$  gauge groups are spontaneously broken down 
at the same energy scale $M_H = v \sqrt{2\lambda}$.

We proceed along the line of argument in the previous section. 
So, we start with an analysis of the unbroken phase 
and present the argument for  
the quantization~(\ref{quantmuij}) of the topological mass $\mu$
in the presence of Dirac monopoles in subsection~\ref{ub12}. 
In subsection~\ref{br12}, we then discuss 
the CS Higgs screening mechanism in the broken phase and 
establish the AB interactions 
between the charges and magnetic fluxes in the spectrum. 
Finally, subsection~\ref{revii} contains  
a study of the type~II $\Z_{N^{(1)}} \times \Z_{N^{(2)}}$ CS theory 
describing the long distance physics in the broken phase of this model.

\subsection{Unbroken phase with Dirac monopoles}
\label{ub12}

In this subsection, we address the implications of the presence 
of Dirac monopoles in the {\em unbroken} phase of the 
model~(\ref{action12}). That is, we set $v=0$ and 
$\mu \neq 0$ for the moment.

Variation of the action~(\ref{action12}) 
w.r.t.\ the gauge fields $A_{\kappa}^{(1)}$ 
and $A_{\kappa}^{(2)}$, respectively, 
gives rise to the following field equations 
\bea       \label{fe1}  \ba{lcl}             
\partial_\nu F^{(1) \;\nu\kappa} + 
\frac{\mu}{2}\epsilon^{\kappa\nu\tau} \partial_{\nu}A^{(2)}_{\tau} &=& 
j^{(1)\;\kappa}+j^{(1)\;\kappa}_H                                     \\
\partial_\nu F^{(2) \;\nu\kappa} +
\frac{\mu}{2}\epsilon^{\kappa\nu\tau} \partial_{\nu}A^{(1)}_{\tau} &=&
j^{(2)\;\kappa}+j^{(2)\;\kappa}_H \, , 
\ea 
\eea
with $j^{(i)}$ the two matter currents in~(\ref{j12mat}) and
$j^{(i)}_H$ the two Higgs currents in this model.
This coupled set of differential equations 
leads to Klein-Gordon equations for the 
dual field strengths $\tilde{F}^{(1)}$ and $\tilde{F}^{(2)}$ 
with mass $|\mu|/2$.
Thus the field strengths fall off exponentially and 
the Gauss' laws take the form
\bea    \label{g1}   \ba{lcl}
Q^{(1)} &=& q^{(1)} +q_H^{(1)} + \frac{\mu}{2} \phi^{(2)} \; = \; 0   \\
Q^{(2)} &=& q^{(2)} +q_H^{(2)} + \frac{\mu}{2} \phi^{(1)} \; = \; 0 \, ,
\ea
\eea
with $Q^{(i)} =  \int\! d\, ^2x\, \nabla \!\cdot\! \mbox{\bf E}^{(i)}=0$, 
$\phi^{(i)} = \int \! d\,^2x \,\epsilon^{jk}\partial_j A^{(i) \; k}$,
$q^{(i)} = \int\! d\, ^2x \, j^{(i) \; 0}$ and 
$q_H^{(i)} \; = \; \int\! d\,^2x \, j^{(i) \; 0}_H$. 
Hence, the screening mechanism operating 
in this theory attaches fluxes which belong to one $U(1)$ gauge group 
to the charges of the other~\cite{hagen}.

The long range interactions that remain between the 
particles in the spectrum of this model 
are the topological AB interactions implied by 
the couplings~(\ref{j12mat}) and~(\ref{CSac12}).       
These can be summarized as~\cite{hagen} 
\bea           \label{ons12}
{\cal R}^2 \; |q^{(1)},q^{(2)}\rangle|q^{(1)'},q^{(2)'}\rangle 
&=&
e^{-\im (\frac{2 q^{(1)} q^{(2)'}}{\mu} \, + \, \frac{2 q^{(1)'} 
q^{(2)}}{\mu})}         \, 
|q^{(1)},q^{(2)}\rangle|q^{(1)'},q^{(2)'}\rangle  \\
{\cal R}  \; |q^{(1)},q^{(2)}\rangle|q^{(1)},q^{(2)}\rangle  
&=&
e^{-\im \frac{2 q^{(1)} q^{(2)}}{\mu}} \,
|q^{(1)},q^{(2)}\rangle|q^{(1)},q^{(2)}\rangle \, .   \label{ons12an}
\eea
From~(\ref{ons12an}), we then conclude that the only particles
endowed with a nontrivial spin are those that carry 
charges w.r.t.\ both $U(1)$ gauge groups. 
In other words, only these particles obey
anyon statistics. The other particles are bosons.

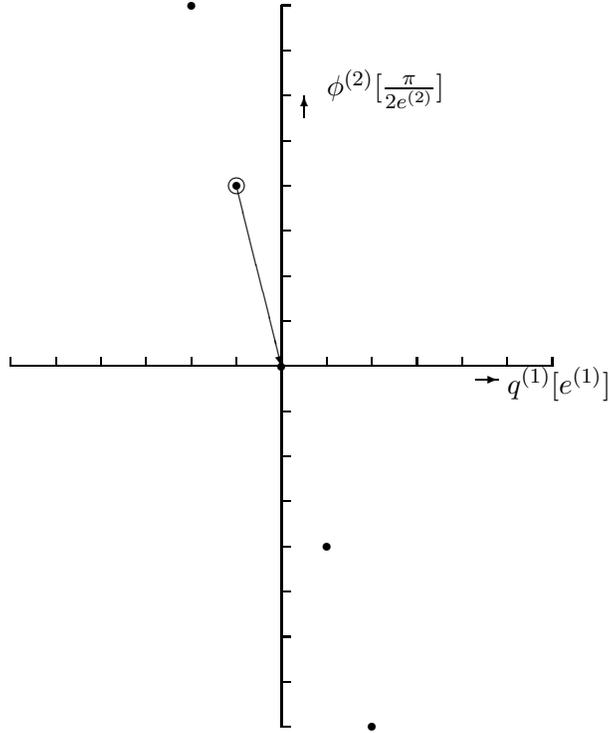
\begin{figure}[tbh] 
\begin{center}
\begin{picture}(125,160)(-60,-80)
\put(-60,0){\line(1,0){120}}
\put(0,-80){\line(0,1){160}}
\thinlines
\multiput(-60,0)(10,0){13}{\line(0,1){2}}
\multiput(0,-80)(0,10){17}{\line(1,0){2}}
\multiput(-20,80)(10,-40){5}{\circle*{2.0}}
\put(-10,40){\circle{3.6}}
\put(-10,40){\vector(1,-4){10}} 
\put(5,55){\vector(0,1){5}}
\put(43,-3){\vector(1,0){5}}
\put(10,60){\small $\phi^{(2)}[\frac{\pi}{2e^{(2)}}]$}
\put(50,-6){\small$q^{(1)} [e^{(1)}]$}
\end{picture}
\end{center}
\vspace{0.5cm}
\caption{\sl Spectrum of unbroken  $U(1) \times U(1)$ 
CS theory of type~II. 
We just depict the $q^{(1)}$ versus $\phi^{(2)}$ diagram. 
The topological mass $\mu$ is set to its minimal nontrivial value 
$\mu=\frac{e^{(1)} e^{(2)}}{\pi}$, i.e.\ $p=1$. 
The arrow represents the tunneling induced by a {\em charged} 
Dirac monopole/instanton $(2)$, which indicates that there are no stable 
particles in this theory for $p=1$.
The charge/flux diagram for $q^{(2)}$ versus $\phi^{(1)}$ 
is obtained from this one by the replacement $(1) \leftrightarrow (2)$.}
\label{u12}
\end{figure}

We proceed with the incorporation of Dirac monopoles/instantons in 
this compact CS gauge theory (see also reference~\cite{diamant}).
There are two different species associated to the two compact 
$U(1)$ gauge groups. The magnetic charges carried by  these  
Dirac monopoles are quantized as
$g^{(i)} = \frac{2\pi m^{(i)}}{e^{(i)}}$ with $m^{(i)} \in \Z$ and 
$i=1,2$.
Given the coupling  between 
the two $U(1)$ gauge fields established by the 
CS term~(\ref{CSac12}), 
the magnetic flux tunnelings induced by these monopoles
in one $U(1)$ gauge group are accompanied by charge tunnelings in the other.
Specifically, as indicated by the Gauss' laws~(\ref{g1}), 
the tunnelings associated with the two minimal Dirac monopoles become
\bea                             \label{inst1}
\mbox{instanton $(1)$ :} & & \left\{   \ba{lcllcl}
\Delta \phi^{(1)}  & = &  -\frac{2\pi}{e^{(1)}} \, , 
& \qquad \Delta \phi^{(2)}  &=&   0 \\
\Delta q^{(1)}  &=&  0 \, , 
& \qquad \Delta q^{(2)}  &=&   \mu \frac{\pi}{e^{(1)}} 
\ea \right.                                            \\
      &&  \nn \\                        
\label{inst2}
\mbox{instanton $(2)$ :} & & \left\{   \ba{lcllcl}
\Delta \phi^{(1)}  &=&   0 \, , 
& \qquad  \;  \Delta \phi^{(2)}  &=&  -\frac{2\pi}{e^{(2)}} \\
\Delta q^{(1)}  &=&  \mu \frac{\pi}{e^{(2)}} \, , 
& \qquad  \; \Delta q^{(2)}  &=&  0 \, . 
\ea \right.
\eea
The presence of the Dirac monopole~(\ref{inst1}) implies 
quantization of the charges $q^{(1)}$ in multiples of~$e^{(1)}$. 
This can be seen by the following simple argument. 
The tunneling event induced by the monopole~(\ref{inst1})
should be invisible to the monodromies involving 
the various charges in the spectrum of this  theory.
Hence, from~(\ref{ons12}) we infer that  the AB phase
$\exp (-\im \frac{ 2 q^{(1)} \Delta q^{(2)}}{\mu})
=\exp (- \im \frac{2\pi q^{(1)}}{e^{(1)}})$  should be trivial. Therefore,
$q^{(1)}=n^{(1)} e^{(1)}$ with $n^{(1)} \in \Z$.
In a similar way, we infer that the presence of the 
Dirac monopole~(\ref{inst2})
leads to quantization of $q^{(2)}$ in multiples of $e^{(2)}$.
Moreover, for consistency,
the tunneling events induced by the monopoles should 
respect these quantization rules for $q^{(1)}$ and $q^{(2)}$.
As follows from~(\ref{inst1}) and~(\ref{inst2}), this means that 
the topological mass is necessarily quantized as~\footnote{This is the same 
quantization condition derived by Diamantani et al.~\cite{diamant} 
using a somewhat different argument.}
\bea                 \label{mu12}
\mu &=& \frac{pe^{(1)}e^{(2)}}{\pi}   \qquad \qquad 
\mbox{with $p \in \Z$} \, ,
\eea
which is the result alluded to in~(\ref{quantmuij}).
It is easily verified that the consistency 
demand requiring the particles
connected by Dirac monopoles to give rise to the same spin factor
or quantum statistics phase~(\ref{ons12an})
does {\em not} lead to a further constraint on $\mu$ in this case.

To conclude, the spectrum of this unbroken 
$U(1) \times U(1)$ CS theory featuring Dirac monopoles
can be presented as in~figure \ref{u12}. 
The modulo calculus for the charges $q^{(1)}$ and $q^{(2)}$
induced by the Dirac monopoles~(\ref{inst2}) and~(\ref{inst1}), respectively,
implies a compactification of the spectrum 
to $(p-1)^2$ different stable particles~\cite{diamant,thesis} with $p$ 
the integral CS parameter in~(\ref{mu12}).

\subsection{Higgs phase}   \label{br12}

We now switch on  the  Higgs mechanism, so $v \neq 0$ in the following.
At energies well below the symmetry breaking 
scale $M_H=v \sqrt{2\lambda}$
both  Higgs fields $\Phi^{(i)}$ are then completely condensed:  
$ \Phi^{(i)}(x) \mapsto v \exp (\im \sigma^{(i)}(x) ) $
for $i=1,2$.
Hence, the dynamics of the CS Higgs medium in this model
is described by the effective action obtained 
from the following simplification in~(\ref{action12})
\bea  \label{ma12}
({\cal D}^{\kappa}\Phi^{(i)})^* {\cal D}_{\kappa}\Phi^{(i)}
-V(|\Phi^{(i)}|)   & \longmapsto &
\frac{M_A^{(i)\,2}}{2} \tilde A^{(i) \; \kappa}\tilde A^{(i)}_{\kappa} \, ,
\label{mhed12}                   \qquad  \qquad                  
\eea
with $\tilde{A}^{(i)}_{\kappa} :=  A^{(i)}_{\kappa} + 
\frac{1}{N^{(i)}e^{(i)}}\partial_{\kappa}
\sigma^{(i)}$ and 
$M_A^{(i)} := N^{(i)}e^{(i)} v\sqrt{2}$
for  $i=1,2$.   A derivation similar to the one for~(\ref{mass})
reveals that the two polarizations  $+$ and $-$ of the 
photon fields $\tilde{A}^{(i)}_\kappa$  acquire  
masses  $M^{(i)}_{\pm}$ which differ by the topological mass $|\mu|/2$.
I refrain from giving the explicit  expressions
of the masses $M^{(i)}_{\pm}$ in terms of  
$\mu$, $M_A^{(1)}$ and $M_A^{(2)}$.

In this broken phase, 
the Higgs currents $j_H^{(i)}$ appearing in the field 
equations~(\ref{fe1}) become 
screening currents. That is, 
$ j_H^{(i)} \mapsto j_{\mbox{\scriptsize scr}}^{(i)} := 
-M_A^{(i)\,2} \tilde{A}^{(i)}$. 
In particular, the Gauss'  laws~(\ref{g1}) now take the form  
\bea    \label{g1higg}   \ba{lcl}
Q^{(1)} &=& q^{(1)} +q_{\mbox{\scriptsize scr}}^{(1)} + 
\frac{\mu}{2} \phi^{(2)} 
\; = \; 0    \\
Q^{(2)} &=& q^{(2)} +q_{\mbox{\scriptsize scr}}^{(2)} + 
\frac{\mu}{2} \phi^{(1)}
\; = \; 0 \, ,
\ea
\eea   
with 
$q_{\mbox{\scriptsize scr}}^{(i)} :=  
\int \! d\,^2 x \, j^{(i)\,0}_{\mbox{\scriptsize scr}}$.
As we have seen in section~\ref{bp}, the emergence 
of these screening charges $q_{\mbox{\scriptsize scr}}^{(i)}$ 
is at the heart of the 
de-identification of charge and flux  occurring  
in the phase transition from the unbroken phase to 
the broken phase in a CS gauge theory. 
They accompany the matter charges $q^{(i)}$ provided
by the currents $j^{(i)}$ as well as the magnetic vortices.

Let us first focus on the magnetic vortices in this model.  
There are two different species 
associated with the winding of the two different 
Higgs fields $\Phi^{(1)}$ and $\Phi^{(2)}$.
These two different vortex species (both of characteristic size $1/M_H$)
carry the  quantized  magnetic fluxes 
\bea                       \label{fluqu12}
\phi^{(1)} \; = \; \frac{2 \pi a^{(1)}}{N^{(1)}e^{(1)}}
\qquad \mbox{and} \qquad
\phi^{(2)} \;=\; \frac{2 \pi a^{(2)}}{N^{(2)}e^{(2)}}  
\qquad  \mbox{with $a^{(1)}, a^{(2)} \in \Z$}\, ,
\eea
and (as indicated by the Gauss' laws~(\ref{g1higg}))  
induce the screening charges  
$q_{\mbox{\scriptsize scr}}^{(1)} = -\mu\phi^{(2)}/2$  and
$q_{\mbox{\scriptsize scr}}^{(2)} = -\mu\phi^{(1)}/2$, respectively, 
in the Higgs medium. These screening charges completely screen 
the Coulomb fields generated by the magnetic fluxes~(\ref{fluqu12}) 
carried by the vortices, but do {\em not} couple to the AB interactions.
Therefore, the long range AB interactions among  the 
vortices implied by the CS coupling~(\ref{CSac12})
are {\em not} screened
\bea           \label{onsflux12}
{\cal R}^2 \; |\phi^{(1)},\phi^{(2)}\rangle|\phi^{(1)'},\phi^{(2)'}\rangle 
&=&
e^{\im \frac{\mu}{2}(\phi^{(1)} \phi^{(2)'}  + \, \phi^{(1)'} \phi^{(2)})} \, 
|\phi^{(1)},\phi^{(2)}\rangle|\phi^{(1)'},\phi^{(2)'}\rangle  \\
{\cal R}  \; |\phi^{(1)},\phi^{(2)}\rangle|\phi^{(1)},\phi^{(2)}\rangle  
&=&
e^{\im \frac{\mu}{2} \phi^{(1)} \phi^{(2)}} \,
|\phi^{(1)},\phi^{(2)}\rangle|\phi^{(1)},\phi^{(2)}\rangle \,.  
\label{onsfl12an}
\eea    
Note that there are no AB phases generated 
among vortices of the same species. Thus there is only a nontrivial spin 
assigned to composites  carrying flux w.r.t.\ 
both  broken $U(1)$ gauge groups.

Finally, the matter charges 
$q^{(i)}$ provided by the currents $j^{(i)}$ induce the  screening charges 
$q_{\mbox{\scriptsize scr}}^{(i)} =-q^{(i)}$ in the Higgs medium, 
screening their Coulomb interactions, 
but not their long range AB interactions 
with the vortices $\phi^{(i)}$ 
implied by the  matter coupling~(\ref{j12mat}):
\bea                                 \label{qphi}
{\cal R}^2 \; |q^{(1)},q^{(2)}\rangle|\phi^{(1)},\phi^{(2)}\rangle 
&=&
e^{\im ( q^{(1)} \phi^{(1)}  +  \, q^{(2)} \phi^{(2)})}         \, 
|q^{(1)},q^{(2)}\rangle|\phi^{(1)},\phi^{(2)}\rangle \, . 
\eea

\subsection{$\Z_{N^{(1)}} \times \Z_{N^{(2)}}$ Chern-Simons theory of type~II}
\label{revii}

The discussion in the previous subsection pertained to
all values of the topological mass $\mu$.
Henceforth, it is again assumed that the model features 
the Dirac monopoles~(\ref{inst1}) and~(\ref{inst2}), 
so $\mu$ is quantized as~(\ref{mu12}).
It will be shown that under these circumstances
the long distance physics of the Higgs phase
is described  by a $\Z_{N^{(1)}} \times \Z_{N^{(2)}}$ 
gauge theory with a  3-cocycle $\omega_{\mbox{\scriptsize II}}$ 
of type~II determined by the homomorphism~(\ref{homoII}).

\begin{figure}[tbh] 
\begin{center}
\begin{picture}(85,80)(-15,-15)
\put(-5,-5){\dashbox(20,40)[t]{}}
\put(-15,0){\line(1,0){60}}
\put(0,-15){\line(0,1){60}}
\thinlines
\multiput(-10,-10)(0,10){6}{\multiput(0,0)(10,0){5}{\circle*{2.0}}}
\multiput(0,-10)(0,10){6}{\multiput(0,0)(20,0){3}{\circle*{2.0}}}
\put(0,40){\circle{3.6}}
\put(0,40){\vector(1,-4){10}} 
\put(-3,42){\vector(0,1){5}}
\put(43,-3){\vector(1,0){5}}
\put(-10,50){\small$\phi^{(2)}[\frac{\pi}{2e^{(2)}}]$}
\put(50,-6){\small$q^{(1)}[e^{(1)}]$}
\put(18,-24){(a)}
\end{picture}
\hspace{2cm}
\begin{picture}(85,80)(-15,-15)
\put(-5,-5){\dashbox(40,30)[t]{}}
\put(-15,0){\line(1,0){60}}
\put(0,-15){\line(0,1){60}}
\thinlines
\multiput(-10,0)(0,20){3}{\multiput(0,0)(10,0){6}{\circle*{2.0}}}
\put(0,40){\circle{3.6}}
\put(0,40){\vector(1,-4){10}} 
\put(-3,42){\vector(0,1){5}}
\put(43,-3){\vector(1,0){5}}
\put(-10,50){\small$\phi^{(1)}[\frac{\pi}{e^{(1)}}]$}
\put(50,-6){\small$q^{(2)}[e^{(2)}]$}
\put(18,-24){(b)}
\end{picture}
\end{center}  
\vspace{0.5cm}
\caption{\sl The spectrum of a Higgs phase with  residual
gauge group $\Z_2 \times \Z_4$ and CS action of type~II 
compactifies to the particles in the dashed boxes. 
We have displayed the flux $\phi^{(2)}$ 
versus the charge $q^{(1)}$ and  
the flux $\phi^{(1)}$ versus the charge $q^{(2)}$.
Here, the topological mass is assumed to take 
its minimal nontrivial value 
$\mu=\frac{e^{(1)}e^{(2)}}{\pi}$, that is, \ $p=1$. 
The arrows in figure~(a) and~(b) visualize the tunnelings corresponding 
to the Dirac monopole~(2)  and the monopole~(1)  respectively.}
\label{figz42} 
\end{figure}
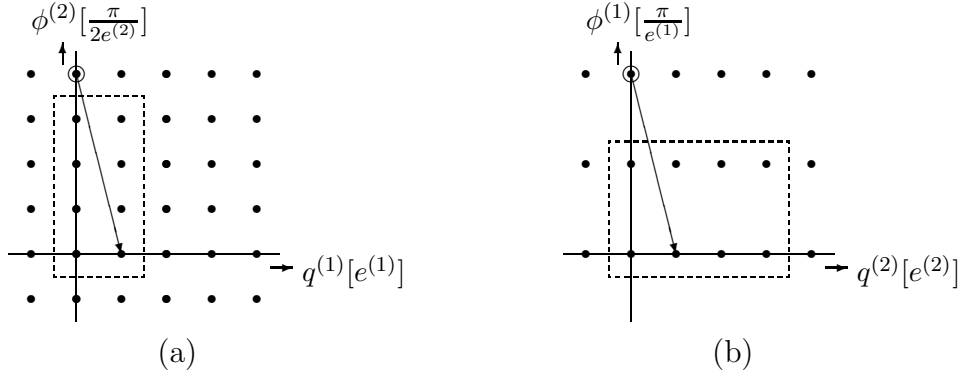

Let me first recall  from the previous subsection that 
the spectrum of the $\Z_{N^{(1)}} \times \Z_{N^{(2)}}$ CS 
Higgs phase consists of the quantized matter charges 
$q^{(i)}=n^{(i)} e^{(i)}$, 
the quantized magnetic fluxes~(\ref{fluqu12}) 
and the dyonic combinations. These 
particles will e labeled as $\left( A,n^{(1)}  n^{(2)}\right)$ with 
$A := (a^{(1)}, a^{(2)})$ and $a^{(i)}, n^{(i)} \in \Z$.
Upon implementing the quantization condition~(\ref{mu12}), 
the AB interactions~(\ref{onsflux12}), (\ref{onsfl12an})
and~(\ref{qphi}) between these particles can then be recapitulated
as
\bea                             
{\cal R}^2 \; |A, n^{(1)}  n^{(2)} \rangle 
| A',n^{(1)'}  n^{(2)'} \rangle      &=&
\alpha' (A) \; \alpha (A') \; |A, n^{(1)}  n^{(2)} \rangle
|A',n^{(1)'}  n^{(2)'} \rangle      \qquad
                            \label{brz2}
\\                               \label{anz123jo}
{\cal R} \; |A, n^{(1)}  n^{(2)} \rangle|A, n^{(1)}  n^{(2)} \rangle &=&
\alpha (A) \;
|A, n^{(1)}  n^{(2)} \rangle|A, n^{(1)}  n^{(2)} \rangle
\\ \label{spin12}
T \; |A, n^{(1)}  n^{(2)} \rangle  &=& 
\alpha(A) \; |A, n^{(1)}  n^{(2)}\rangle \, ,
\eea 
with $
\alpha (A') := \varepsilon_A(A') \; \Gamma^{n^{(1)}n^{(2)}} (A') $ and 
$\alpha' (A) := \varepsilon_{A'} (A) \; \Gamma^{n^{(1)'}n^{(2)'}} (A)$. 
The epsilon factors appearing here are identical to~(\ref{epii}), i.e.\ 
$
\varepsilon_A(A') = \exp \left( \frac{2\pi \im p}{N^{(1)}N^{(2)}} 
\, a^{(1)}a^{(2)'} \right) 
$ with $p$ the integral CS parameter in~(\ref{mu12}), whereas 
$
\Gamma^{n^{(1)} n^{(2)}} (A) 
=   \exp \left( \frac{2 \pi \im}{N^{(1)}} \, n^{(1)} a^{(1)} +
\frac{2 \pi \im}{N^{(2)}} \, n^{(2)} a^{(2)} \right)$   
denotes an UIR of the group $\Z_{N^{(1)}} \times \Z_{N^{(2)}}$.
Under the remaining long range 
AB interactions~(\ref{brz2}) and~(\ref{anz123jo}), 
the charge labels $n^{(i)}$ 
clearly become $\Z_{N^{(i)}}$ quantum numbers. Moreover, in the presence 
of the Dirac monopoles~(\ref{inst1}) and~(\ref{inst2})
the fluxes $a^{(i)}$ are conserved modulo $N^{(i)}$.
Specifically, in terms of the integral 
charge and flux quantum numbers $n^{(i)}$ and
$a^{(i)}$ the tunneling events corresponding to these 
minimal monopoles read
\bea                             \label{instb1}
\mbox{instanton $(1)$ :} & & \left\{   \ba{lcl}
a^{(1)} & \mapsto & a^{(1)} -N^{(1)}  \\
n^{(2)} & \mapsto & n^{(2)}  + p 
\ea \right.                                            \\
      &&  \nn \\                        
\label{instb2}
\mbox{instanton $(2)$ :} & & \left\{   \ba{lcl}
a^{(2)} & \mapsto &  a^{(2)} -N^{(2)} \\
n^{(1)} & \mapsto & n^{(1)} +p \, .
\ea \right.
\eea   
Here, I substituted~(\ref{mu12}) in~(\ref{inst1}) and~(\ref{inst2})
respectively. Hence, the decay of an unstable
flux corresponding to one residual cyclic gauge group is accompanied 
by the creation of the charge $p$ w.r.t.\ the other cyclic gauge group,
as displayed in figure~\ref{figz42}.
It is again easily verified that these local 
tunneling events are invisible to the long range AB 
interactions~(\ref{brz2}) and that the particles connected by the 
monopoles exhibit the same spin factor~(\ref{spin12}).
The conclusion then becomes that the spectrum of a 
$\Z_{N^{(1)}} \times \Z_{N^{(2)}}$ Higgs phase corresponding to an
integral CS parameter $p$ compactifies to
\bea                          \label{compsp12}
\left( A,n^{(1)}  n^{(2)}\right)    \qquad\qquad 
\mbox{with $A=(a^{(1)}, a^{(2)})$ and  $a^{(i)}, n^{(i)}
\in 0,1, \ldots, N^{(i)}-1$} \, , \qquad
\eea          
where the modulo calculus for the flux quantum numbers $a^{(i)}$ 
involves the charge jumps displayed in~(\ref{instb1}) and~(\ref{instb2}).

It is now readily checked that in accordance with the 
homomorphism~(\ref{homoII}) for this case, the 
$\Z_{N^{(1)}} \times \Z_{N^{(2)}}$ gauge theory 
labeled by the integral CS parameter $p$ corresponds
to the 3-cocycle of type~II given in~(\ref{type2do}), which we repeat for
convenience
\bea                  \label{ronaldo}
\omega_{\mbox{\scriptsize II}} (A,B,C) &=&             
\exp \left( \frac{2 \pi \im p} {N^{(1)}N^{(2)}}  \;
a^{(1)}(b^{(2)} +c^{(2)} - [b^{(2)}+c^{(2)}]) \right).
\eea 
In other words, the spectrum~(\ref{compsp12}) with the topological 
interactions summarized in the expressions~(\ref{brz2}), (\ref{anz123jo}) 
and~(\ref{spin12}) 
is governed by the quasi-quantum 
double $D^{\omega_{\mbox{\tiny II}}} (\Z_{N^{(1)}} \times \Z_{N^{(2)}})$ 
with $\omega_{\mbox{\scriptsize II}}$ the 3-cocycle~(\ref{ronaldo}). 
In particular, the fusion rules following from~(\ref{Ncoef}) 
\bea \label{CSfusion12}
\left( A,n^{(1)}  n^{(2)} \right) \times
\left( A',n^{(1)'}  n^{(2)'} \right)  &=&   
\left( [A+A'], n_{\mbox{\scriptsize sum}}^{(1)}  
n_{\mbox{\scriptsize sum}}^{(2)} \right),
\eea
with
\beas 
[A+A'] &=& ([a^{(1)}+a^{(1)'}],  [a^{(2)}+a^{(2)'}]) \\
n_{\mbox{\scriptsize sum}}^{(1)}
&=&[n^{(1)}+n^{(1)'}+\frac{p}{N^{(2)}}(a^{(2)}+a^{(2)'}-
[a^{(2)}+a^{(2)'}])]       \\
n_{\mbox{\scriptsize sum}}^{(2)}&=& 
[n^{(2)}+n^{(2)'}+\frac{p}{N^{(1)}}(a^{(1)}+a^{(1)'}-[a^{(1)}+a^{(1)'}])] \, ,
\eeas
are again a direct reflection of the tunneling properties 
induced by the monopoles~(\ref{inst1}) and~(\ref{inst2}). 
Note that these `twisted' tunneling properties 
actually imply that the  complete spectrum~(\ref{compsp12}) of this theory 
is  just generated by the two fluxes $a^{(1)}=1$ 
and $a^{(2)}=1$, if the CS parameter $p$ is set to $1$.

To conclude, at first sight the periodicity $\gcd (N^{(1)}, N^{(2)})$ 
in the CS parameter $p$ as indicated by the mapping~(\ref{homoII}) 
is not completely obvious from the fusion rules~(\ref{CSfusion})
and the topological interactions~(\ref{brz2}), (\ref{anz123jo}) 
and~(\ref{spin12}). To make this periodicity explicit, we 
have to appeal to the chinese remainder theorem~(\ref{theorem}), which was
the crucial ingredient in the proof that the 
3-cocycle~(\ref{ronaldo}) boils down to a 3-coboundary 
for $p=\gcd (N^{(1)}, N^{(2)})$. 
With this theorem, we simply infer that setting $p$ to 
$\gcd (N^{(1)}, N^{(2)})$ amounts to the automorphism 
$
\left( A,n^{(1)}  n^{(2)} \right) \mapsto 
\left( A, [n^{(1)} +xa^{(2)}]  [n^{(2)}+ya^{(1)}] \right)
$
of the spectrum~(\ref{compsp12}) for $p=0$,
where  $x$ and $y$ are the integers appearing at the r.h.s.\ 
of~(\ref{theorem}).
Hence, the theories for $p=0$ and $p=\gcd (N^{(1)}, N^{(2)})$ are the same 
up to a relabeling of the dyons.

\sectiona{$\Z_2 \times \Z_2 \times \Z_2$ Chern-Simons theory}
\label{typeIII}

CS actions~(\ref{type3do}) of type~III are conceivable 
for finite abelian gauge groups ${ H}$ being direct products  
of three or more cyclic groups. 
As indicated by the homomorphism (\ref{homo}), 
such type~III CS theories do {\em not} occur as the long distance
remnant of spontaneously broken $U(1)^k$ CS theories.
At present, it is not clear to me whether there actually 
exist symmetry breaking schemes
giving rise to 3-cocycles of type~III for a residual finite abelian gauge 
group in the Higgs phase. 
This point deserves further scrutiny since adding a type~III CS action to 
an abelian discrete $H$ gauge theory has a drastic consequence:
it renders such a theory nonabelian. 
In general these type~III CS theories are, in fact, 
dual versions of gauge theories featuring a  
{\em nonabelian} finite gauge group.

In this section, I illustrate these phenomena with the simplest example 
of a type~III CS theory, namely that with gauge 
group ${ H} \simeq \Z_2 \times \Z_2  \times \Z_2 $. 
The generalization to other abelian groups 
allowing for 3-cocycles of type~III is straightforward.
The outline is as follows. 
In  subsection~\ref{thes},  it will be shown 
that the incorporation of the 
type~III CS action in a 
$\Z_2 \times \Z_2  \times \Z_2 $ gauge theory involves 
a `collapse' of the spectrum. 
Whereas the ordinary $\Z_2 \times \Z_2  \times \Z_2$ theory
features 64 different singlet particles, the spectrum 
just consists of 22 different particles in the presence of 
the 3-cocycle of type~III. 
Specifically, the dyon charges, 
which formed one dimensional UIR's of $\Z_2 \times \Z_2 \times \Z_2$, 
are  reorganized into two dimensional or doublet projective 
representations  of $\Z_2 \times \Z_2 \times \Z_2$.
This abelian gauge theory
then describes nonabelian topological interactions between these 
doublet dyons, which will be discussed in some detail in 
subsection~\ref{z23ab}. 
In subsection~\ref{emduality}, I finally establish that this theory
is actually a  dual version of the ordinary discrete gauge theory 
with nonabelian gauge group the dihedral group $D_4$. Furthermore, 
I show that upon adding a type~I CS action, the theory  
becomes dual to the ordinary discrete gauge theory with gauge group the 
double dihedral (or quaternion) group $\bar{D}_2$.

\subsection{Spectrum}
\label{thes}

The type~III  CS action~(\ref{type3}) 
for the gauge group $\Z_2 \times \Z_2 \times \Z_2 $ takes the form
\bea                                  \label{trico}
\omega_{\mbox{\scriptsize III}} (A,B,C) &=& 
\exp \left( \pi \im \,  a^{(1)} b^{(2)} c^{(3)} \right),
\eea
where I have set the integral cocycle parameter to its nontrivial 
value, i.e.\  $p_{\mbox{\scriptsize III}} = 1$.  
From the slant product~(\ref{c}) applied to the 3-cocycle~(\ref{trico}), 
we infer that the 2-cocycle $c_A$ entering the definition of the 
projective  $\Z_2 \times \Z_2 \times \Z_2 $ dyon charge
representations~(\ref{project}) for the magnetic flux $A$ 
in this CS theory reads 
\bea                     \label{cIII}
c_A (B, C) &=& \exp \left(
\pi \im \{ a^{(1)} b^{(2)} c^{(3)} +b^{(1)}c^{(2)}a^{(3)}
              -b^{(1)}a^{(2)}c^{(3)} \} \right).
\eea
For the trivial magnetic flux sector $A=(0,0,0):=0$, this 2-cocycle 
naturally vanishes. So the pure charges are given by the ordinary 
UIR's of $\Z_2 \times \Z_2 \times \Z_2$.
For the nontrivial magnetic flux sectors 
$A \neq 0$, the 2-cocycle  $c_A$ is nontrivial. That is,
it can not be decomposed as~(\ref{repphase}). 
Hence,  we are dealing with projective 
representations that can not be obtained from ordinary representations
by the inclusion of extra AB phases $\varepsilon$ 
as in~(\ref{rei}). 

An important result in projective representation theory now 
states that for a given finite group ${ H}$ 
the number of inequivalent irreducible projective 
representations~(\ref{project}) 
associated with a 2-cocycle $c$ equals the number of $c$-regular
classes in ${ H}$~\cite{karpov}. An element $h \in { H}$ is called
c-regular iff $c(h,g)=c(g,h)$ for all $g \in { H}$. If $h$ 
is c-regular, so are all its conjugates. 
In our abelian example with the 2-cocycle $c_A$ for  $A \neq 0$, 
it is easily verified that there are only two $c_A$ 
regular classes in $\Z_2 \times \Z_2 \times \Z_2$, 
namely the trivial flux $0$ and $A$ itself.
Hence, there are only two inequivalent irreducible projective representations
associated with $c_A$. Just as for ordinary UIR's,
the sum of the squares of the dimensions 
of these  projective UIR's should equal the order 8 of 
the group $\Z_2 \times \Z_2 \times \Z_2$ 
and we find that both representations are two dimensional.
An explicit construction of these representations can be found 
in~\cite{karpov}.

Let me illustrate these general remarks by considering the effect
of the presence of the 2-cocycle $c_A$ for the particular
magnetic flux $A=100$.~\footnote{For notational convenience, I use the 
abbreviation $a^{(1)}a^{(2)}a^{(3)} := (a^{(1)},a^{(2)},a^{(3)})$ 
to denote the elements of $\Z_2 \times \Z_2 \times \Z_2$ from now on.}
Substituting~(\ref{cIII}) in~(\ref{project}) yields the following 
set of defining relations for the generators 
of $\Z_2 \times \Z_2 \times \Z_2$ in 
the projective representation $\alpha$
\bea           \label{defre}   \ba{rcl}
\alpha(100)^2 &=& \alpha(010)^2 \; = \; \alpha(001)^2 \; = \; {\mbox{\bf 1}} \\
\alpha(100) \, \cdot \, \alpha(010) &=& 
\alpha(010) \, \cdot \, \alpha(100)  \\
\alpha(100) \, \cdot \, \alpha(001) &=& 
\alpha(001) \, \cdot \, \alpha(100)         \\
\alpha(010) \, \cdot \, \alpha(001) &=& 
-\alpha(001) \, \cdot \, \alpha(010) \, ,
\ea
\eea
where ${\mbox{\bf 1}}$ denotes the unit matrix.
In other words, the generators $\alpha(010)$ and 
$\alpha(001)$ become anti-commuting, which indicates that 
the  projective representation $\alpha$ is necessarily higher dimensional.
Specifically, the two inequivalent two dimensional 
projective UIR's associated to the 2-cocycle $c_{100}$ 
are given by~\cite{karpov}
\be        \label{matrixa}    
\alpha^1_{\pm} (100) =  \pm \left( \ba{rr} 1 & 0 \\ 0 & 1 \ea \right),
\qquad \alpha^1_{\pm} (010) = \left( \ba{rr} 0 & 1 \\ 1 & 0 \ea \right),
\qquad \alpha^1_{\pm} (001) = \left( \ba{rr} 1 & 0 \\ 0 & -1 \ea \right).
\ee
Here, the subscript $+$ and $-$ labels the two inequivalent representations,
whereas the superscript $1$ refers to 
the fact that $A=100$ denotes the nontrivial magnetic 
flux associated to the first gauge group $\Z_2$ in the 
product $\Z_2 \times \Z_2 \times \Z_2$. 
In passing, I note that the set of matrices~(\ref{matrixa}) 
generates the two dimensional UIR of 
the dihedral point group $D_4$ displayed in the character table~\ref{charis}
of appendix~\ref{dvier}.

It is instructive to examine the projective representations 
in~(\ref{matrixa}) a little closer.
In an ordinary $\Z_2 \times \Z_2 \times \Z_2$ gauge theory, 
the three global  $\Z_2$ symmetry generators commute with each other and 
with the flux projection operators.
Thus the total internal Hilbert space of this gauge theory allows for 
a basis of mutual eigenvectors $|A,n^{(1)}n^{(2)}n^{(3)}\rangle$, 
where the labels $n^{(i)} \in 0,1$ 
denote the $\Z_2$ representations and $A \in \Z_2 \times \Z_2 \times \Z_2$
the different magnetic fluxes. In other words, the spectrum 
consists of 64 different particles each carrying a one dimensional internal
Hilbert space labeled by a flux and a charge.
Upon introducing the type~III CS action~(\ref{trico})
in this abelian discrete gauge theory, the global $\Z_2$ symmetry generators 
cease to commute with each other as we have seen explicitly for the flux 
sector $A=100$ in~(\ref{defre}). 
In this sector, the eigenvectors of the two non-commuting $\Z_2$ generators
are rearranged into an irreducible doublet representation. We can, however,
still diagonalize the generators in this doublet representation 
separately to uncover the $\Z_2$ eigenvalues $1$ and $-1$.
Hence, the $\Z_2$ charge quantum numbers $n^{(i)} \in 0,1$ 
remain unaltered in the presence of a CS action 
of type~III.

The analysis is completely similar for the other flux sectors. 
First of all, the two 2-dimensional 
projective dyon charge representations $\alpha^2_{\pm}$ associated with
the magnetic flux $A=010$ follow from a cyclic 
permutation of the set of matrices in~(\ref{matrixa}) such that
the diagonal matrix $\pm {\mbox{\bf 1}}$ ends up at the second position.
That is,   $\alpha^2_{\pm} (010)=\pm {\mbox{\bf 1}}$. 
Here, the superscript $2$ indicates that we are dealing with a nontrivial
flux w.r.t.\ the second cyclic gauge group in the product
$\Z_2 \times \Z_2 \times \Z_2$. 
The two projective 
representations $\alpha^3_{\pm}$ for $A=001$ are then defined 
by the cyclic permutation of the matrices in~(\ref{matrixa})
fixed by demanding $\alpha^3_{\pm}(001) =\pm {\mbox{\bf 1}}$.
To proceed, the two 2-dimensional 
projective representations $\beta_{\pm}^1$
for the flux $A=011$ are determined by 
\be        \label{matrixb}
\beta^1_{\pm} (100) =   \left( \ba{rr} 1 & 0 \\ 0 & -1 \ea \right),
\qquad \beta^1_{\pm} (010) =  \left( \ba{rr} 0 & 1 \\ 1 & 0 \ea \right),
\qquad \beta^1_{\pm} (001) = \pm \left( \ba{rr} 0 & 1 \\ 1 & 0 \ea \right).
\ee
Here, the subscript $+$ and $-$ again labels the two inequivalent 
representations, while the superscript now reflects the fact that $A=011$ 
corresponds to a trivial flux w.r.t.\ 
the first gauge group $\Z_2$ in the product 
$\Z_2 \times \Z_2 \times \Z_2$. 
The two representations $\beta_{\pm}^2$ associated to the flux $A=101$
are defined by the same set of matrices~(\ref{matrixb}) 
moved one step to the right with cyclic boundary conditions, 
whereas the representations $\beta_{\pm}^3$ for $A=110$ are given 
by the same set moved two steps to the right with cyclic boundary conditions.
Finally, the two inequivalent dyon charge representations $\gamma_{\pm}$ 
for the magnetic flux $A=111$ are generated by  
the Pauli matrices  
\be        \label{matrixg}
\gamma_{\pm} (100) =   \pm \left(  \ba{rr} 0 & 1 \\ 1 & 0 \ea \right),\;
\;\; \gamma_{\pm} (010) = \pm \left( \ba{rr} 0 & -\im \\ \im & 0 \ea \right),
\;\;\; \gamma_{\pm} (001) = \pm \left( \ba{rr} 1 & 0 \\ 0 & -1 \ea \right).
\ee
In contrast with the sets of matrices contained in~(\ref{matrixa}) 
and~(\ref{matrixb}), which generate the 2-dimensional representation
of the dihedral group $D_4$, the two sets in~(\ref{matrixg}) 
generate the two dimensional UIR's of the truncated pure braid group
$P(3,4)$ displayed in the character table~\ref{char} 
of appendix~\ref{trubra}.

The complete spectrum of this $\Z_2 \times \Z_2 \times \Z_2$ CS 
theory of type~III can now be summarized as
\be             \label{spectz23}
\ba{cc}
\mbox{particle}                &\qquad   \exp (2\pi \im  s)    \\
(0,n^{(1)}n^{(2)}n^{(3)})  &\qquad  1    \\
(100, \alpha^1_{\pm}), \, 
(010, \alpha^2_{\pm}), \,
(001, \alpha^3_{\pm})      &\qquad   \pm 1   \\
(011,\beta^1_{\pm}), \,
(101,\beta^2_{\pm}), \, 
(110,\beta^3_{\pm})        &\qquad    \pm 1  \\
(111, \gamma_{\pm})        &\qquad    \mp \im \, , 
\ea
\ee
where the spin factors for the particles are obtained from the action of 
the flux of the particle on its own dyon charge as indicated by 
expression~(\ref{anp}). Hence, there are 7 nontrivial pure charges 
$(0,n^{(1)}n^{(2)}n^{(3)})$ labeled by the ordinary nontrivial one 
dimensional $\Z_2 \times \Z_2 \times \Z_2$ representations~(\ref{hrepz}).
The trivial representation naturally corresponds to the vacuum.
In addition, there are 14 dyons carrying a nontrivial 
abelian magnetic flux and a doublet charge.
The conclusion then becomes that the introduction of a CS action 
of type~III leads  to a compactification or `collapse' of the spectrum.
Whereas an ordinary $\Z_2 \times \Z_2 \times \Z_2$ gauge theory
features 64 different singlet particles, 
we only have 22 distinct particles in the presence 
of  a type~III CS action~(\ref{trico}). 
To be specific,  the  singlet dyon charges
are rearranged into doublets so that the squares 
of the dimensions of the internal Hilbert spaces for the particles in the 
spectrum still add up to the order of the quasi-quantum double 
$D^{\omega_{\mbox{\tiny III}}} (\Z_2 \times \Z_2 \times \Z_2)$,
that is, $8^2= 8\cdot 1^2 + 14 \cdot 2^2$.
Let me close with the remark that this collapse of the spectrum can also 
be seen directly by evaluating the Dijkgraaf-Witten invariant for the 
3-torus $S^1 \times S^1\times S^1$ with the 3-cocycle~(\ref{trico}). 
See section~\ref{dijwit} in this connection.

\subsection{Nonabelian topological interactions}
\label{z23ab}

Here, I highlight the {\em non}abelian 
nature of the topological interactions
in the type~III CS theory with {\em abelian}
gauge group $\Z_2 \times \Z_2 \times \Z_2$ of the previous subsection.

Let me start by considering an AB scattering 
experiment~\cite{ahabo,ver1,preslo} with an incoming beam of dyons 
$(100, \alpha_+^1)$ and scatterer the dyon $(011, \beta^1_+)$.  
I choose the following natural flux/charge eigenbasis  
for the associated four dimensional 2-particle internal Hilbert space 
$V^{100}_{\alpha^1_+} \times V^{011}_{\beta^1_+}$ 
\bea        \label{cbasis}
e_1 
&=& |100, \left( \ba{c} 1 \\ 0  \ea \right) \rangle \ot
        |011, \left( \ba{c} 1 \\ 0  \ea \right) \rangle  
\; := \; {|\uparrow \, \rangle |\uparrow \, \rangle} \\
e_2 
&=& |100, \left( \ba{c} 0 \\ 1  \ea \right) \rangle \ot
        |011, \left( \ba{c} 1 \\ 0  \ea \right) \rangle 
\; := \; {| \downarrow \, \rangle |\uparrow \, \rangle} \nn  \\
e_3 
&=& |100, \left( \ba{c} 1 \\ 0  \ea \right) \rangle \ot
        |011, \left( \ba{c} 0 \\ 1  \ea \right) \rangle 
\; := \; |\uparrow \, \rangle | \downarrow \, \rangle  \nn  \\
e_4 
&=& |100, \left( \ba{c} 0 \\ 1  \ea \right) \rangle \ot  
        |011, \left( \ba{c} 0 \\ 1  \ea \right) \rangle
\; := \; |\downarrow \, \rangle |\downarrow \, \rangle \, .     \nn
\eea
From relations~(\ref{braidaction}),~(\ref{matrixa}) and~(\ref{matrixb}), one 
infers that the monodromy matrix takes the following block 
diagonal form in this basis
\bea                                   \label{alles}
{\cal R}^2 &=&  \left( \ba{rrrr} 0 & 1 & 0 & 0\\ -1 & 0 & 0 & 0 \\
                               0 & 0 & 0 & -1 \\ 
                               0 & 0 & 1 & 0 \ea \right),
\eea  
where I used 
\beas
\alpha_+^1 (011) \; = \; 
c_{100}^{-1} (010,001)\; \alpha_+^1(010) \, \cdot \, \alpha_+^1 (001)
\; = \; 
        \left( \ba{rr} 0 & 1  \\ -1 & 0  \ea \right),
\eeas
which follows from relation~(\ref{project}), (\ref{cIII}) and~(\ref{matrixa}).
The monodromy matrix~(\ref{alles})  expresses the fact that the magnetic
flux $A=011$ acts as an Alice flux on the doublet dyon charge $\alpha_+^1$.
That is, after transporting the dyon $(100, \alpha_+^1)$
in a counterclockwise fashion around  the dyon $(011, \beta^1_+)$, 
it returns with the orientation ($\uparrow$ or $\downarrow$) of its 
charge $\alpha_+^1$ flipped ($\downarrow$ or $\uparrow$).
Furthermore, the orientation of the doublet dyon charge $\beta^1_+$ 
is unaffected by this process, as witnessed by the block diagonal 
form of~(\ref{alles}). 
As a matter of fact, the monodromy matrix~(\ref{alles}) is 
identical to the one displayed in Eq.\ (3.2.2) 
of reference~\cite{banff}
for a system of a pure doublet charge $\chi$ and a pure doublet flux
$\sigma_2^+$ in a $\bar{D}_2$ gauge theory without CS action.
So, this AB scattering problem is equivalent to the one discussed 
in section~3.2 of reference~\cite{banff} and gives rise to the same  
cross sections, which we repeat for convenience 
\bea             
\frac{{\rm d} \sigma_+}{{\rm d} \theta}   &=&
\frac{1+\sin{(\theta/2)}}{8 \pi p \sin^2{(\theta/2)}}   \label{c+}  \\
\frac{{\rm d} \sigma_-}{{\rm d} \theta} &=& 
\frac{1-\sin{(\theta/2)}}{8 \pi p \sin^2{(\theta/2)}}    \label{c-}  \\
\frac{{\rm d} \sigma}{{\rm d}\theta} &=& 
\frac{{\rm d}\sigma_-}{{\rm d}\theta}  + 
\frac{{\rm d}\sigma_+}{{\rm d}\theta} \; = \; 
 \frac{1}{4 \pi p \sin^2{(\theta/2)}} \; .             \label{singvac}
\eea
Here, $\theta$ denotes the scattering angle and $p$ the momentum of 
the incoming projectiles $(100, \alpha_+^1)$. 
The  (multi-valued) exclusive Lo-Preskill~\cite{preslo} 
cross section~(\ref{c+}) is measured by 
a detector which only signals scattered dyons $(100, \alpha_+^1)$ with 
the same charge orientation as the incoming beam of  
projectiles.  A device just detecting 
dyons $(100, \alpha_+^1)$ with charge orientation opposite to the 
charge orientation of the projectiles, in turn,  measures the 
multi-valued charge flip cross section~(\ref{c-}).
Finally, Verlinde's~\cite{ver1} single-valued inclusive
cross section~(\ref{singvac}) for this case  
is measured by a detector which signals scattered 
dyons $(100, \alpha_+^1)$  irrespective of the orientation of their
charge.

The fusion rules for the particles in the spectrum~(\ref{spectz23})
are easily obtained  from expression~(\ref{Ncoef}). 
I refrain from presenting the complete set and confine ourselves 
to the fusion rules that will enter the discussion later on.
First of all, the pure charges naturally add modulo $2$ 
\bea                                       \label{chesf}
(0,n^{(1)}n^{(2)}n^{(3)}) \times (0,n^{(1)'}n^{(2)'}n^{(3)'})
\:=\: (0,[n^{(1)}+n^{(1)'}] [n^{(2)}+n^{(2)'}][n^{(3)} + n^{(3)'}]) \, .  
\eea 
The same holds for the magnetic fluxes of the dyons, whereas
the composition rules for the dyon charges are less trivial, e.g.
\bea
(100, \alpha^1_s) \times (100, \alpha^1_s) &=&
(0) + (0,010) + (0,001) + (0,011)   \label{1al} \\
(011, \beta^1_s) \times (011, \beta^1_s)
    &=&(0) + (0,100) + (0,111) + (0,011)   \label{1bet} \\
(111, \gamma_s) \times (111, \gamma_s)
    &=&(0) + (0,011) + (0,101) + (0,110)   \label{1gamm} \\
(010, \alpha^2_s) \times (001, \alpha^3_s)    \label{hetv}
    &=& (011, \beta^1_{+}) + (011, \beta^1_{-})    \\
(100, \alpha^1_+) \times (011, \beta^1_s)
  &=& (111, \gamma_{+}) + (111, \gamma_{-}) \, ,   \label{hetvo}
\eea
with $s \in +,-$ and $(0)$ denoting the vacuum. The occurrence 
of the vacuum in the fusion rules~(\ref{1al}), (\ref{1bet}) 
and~(\ref{1gamm}), respectively, then indicates that the dyons 
$(100, \alpha^1_{\pm})$, $(011, \beta^1_{\pm})$ and $(111, \gamma_{\pm})$
are their own anti-particles. In fact, it is easily verified that 
this observation holds for all particles in the spectrum, i.e.\
the charge conjugation operator~(\ref{charconj}) acts diagonal on the 
spectrum~(\ref{spectz23}): $\cal{C}= S^2 = \mbox{\bf 1}$.

\begin{figure}[htb] 
   \epsfxsize=8cm
\centerline{\epsffile{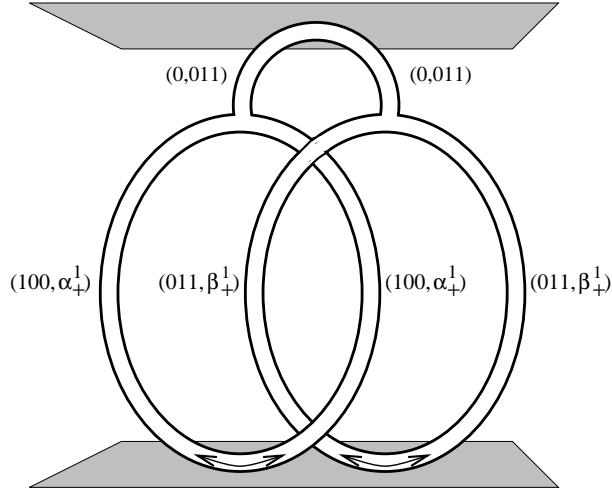}} 
\caption{\sl  A  dyon/anti-dyon pair $(100, \alpha^1_{+})$ and a 
dyon/anti-dyon pair $(011, \beta^1_+)$  are created from the vacuum
at a certain time slice. 
The ribbons represent the worldlines of the particles.
After the dyon $(100, \alpha^1_{+})$  has encircled the flux 
$(011, \beta^1_+)$, 
both particle/anti-particle pairs carry Cheshire charge 
$(0,011)$. These Cheshire charges become  localized upon bringing the members
of the pairs together again. Subsequently, the two charges 
$(0,011)$ annihilate each other.}
\label{ccatd2} 
\end{figure}

From the fusion rules~(\ref{1al}) and~(\ref{1bet}), respectively,
we learn that a  pair of dyons $(100, \alpha^1_{+})$ as well as a  
pair of dyons $(011, \beta^1_{+})$ can carry three different types 
of nontrivial Cheshire charge~\cite{banff,preskra,spm,sm,alice}. 
The nondiagonal form of the 
matrix~(\ref{alles}) then implies that these two different 
pairs exchange Cheshire charge when a particle in one pair encircles a 
particle in the other pair. Consider, for instance, the process 
depicted in figure~\ref{ccatd2} in which
a certain timeslice sees the creation
of a $(100, \alpha^1_+)$ 
dyon/anti-dyon pair and a $(011, \beta^1_+)$ 
dyon/anti-dyon pair from the vacuum. 
Hence, both pairs carry a trivial Cheshire charge at this stage, 
i.e.\ both pairs are in the vacuum channel $(0)$ of their fusion rule.
After the monodromy involving a dyon in the pair 
$(100, \alpha^1_+)$  and a dyon in the pair $(011, \beta^1_+)$,
both pairs carry the nonlocalizable Cheshire charge $(0, 011)$
which become localized charges upon fusing the members of the pairs. 
As follows from the rule~(\ref{chesf}), these localized charges
annihilate each other when they are brought together. Hence,
as it should, global $\Z_2 \times \Z_2 \times \Z_2$ charge is conserved 
in this process. All this becomes clear upon writing the 
process described above in terms of the associated internal quantumstates:
\bea   \label{uitz}
|0 \rangle 
 & \longmapsto & \frac{1}{2} 
\{ |100, \left( \ba{c} 1 \\ 0  \ea \right) \rangle 
|100, \left( \ba{c} 1 \\ 0  \ea \right) \rangle
\; + \; |100, \left( \ba{c} 0 \\ 1  \ea \right) \rangle
|100, \left( \ba{c} 0 \\ 1  \ea \right) \rangle \} \ot   
\qquad \label{eennul} \\  & & 
\ot 
\{ |011, \left( \ba{c} 1 \\ 0  \ea \right) \rangle
|011, \left( \ba{c} 0 \\ 1  \ea \right) \rangle  \; + \;
|011, \left( \ba{c} 0 \\ 1  \ea \right) \rangle
|011, \left( \ba{c} 1 \\ 0  \ea \right) \rangle  \}  \label{tweenul} 
\qquad   \\
  & \stackrel{{\mbox{\scriptsize \bf 1}} \ot {\cal R}^2 \ot 
{\mbox{\scriptsize\bf 1}}}{\longmapsto} &
\frac{1}{2}
\{ |100, \left( \ba{c} 1 \\ 0  \ea \right) \rangle
|100, \left( \ba{c} 0 \\ 1  \ea \right) \rangle
\; - \; |100, \left( \ba{c} 0 \\ 1  \ea \right) \rangle
|100, \left( \ba{c} 1 \\ 0  \ea \right) \rangle\}  
\ot  \qquad  \nn \\  & & 
\ot
\{ |011, \left( \ba{c} 0 \\ 1  \ea \right) \rangle 
|011, \left( \ba{c} 1 \\ 0  \ea \right) \rangle  \; - \; 
|011, \left( \ba{c} 1 \\ 0  \ea \right) \rangle 
|011, \left( \ba{c} 0 \\ 1  \ea \right)\rangle \}  \nn 
\qquad  \\
& \longmapsto & |0,011 \rangle \ot |0, 011 \rangle \nn \\
  & \longmapsto & |0 \rangle \, .            \nn
\eea 
The quasi-quantum double 
$D^{\omega_{\mbox{\tiny III}}} (\Z_2 \times \Z_2 \times \Z_2)$
acts on the two particle state~(\ref{eennul})
for the dyons $(100, \alpha^1_+)$ through
the comultiplication~(\ref{coalgebra}) with the 2-cocycle~(\ref{cIII}).
From the action of the flux projection operators, we 
obtain that this state carries trivial total flux.
Further, the global symmetry transformations, which act by means of 
the matrices~(\ref{matrixa}), 
leave  this two particle state invariant. So, this state
indeed carries trivial total $\Z_2 \times \Z_2 \times \Z_2$ charge.
In a similar fashion, we infer 
that the two particle state~(\ref{tweenul}) for
the dyons $(011, \beta^1_{+})$ corresponds to trivial total flux 
and charge. After the monodromy, 
which involves the matrix~(\ref{alles}), both 
two particle states then carry the global charge $(0,011)$. 
Note that just as in the analogous process in the $\bar{D}_2$ theory discussed 
in~\cite{banff} this exchange of Cheshire charge is accompanied 
by an exchange of quantum statistics. That is to say,
the two particle states~(\ref{eennul})
and~(\ref{tweenul}) are bosonic in accordance with the trivial
spin~(\ref{spectz23}) assigned to the dyons $(100, \alpha^1_+)$ and 
$(011, \beta^1_{+})$ respectively. Both two particle states
emerging after the monodromy, in turn, are fermionic.

As has been explained in section~\ref{symalg}, the internal Hilbert space 
for a multi-particle system in an abelian discrete (CS) theory carries 
a representation of the direct product of the associated (quasi-) 
quantum double and some truncated braid group defined in 
appendix~\ref{trubra}. In the remainder of this subsection,
I identify the truncated braid groups showing up in this particular 
type~III $\Z_2 \times \Z_2 \times \Z_2$ CS theory and elaborate on the 
decomposition of the internal Hilbert space for some representative 
3-particle systems into a direct some of irreducible subspaces 
under the action of the direct product of the quasi-quantum double 
$D^{\omega_{\mbox{\tiny III}}} (\Z_2 \times \Z_2 \times \Z_2)$ and the 
related truncated braid group. I start with the identical particle 
systems.   

In fact, the only identical particle systems 
that obey nontrivial braid statistics in this model are those that consist 
either of the dyons $(111, \gamma_{+})$ or of the dyons $(111, \gamma_{-})$. 
It is easily verified that the braid operator~(\ref{braidaction}) for these 
systems is of order $4$. Hence, the internal Hilbert space for a system 
containing $n$ of such identical dyons carries a representation of the 
truncated braid group $B(n,4)$ which is in general reducible, i.e.\ it  
decomposes into a direct sum of UIR's. The one dimensional UIR's that may 
occur in this decomposition then correspond to semion statistics, whereas 
the higher dimensional UIR's correspond to nonabelian braid statistics. 
All other identical particle systems realize permutation statistics. 
That is, the associated truncated braid groups boil down to the permutation
group. Specifically, the pure charges $(0,n^{(1)}n^{(2)}n^{(3)})$ 
are bosons, which is in accordance with the canonical spin statistics 
connection~(\ref{spist}), since there is a trivial spin 
factor~(\ref{spectz23}) assigned to these particles. Finally, 
the remaining dyons in general obey bose, fermi or parastatistics.  

As an example of an identical particle system, I consider
a system consisting of 3 dyons  $(111, \gamma_+)$. 
From the fusion rule~(\ref{1gamm}) and the rules 
\beas 
(111, \gamma_{+}) \times (0,011) \;=\; 
(111, \gamma_{+}) \times (0,101) \;=\;
(111, \gamma_{+}) \times (0,101) \;=\; 
(111, \gamma_{-}) \, , 
\eeas 
one obtains that the associated 3-particle internal Hilbert space
decomposes into the following irreducible pieces under the action
of $D^{\omega_{\mbox{\tiny III}}} (\Z_2 \times \Z_2 \times \Z_2)$  
\bea \label{ingr1}
(111, \gamma_{+}) \times (111, \gamma_{+}) \times 
(111, \gamma_{+}) & = & 4 \; (111, \gamma_-) \, .  
\eea 
According to the general discussion at the end of section~\ref{fuspbr},
the occurrence of four isotypical fusion channels indicates that 
nonabelian braid statistics is conceivable for this identical particle
system. Indeed, higher dimensional UIR's of the related truncated 
braid group $B(3,4)$ appear.
The 96 elements of $B(3,4)$ are divided into 16 conjugacy classes as displayed
in relation~(\ref{B34}) of appendix~\ref{trubra}. The matrices assigned to 
these elements in the representation of $B(3,4)$ realized by this 
distinguishable particle system follow from the 
relations~(\ref{brareco}), (\ref{project}), 
(\ref{cIII}) and~(\ref{matrixg}).  A lengthy but straightforward 
calculation of the trace of an arbitrary 
representative for each of the 16 conjugacy classes in~(\ref{B34}) and 
a subsequent evaluation of the innerproduct of the resulting character vector 
with those for the UIR's of $B(3,4)$ displayed in table~\ref{tab:tab4} of 
appendix~\ref{trubra} then reveals that the 
$B(3,4)$ representation carried by the internal Hilbert space of 
this system breaks up into the following irreducible pieces
\bea \label{ingr2}
\Lambda_{B(3,4)} & = & 4 \; \Lambda_{3} + 2 \; \Lambda_5 \, .   
\eea  
Here, $\Lambda_{3}$ denotes the 1-dimensional UIR 
and $\Lambda_5$ the 2-dimensional UIR of $B(3,4)$ defined 
in character table~\ref{tab:tab4}. 
From~(\ref{ingr1}) and~(\ref{ingr2}), we now conclude that under 
the action of the direct product 
$D^{\omega_{\mbox{\tiny III}}} (\Z_2 \times \Z_2 \times \Z_2)
\times B(3,4)$ the internal Hilbert space for a system of 
three dyons $(111, \gamma_{+})$ decomposes into the following 
irreducible pieces
\bea  \label{vlag}
2 \; ((111, \gamma_{-}), \Lambda_{3}) 
\: +  \: ((111, \gamma_{-}), \Lambda_{5}) \, , 
\eea  
where $((111, \gamma_{-}), \Lambda_{3})$ stands for a 2-dimensional
representation and $((111, \gamma_{-}), \Lambda_{5})$
for a 4-dimensional representation. Further,
from relation~(\ref{B34}) and table~\ref{tab:tab4}
in appendix~\ref{trubra}, we learn
that the generators $\tau_1$ and $\tau_2$ of $B(3,4)$ are represented 
by the scalar $-\im$ in the representation $\Lambda_{3}$, which coincides with 
the spin factor~(\ref{spectz23}) assigned to the dyon $(111, \gamma_{+})$.
Thus the 3-particle states in the irreducible components  
$((111, \gamma_{-}), \Lambda_{3})$ in~(\ref{vlag}) satisfy the 
canonical spin-statistics connection~(\ref{spist}). 
In passing,  I just mention that
an analogous calculation shows that the internal Hilbert space for a 
system of three dyons $(111, \gamma_{-})$ decomposes into the following 
irreducible subspaces under the action of 
$D^{\omega_{\mbox{\tiny III}}} (\Z_2 \times \Z_2 \times \Z_2) \times B(3,4)$ 
\bea 
2 \; ((111, \gamma_{+}), \Lambda_{1}) 
\: +  \: ((111, \gamma_{+}), \Lambda_{5}) \, , 
\eea  
with $\Lambda_{1}$ the complex conjugate of the 1-dimensional 
UIR $\Lambda_{3}$ of $B(3,4)$ in~(\ref{ingr2}).

Let me finally also briefly comment on the distinguishable particle systems
in this theory. It is readily checked that the maximal order 
of the monodromy operator for distinguishable particles in this theory is 4. 
So, the  distinguishable particles configurations in this theory
realize representations of the truncated colored braid group $P(n,8)$ and 
its subgroups. The order of the two monodromy operators for a system 
consisting  of the three dyons $(100, \alpha^1_{+})$, 
$(010, \alpha^2_{+})$ and $(001, \alpha^3_{+})$, for instance, is of order 2. 
Hence, the associated truncated colored braid group is 
$P(3,4) \subset P(3,8)$ discussed in appendix~\ref{trubra}.
It consists of 16 elements organized into 10 conjugacy classes 
as displayed in relation~(\ref{p34}). The matrices assigned to these 
elements in the representation of $P(3,4)$ realized by this 3-particle 
system now follow from the relations~(\ref{p34mongen}), (\ref{brareco}), 
and~(\ref{matrixa}).  
In a similar fashion as before, it is readily inferred   
that the $P(3,4)$ representation carried by the internal Hilbert space of 
this system breaks up into the following irreducible pieces
\bea                \label{eindt}
\Lambda_{P(3,4)} &=& 2 \; \Omega_8 + 2 \; \Omega_9 \, ,
\eea
with $\Omega_8$ and $\Omega_9$ the two dimensional UIR's
in the character table~\ref{char} of appendix~\ref{trubra}.
To proceed, from the fusion rules~(\ref{hetv}) and~(\ref{hetvo}), 
we learn that under the action of 
$D^{\omega_{\mbox{\tiny III}}} (\Z_2 \times \Z_2 \times \Z_2)$, 
the internal Hilbert space for this  3-particle system 
decomposes into the following direct sum of 4 irreducible 
representations
\bea         \label{ein}
(100, \alpha^1_{+}) \times (010, \alpha^2_{+}) \times (001, \alpha^3_{+})
    &=& 2 \; (111, \gamma_{+}) + 2 \; (111, \gamma_{-}) \, .
\eea 
To conclude, by constructing a basis adapted to the simultaneous 
decomposition of this 3-particle internal Hilbert space, it can 
be verified that the two UIR's $\Omega_8$
of $P(3,4)$  in~(\ref{eindt}) combine with the two UIR's  $(111, \gamma_{+})$ 
of $D^{\omega_{\mbox{\tiny III}}} (\Z_2 \times \Z_2 \times \Z_2)$ 
in~(\ref{ein}) and the two UIR's $\Omega_9$  with
the two UIR's $(111, \gamma_{-})$. 
So, this internal Hilbert space 
system decomposes into the following irreducible subspaces
\bea 
\left( (111, \gamma_{+}), \Omega_8 \right) \: + \: 
\left( (111, \gamma_{-}) , \Omega_9 \right),
\eea 
under the action of  the direct product $D^{\omega_{\mbox{\tiny III}}} 
(\Z_2 \times \Z_2 \times \Z_2) \times P(3,4)$, where 
$\left( (111, \gamma_{+}), \Omega_8 \right)$ and 
$\left( (111, \gamma_{-}) , \Omega_9 \right)$ both label 
a four dimensional representation.

\subsection{Electric/magnetic duality}
\label{emduality}

The analysis of the previous subsections revealed some 
striking similarities between the type~III CS theory with 
gauge group $\Z_2 \times \Z_2 \times \Z_2$ 
and  the ordinary discrete gauge theory with nonabelian finite gauge group
the double dihedral group $\bar{D}_2$ discussed in
full detail in~\cite{banff,thesis}. See also~\cite{spm,spm1,sm}.
First of all, the orders of these gauge groups
are the same: $|\Z_2 \times \Z_2 \times \Z_2|=|\bar{D}_2|=8$.
Moreover, including the vacuum, the spectrum of both 
theories consists  of $8$  singlet particles 
and $14$ doublet particles which adds up to a 
total number of 22 distinct particles. 
Also, the charge conjugation operation 
acts trivially 
on these spectra: ${\cal C}= S^2 = {\mbox{\bf 1}}$.  
That is, the particles in both theories are 
their own anti-particle. 
Finally, the truncated braid groups that govern the particle exchanges 
in these discrete gauge theories are similar.
Hence, it seems that these theories are dual. 
As it stands, however, this is not the case. 
This becomes clear upon comparing the spins of the  
particles in the two different theories. The spin factors assigned to the 
particles in the $\bar{D}_2$ gauge theory can be found 
in Eq.~(3.1.2) of reference~\cite{banff}. In particular, there are three
particles with spin factor $\im$ and three particles with
spin factor $-\im$.  
As displayed in~(\ref{spectz23}),
the spectrum of the $\Z_2 \times \Z_2 \times \Z_2$ CS theory of type~III, 
in contrast, contains
just one particle with spin factor $\im$ and one with $-\im$.
Hence, the modular $T$ matrices  associated 
to these models differ.
Moreover, it can be verified that the modular $S$ matrices
classifying the monodromy properties of the particles in these theories 
are also distinct.

Let me now recall from~(\ref{conj3e}) that 
the full set of CS actions for the 
gauge group  $\Z_2 \times \Z_2 \times \Z_2$
consists of three nontrivial 3-cocycles~(\ref{type1}) 
of type~I, three 3-cocycles~(\ref{type2}) of type~II, 
one 3-cocycle~(\ref{trico}) of type~III and products thereof. 
It turns out that the 
$\Z_2 \times \Z_2 \times \Z_2$ CS theories corresponding to 
the product of the 3-cocycle of type~III and either one of the three 
3-cocycles of type~I are, in fact,
dual to a $\bar{D}_2$ gauge theory. 
Here, I just explicitly show  this duality for the 
$\Z_2 \times \Z_2 \times \Z_2$ CS theory defined by
\bea                                  
\omega_{\mbox{\scriptsize{I+III}}} (A,B,C) &=& 
\exp \left( 
\frac{\pi \im }{2} a^{(1)}(b^{(1)} +c^{(1)} -[b^{(1)}+c^{(1)}]) 
+ \pi \im \,  a^{(1)} b^{(2)} c^{(3)} \right). 
\qquad
\eea
So the total CS action is the product of the 3-cocycle~(\ref{trico}) 
of type~III and the nontrivial 3-cocycle~(\ref{type1}) of type~I for 
the first $\Z_2$ gauge group in $\Z_2 \times \Z_2 \times \Z_2$.
As indicated by~(\ref{anp}),~(\ref{rei}) and~(\ref{epi}), adding 
this type~I 3-cocycle to the 
$\{\Z_2 \times \Z_2 \times \Z_2, \omega_{\mbox{\scriptsize{III}}}\}$ 
CS theory  
implies the assignment of  an additional imaginary spin factor $\im$ 
to those dyons in the spectrum~(\ref{spectz23}) that carry nontrivial 
flux w.r.t.\  the first $\Z_2$ gauge group 
of the product $\Z_2 \times \Z_2 \times \Z_2$.
The spin factors of the other particles are unaffected.
Thus the spin factors associated to the different 
particles in the 
$\{\Z_2 \times \Z_2 \times \Z_2, \omega_{\mbox{\scriptsize{I+III}}}\}$ CS 
theory become
\be             \label{spz23de}
\ba{cc}
\mbox{particle}                &\qquad   \exp (2\pi \im  s)    \\
(0,n^{(1)}n^{(2)}n^{(3)})  &\qquad  1    \\
(011,\beta^1_{\pm}), \, 
(010, \alpha^2_{\pm}), \,
(001, \alpha^3_{\pm})      &\qquad   \pm 1   \\
(100, \alpha^1_{\pm}), \,
(101,\beta^2_{\pm}), \, 
(110,\beta^3_{\pm})        &\qquad    \pm \im  \\
(111, \gamma_{\pm})        &\qquad    \pm 1 \, .
\ea
\ee      
Note that the spin structure of the spectrum of this theory indeed 
matches that of the $\bar{D}_2$ gauge theory exhibited in 
Eq.~(3.1.2) of reference~\cite{banff}.
Moreover, it is readily checked that the modular 
$S$ matrix~(\ref{fusion}) for this 
$\Z_2 \times \Z_2 \times \Z_2$ CS theory is also equivalent to that
for the $\bar{D}_2$ theory given in table~3.3 of 
reference~\cite{banff}. To be explicit,  the exchange 
\bea
\bar{D}_2   &\longleftrightarrow &
\{\Z_2 \times \Z_2 \times \Z_2, \omega_{\mbox{\scriptsize{I+III}}} \} \, ,
\eea 
which involves the following interchange 
of the particles in the $\bar{D}_2$ 
theory~\footnote{We use the labeling 
of the particles in the  $\bar{D}_2$ gauge theory explained 
in reference~\cite{banff}.} with the particles~(\ref{spz23de}) 
in the 
$\{ \Z_2 \times \Z_2 \times \Z_2, \omega_{\mbox{\scriptsize{I+III}}} \}$
CS theory
\bea            \label{duality}
\ba{rclrcl}
1                 &\longleftrightarrow &   (0) \, ,                &
\qquad \bar{1}           &\longleftrightarrow &   (0,100)               \\
J_1               &\longleftrightarrow &   (0,011) \, ,                &
\qquad \bar{J}_1         &\longleftrightarrow &   (0,111)               \\
J_2               &\longleftrightarrow &   (0,101) \, ,                &
\qquad \bar{J}_2         &\longleftrightarrow &   (0,001)               \\
J_3               &\longleftrightarrow &   (0,110)\, ,                &
\qquad \bar{J}_3         &\longleftrightarrow &   (0,010)               \\
\chi              &\longleftrightarrow &   (111, \gamma_+) \, ,        &
\qquad \bar{\chi}        &\longleftrightarrow &   (111, \gamma_-)       \\
\sigma^{\pm}_1    &\longleftrightarrow &   (011, \beta^1_{\pm}) \, ,   &
\qquad \tau^{\pm}_1      &\longleftrightarrow &   (100, \alpha^1_{\pm}) \\
\sigma^{\pm}_2    &\longleftrightarrow &   (010, \alpha^2_{\pm}) \, ,  &
\qquad \tau^{\pm}_2      &\longleftrightarrow &   (101, \beta^2_{\pm})  \\
\sigma^{\pm}_3    &\longleftrightarrow &   (001, \alpha^3_{\pm}) \, ,        &
\qquad \tau^{\pm}_3      &\longleftrightarrow &   (110, \beta^3_{\pm}) \, ,
\ea   
\eea
corresponds to an invariance of the modular matrices: 
\bea
S_{\bar{D}_2}  \; = \;    S_{ \{ \mbox{\scriptsize \bf Z}_2 \times 
\mbox{\scriptsize \bf Z}_2 \times \mbox{\scriptsize \bf Z}_2, \,
\omega_{\mbox{\tiny{I+III}}} \} } \, , \qquad
T_{\bar{D}_2}  \; = \;  T_{ \{ \mbox{\scriptsize \bf Z}_2 \times 
\mbox{\scriptsize \bf Z}_2 \times \mbox{\scriptsize \bf Z}_2, \,
\omega_{\mbox{\tiny{I+III}}} \} } \, .
\eea 
Hence, these two theories are dual; they describe 
the same spectrum and the same topological interactions. 

A couple of remarks concerning the foregoing duality are in order.
To start with, the duality transformation~(\ref{duality}) exchanges the 
nonabelian $\bar{D}_2$ magnetic flux doublets 
with the projective $\Z_2 \times \Z_2 \times \Z_2$  
doublet dyon charges, while the $\Z_4$ singlet dyon charges associated
to these $\bar{D}_2$ doublet fluxes are exchanged with 
the abelian $\Z_2 \times \Z_2 \times \Z_2$  magnetic fluxes. 
Also, the pure $\bar{D}_2$ singlet flux $\bar{1}$ is exchanged with 
the pure $\Z_2 \times \Z_2 \times \Z_2$ charge $(0,100)$.
In short, we are dealing with some kind of nonabelian 
electric/magnetic duality. It should be stressed though that the interchange
of electric and magnetic quantum numbers does not extend to the other 
particles. That is, the pure $\bar{D}_2$ singlet 
{\em charges} $J_1$, $J_2$ and $J_3$ and the singlet {\em dyons} $\bar{J}_1$, 
$\bar{J}_2$ and $\bar{J}_3$ are exchanged with pure 
$\Z_2 \times \Z_2 \times \Z_2$  singlet {\em charges}, 
the pure doublet {\em charge} $\chi$ with the doublet {\em dyon} 
$(111,\gamma_+)$, while the singlet {\em flux} of the 
dyon $\bar{\chi}$ is exchanged with the singlet {\em flux} of the 
dyon $(111, \chi_-)$ and the doublet dyon 
{\em charge} of $\bar{\chi}$ with the doublet dyon {\em charge} $\gamma_-$.   
As a next remark, the duality of  
the $\Z_2 \times \Z_2 \times \Z_2$ gauge theories with CS 
action the product of the 3-cocycle of type~III and either one of 
the other two 3-cocycles~(\ref{type1}) of type~I with the $\bar{D}_2$ theory
is inferred in a similar way. 
The duality transformation for these cases is given by a natural 
permutation of that in~(\ref{duality}).
Furthermore, it can be checked that duality with the $\bar{D}_2$ theory 
also emerges 
for the $\Z_2 \times \Z_2 \times \Z_2$ gauge theory featuring the 
CS action $\omega_{\mbox{\scriptsize I+I+I+III}}$, 
but is lost for the case
$\omega_{\mbox{\scriptsize I+I+III}}$. 
Here, $\omega_{\mbox{\scriptsize  I+I+I+III}}$ denotes
the product of the three distinct 3-cocycles of type~I 
and the 3-cocycle of type~III, while
$\omega_{\mbox{\scriptsize I+I+III}}$ stands for a product of two distinct 
3-cocycles of type~I and the 3-cocycle of type~III.
It is easily verified that the spin structure of the spectrum
for the latter theory is not matching that of the $\bar{D}_2$ theory, 
since the spectrum for $\omega_{\mbox{\scriptsize  I+I+III}}$
contains  five dyons with spin factor $\im$ and five dyons with 
spin factor $-\im$.
Finally, besides the double dihedral group $\bar{D}_2$ also the 
other nonabelian group of order 8 enters the scene, namely the dihedral
group $D_4$. As we will argue next, the 
$\Z_2 \times \Z_2 \times \Z_2$ gauge theory with 
the type~III CS action~(\ref{trico}) itself, for instance,
is dual to the ordinary 2+1 dimensional $D_4$ gauge theory 
discussed in appendix~\ref{dvier}. In fact, 
our earlier observation that the sets of matrices~(\ref{matrixa}) 
and~(\ref{matrixb}) associated to the dyon charges $\alpha^i_{\pm}$ 
and $\beta^i_{\pm}$, respectively, generate the two dimensional UIR of 
$D_4$ already formed circumstantial evidence supporting this result.

As displayed in relation~(\ref{spd4}) of appendix~\ref{dvier}, 
the $D_4$ theory features fourteen particles with trivial spin factor 1,
six particles with spin factor $-1$, one with spin factor $\im$ and 
one with $-\im$. So, the spin structure (and thus the modular $T$ matrix)
of the $D_4$ theory is the same as that given in relation~(\ref{spectz23}) 
for the 
$\{\Z_2 \times \Z_2 \times \Z_2, \omega_{\mbox{\scriptsize III}} \}$
CS theory. In addition, it can be verified that the modular 
$S$ matrix~(\ref{fusion}) for the 
$\{\Z_2 \times \Z_2 \times \Z_2, \omega_{\mbox{\scriptsize III}} \}$
CS theory is equivalent to that for the $D_4$ theory exhibited in
table~\ref{modsd4} of appendix~\ref{dvier}. Specifically, the duality 
transformation   
$
D_4   \leftrightarrow 
\{\Z_2 \times \Z_2 \times \Z_2, \omega_{\mbox{\scriptsize III}} \}   
$
for this case involves the following exchange of the particles in the 
spectrum~(\ref{specd4}) of the $D_4$ theory with the particles~(\ref{spectz23})
featuring in the 
$\{\Z_2 \times \Z_2 \times \Z_2, \omega_{\mbox{\scriptsize III}} \}$
CS theory
\bea            \label{duald4}
\ba{rclrcl}
(0,++)               &\longleftrightarrow &   (0) \, ,                  &
\qquad  (2,++)         &\longleftrightarrow &   (0,111)            \\
(0,+-)               &\longleftrightarrow &   (0,011) \, ,              &
\qquad  (2,+-)         &\longleftrightarrow &   (0,100)            \\
(0,-+)               &\longleftrightarrow &   (0,010) \, ,              &
\qquad  (2,-+)         &\longleftrightarrow &   (0,101)            \\
(0,--)               &\longleftrightarrow &   (0,001) \, ,              &
\qquad  (2,--)         &\longleftrightarrow &   (0,110)            \\
(0,1)                &\longleftrightarrow &   (100, \alpha^1_+) \, ,     &
\qquad  (2,1)          &\longleftrightarrow &   (100, \alpha^1_-)   \\
(1,0)                &\longleftrightarrow &   (011, \beta^1_+) \, ,      &
(X,\pm +)            &\longleftrightarrow &   (101, \beta^2_{\pm})  \\
(1,1)                &\longleftrightarrow &   (111, \gamma_-) \, ,       &
(X,\pm -)            &\longleftrightarrow &   (001, \alpha^3_{\pm}) \\
(1,2)                &\longleftrightarrow &   (011, \beta^1_-) \, ,      &
(\bar{X},\pm +)      &\longleftrightarrow &   (110, \beta^3_{\pm}) \\
(1,3)                &\longleftrightarrow &   (111, \gamma_+) \, ,       &
(\bar{X},\pm -)      &\longleftrightarrow &   (010, \alpha^2_{\pm}) \, .
\ea   
\eea
To proceed, a straightforward calculation shows that 
adding either one of the three 3-cocycles~(\ref{type2}) of 
type~II to the $\{\Z_2 \times \Z_2 \times \Z_2, 
\omega_{\mbox{\scriptsize III}} \}$ CS theory
does not destruct the duality with the $D_4$ theory.
That is, we also have the duality
$
D_4   \leftrightarrow 
\{\Z_2 \times \Z_2 \times \Z_2, \omega_{\mbox{\scriptsize II+III}} \} , 
$ 
where the associated duality transformation between the two spectra again 
corresponds to a permutation of the one in~(\ref{duald4}).

To conclude, a complete discussion, which is beyond the scope of 
this paper, also involves the finite set of CS actions for the two 
nonabelian gauge groups $D_4$ and $\bar{D}_2$.
This and the generalization of the foregoing nonabelian dualities 
to higher order finite abelian gauge groups 
allowing for CS actions of type~III is left for future work.
An interesting question concerning the latter generalization
is whether the nonabelian dual gauge groups are restricted to the 
dihedral and double dihedral series or also involve other nonabelian 
finite groups.

\sectiona{Dijkgraaf-Witten invariants}
\label{dijwit}

In reference~\cite{diwi}, Dijkgraaf and Witten defined a topological invariant
for a compact, closed oriented  three manifold ${\cal M}$ in terms of a 
3-cocycle $\omega \in H^3({ H}, U(1))$ for a finite group $H$.
They represented this invariant as the partition function
$Z({\cal M})$ of a lattice gauge theory with gauge group $H$ and 
CS action $\omega$. It was shown explicitly that $Z(M)$ is indeed a 
combinatorial invariant of the manifold ${\cal M}$.
In this section, I present some results on the Dijkgraaf-Witten 
inavariant for lens spaces using the 3-cocycles of type~II and of 
type~III for a finite abelian group $H$, which to my knowledge have 
not appeared in the literature before.
In addition, the value of the Dijkgraaf-Witten invariant for the 
3-torus ${\cal M} = S^1 \times S^1 \times S^1$ associated with 
the three types of 3-cocycles for 
${ H} \simeq \Z_2 \times \Z_2 \times \Z_2$ is derived.

The Dijkgraaf-Witten invariant for the 
lens space $L(p,q)$ associated with an abelian finite 
group $H$ and 3-cocycle $\omega$ takes the following form~\cite{diwi,altsc1}  
\bea \label{diwiinv}
Z(L(p,q)) &=& \frac{1}{|{ H}|} \sum_{\{A \in { H}|[A^p]=0\}} \;\; 
\prod_{j=1}^{p-1}  \omega (A, A^j, A^n) \, ,
\eea
with $|H|$ the order of $H$ and $n$ the inverse of $q$ mod $p$.
It is known~\cite{altsc1, altsc} that the Dijkgraaf-Witten invariant
for the  3-cocycles~(\ref{type1}) of type~I for ${ H} \simeq \Z_5$
can distinguish the lens spaces 
$L(5,1)$ and $L(5,2)$, which are homeomorphic but of different
homotopy type:   
\bea
Z(L(5,1)) &=& \left\{ \ba{ll}
         1& \mbox{for  $p_{\mbox{\scriptsize I}} =0$} \\
       \frac{1}{\sqrt{5}} &  \mbox{for  $p_{\mbox{\scriptsize I}}=1,4$} \\   
      -\frac{1}{\sqrt{5}} &  
\mbox{for  $p_{\mbox{\scriptsize I}} =2,3$} \ea \right.\\
Z(L(5,2)) &=& \left\{ \ba{cl}
               1& \mbox{for  $p_{\mbox{\scriptsize I}} =0$} \\
       -\frac{1}{\sqrt{5}} & \mbox{for  $p_{\mbox{\scriptsize I}}=1,4$} \\
\frac{1}{\sqrt{5}} &  \mbox{for  $p_{\mbox{\scriptsize I}} =2,3$} \, . 
\ea \right.
\eea
A simple numerical evaluation using for example Mathematica shows that
this nice property of the Dijkgraaf-Witten 
invariant~(\ref{diwiinv}) is lost for 3-cocycles 
of type~II and~III. Specifically, for ${ H} \simeq \Z_5 \times \Z_5$ 
and a 3-cocycle~(\ref{type2}) of type~II, one arrives at
\bea
Z(L(5,1)) \; = \; Z(L(5,2)) \; = \; \left\{ \ba{ll}
               1 & \mbox{for  $p_{\mbox{\scriptsize II}} = 0$} \\
            \frac{1}{5} &  \mbox{for  $p_{\mbox{\scriptsize II}}
 = 1,\dots, 4$} \, , \ea \right.
\eea
while for ${ H} \simeq \Z_5 \times \Z_5 \times \Z_5$  and a 
3-cocycle~(\ref{type3})  of  type~III  the situation becomes completely 
trivial
\bea
Z(L(5,1)) \; = \; Z(L(5,2)) \; = \; 1   
\qquad \mbox{for  $p_{\mbox{\scriptsize III}} = 0,1, \ldots, 4$} \, . 
\eea
In fact, it turns out that in general the invariant~(\ref{diwiinv}) 
based on the 3-cocycles of type~II and~III have less distinctive power 
then the one based on 3-cocycles of type~I.  
Further, the result for the nontrivial 3-cocycle of type~I
for ${ H} \simeq \Z_2$ 
\bea
Z(L(p,1)) &=& \left\{ \ba{ll}
                 \frac{1}{2} & \mbox{for odd  $p$} \\
                \frac{1}{2}(1+(-1)^{p/2}) & \mbox{for even $p$} \, , 
\ea \right. 
\eea
established in~\cite{diwi}, generalizes in the  following manner to the 
nontrivial 3-cocycle~(\ref{type1}) of type~II for 
${ H} \simeq \Z_2 \times \Z_2$ 
\bea
Z(L(p,1)) &=&  \left\{ \ba{ll}
               \frac{1}{4} & \mbox{for odd  $p$} \\
               \frac{1}{4}(3+(-1)^{p/2}) & \mbox{for even $p$} \, , \ea 
\right.
\eea 
and to 
\bea
Z(L(p,1))&=&  \left\{ \ba{ll}
            \frac{1}{8} &      \mbox{for odd  $p$} \\
         \frac{1}{8}(7+(-1)^{p/2}) & \mbox{for even $p$} \, , \ea \right.
\eea
for the nontrivial 3-cocycle~(\ref{type3}) of type~III for 
${ H} \simeq \Z_2 \times \Z_2 \times \Z_2$.

The Dijkgraaf-Witten
invariant for the 3-torus $S^1 \times S^1 \times S^1$
is of particular interest, since it counts the number of 
particles in the spectrum of 
a discrete ${ H}$ CS gauge theory~\cite{diwi}.
For abelian groups ${ H}$ it takes the form 
\bea
Z(S^1 \times S^1 \times S^1) &=&\frac{1}{|{ H}|} \sum_{A,B,C \in { H}} 
W(A,B,C) \, ,
\eea
with
\bea 
W(A,B,C) &=& \frac{\omega(A,B,C) \; \omega(B,C,A) \; \omega(C,A,B)}{
\omega(A,C,B) \; \omega(B,A,C) \; \omega(C,B,A)} \, .
\eea 
It is not difficult to check that for the three different types 
of 3-cocycles for the direct product group 
${ H} \simeq \Z_2 \times \Z_2 \times \Z_2$, the invariant takes the values
\bea
Z(S^1 \times S^1 \times S^1) &=&  \left\{ \ba{cl}
               64 & \mbox{for type~I and~II} \\
               22 & \mbox{for type~III} \, , \ea \right.
\eea 
expressing the collapse of the spectrum we found for 
3-cocycles of type~III in section~\ref{typeIII}.

As an aside, with the same data entering the Dijkgraaf-Witten invariant, 
namely a finite group ${ H}$ and a 3-cocycle~$\omega$, 
Altschuler and Coste~\cite{altsc1,altsc} constructed a surgery 
invariant ${\cal F}({\cal M})$ from a surgery presentation of the 
manifold ${\cal M}$. They conjectured that up to normalization these two 
invariants are the same. That is, ${\cal F}({\cal M)} = Z({\cal M}) /Z(S^3)$
with $Z(S^3)= 1/|{H}|$. Altschuler and Coste verified their conjecture
for lens spaces using the 3-cocycles of type~I for  cyclic groups
${ H} \simeq \Z_N$. In~\cite{thesis}, I subsequently 
reported that extending this analysis to the 3-cocycles of type~II 
and of type~III, which  were not treated in~\cite{altsc1,altsc}, 
did not lead to any counter--examples.
In the meanwhile, I have become aware of reference~\cite{freed}
which contains the proof of the conjecture of Altschuler and Coste.

\section{Concluding remarks and outlook} \label{concl}

Some time ago Dijkgraaf and Witten~\cite{diwi} pointed out that the 
Chern-Simons (CS) actions for a compact gauge group $G$ are 
in one to one correspondence with the different elements of the cohomology 
group $H^4(BG,\Z)$ with $BG$ the classifying space for $G$. 
They also noted that this classification includes the case of finite gauge 
groups $H$. The isomorphism $H^4(BH,\Z) \simeq H^3(H,U(1))$, which only 
holds for finite $H$, then indicates that the different CS actions for a 
finite gauge group $H$ correspond to the inequivalent 3-cocycles 
$\omega \in H^3(H,U(1))$. One of the key results of the present paper has 
been that effective the long distance physics of a 
CS theory in which the gauge group $G$ is broken down to a subgroup $K$ via 
the Higgs mechanism is described by a CS theory with residual 
gauge group $K$ and CS action $S_{\mbox{\scriptsize CS}}' \in H^4(BK,\Z) $ 
determined by the original CS action $S_{\mbox{\scriptsize CS}} \in 
H^4(BG,\Z)$ for the broken gauge group $G$ through the natural 
homomorphism $H^4(BG,\Z) \rightarrow H^4(BK,\Z) $ induced by the 
inclusion $K \subset G$. In case $G$ is broken down to a finite residual 
gauge group $H$, the foregoing homomorphism, also known as the restriction, 
and the aforementioned isomorphism $ H^4(BH,\Z) \simeq H^3(H,U(1))$ then 
combines into a natural homomorphism $H^4(BG,\Z) \rightarrow H^3(H,U(1))$.
The 3-cocycle $\omega  \in H^3(H,U(1))$ being the image of some 
$S_{\mbox{\scriptsize CS}} \in H^4(BG,\Z)$ under the latter mapping
simply summarizes the additional Aharonov-Bohm (AB) interactions cast upon the 
magnetic vortices featuring in this model by the original CS action
$S_{\mbox{\scriptsize CS}}$ for the broken gauge group $G$.
This general scheme was illustrated with CS theories
in which some continuous compact abelian gauge group, typically a direct 
product $G \simeq U(1)^k$ of $k$ compact $U(1)$ gauge groups, is 
spontaneously broken down to a finite subgroup being a direct product 
$H \simeq \Z_{N^{(1)}} \times \Z_{N^{(2)}} \times \cdots \times \Z_{N^{(k)}}$ 
with $\Z_{N^{(i)}}$ the cyclic group of order  ${N^{(i)}}$.
Among other things, it has been argued that the restriction 
$H^4(B(U(1)^k),\Z) \rightarrow 
H^3(\Z_{N^{(1)}} \times \cdots \times \Z_{N^{(k)}} ,U(1))$ accompanying 
this case is not onto. Specifically, there are two types 
of  CS terms for $U(1)^k$. One type describes self couplings of the 
distinct $U(1)$ gauge fields, whereas the other type establishes pairwise 
couplings between the different $U(1)$ gauge fields.  Further,
in the presence of Dirac monopoles the topological masses characterizing 
these CS terms are necessarily quantized which is in agreement with 
the fact that the different CS actions for a compact gauge group $U(1)^k$  
are labeled by the integers: $H^4(B(U(1)^k),\Z) \simeq      
\Z^{ k + \frac{1}{2}k(k-1)}$. The 3-cocycles for the direct product group
$\Z_{N^{(1)}} \times \cdots \times \Z_{N^{(k)}}$, on the other hand, were
shown to split up into three different types. The first type describes 
additional  AB interactions between vortices carrying flux
w.r.t.\ the same cyclic gauge group in the direct product, 
the second type between vortices belonging to two different cyclic 
gauge group in the direct product, and the third type realizes couplings
between fluxes associated with three distinct cyclic gauge groups.
It has been demonstrated that only the first two types 
of 3-cocycles can be obtained from a spontaneously broken $U(1)^k$ CS theory.

In fact, the 3-cocycles of the third type that can not be reached from the 
spontaneous breakdown of a $U(1)^k$ CS theory turned out to be the most 
interesting. Adding such a 3-cocycle or CS action to an {\em abelian} discrete
$H$ gauge theory renders the theory {\em nonabelian}.
Moreover, the $\Z_2 \times \Z_2 \times \Z_2$ CS theory defined by such a 
3-cocycle, for instance, was shown to be dual to an ordinary $D_4$ gauge 
theory with $D_4$ the dihedral group of order 8, while the 
$\Z_2 \times \Z_2 \times \Z_2$ CS theory given by the product of 
a 3-cocycle of the first and third type was demonstrated to be dual
to an ordinary $\bar{D}_2$ gauge theory with $\bar{D}_2$ the double dihedral
group of order 8. The corresponding duality transformations involve 
the exchange of electric and magnetic quantum numbers. Future research
should point out how these nonabelian electric/magnetic dualities
generalize to higher order abelian finite gauge groups $H$.
Another question that deserves further scrutiny is whether abelian 
discrete $H$ theories with a 3-cocycle of the third type can be  
embedded in CS theories with a nonabelian broken continuous gauge group.

The focus of the discussion of the type~II $U(1)\times U(1)$ CS theory 
in section~\ref{typeII} was on the case in which 
both  $U(1)$ gauge groups are simultaneously broken down to a cyclic group.
Of course , it is also conceivable that just one $U(1)$ gauge group is 
broken to a cyclic group. A group cohomogical derivation for the latter case
analogous to those in appendix~\ref{gc} yields
$H^4(B(U(1) \times \Z_N), \Z) \simeq \Z \oplus \Z_N \oplus \Z_N$. 
Here, $\Z$ naturally labels the different type~I CS actions for the 
compact $U(1)$ gauge group, one of the $\Z_N$ terms the 3-cocycles of type~I
for the finite cyclic gauge group $\Z_N$ and the other $\Z_N$ term the 
type~II CS action for $U(1) \times \Z_N$. This results indicates that
if one of the $U(1)$ groups of a type~II $U(1) \times U(1)$ CS theory is 
spontaneously broken down to $\Z_N$ the integral type~II CS parameter becomes 
periodic with period $N$. 
The characteristics of this model are currently under investigation.

Also, the vortices, pure charges and dyons  in these 
spontaneously broken models have been treated as point particles 
in the first quantized description in this paper. 
Rerunning the discussion in the framework of canonical quantization
involves the construction of magnetic vortex creation operators
and charge creation operators and an analysis of their 
nontrivial commutation relations~\cite{moresama}.

Another obvious next step is to consider CS theories in which 
some continuous {\em nonabelian} gauge group is spontaneously broken 
down to a finite {\em nonabelian} gauge group.
For a concise discussion of these models, the reader is referred 
to~\cite{thesis} and references therein. A more detailed study 
will be presented elsewhere.

Finally, it has recently been suggested~\cite{hagen, diamant} that $U(1)^k$ CS 
theories may play a role in multi-layered fractional quantum Hall systems. 
At present, it is not clear to me whether the broken version of these models 
considered in this paper are also relevant in this setting.

\section*{Acknowledgements}
Most of the work reported here has been done while I was a PhD student 
at the Instituut voor Theoretische Fysica of the Universiteit van 
Amsterdam~\cite{thesis}. It is a pleasure to thank my supervisor 
at the time, Sander Bais, for many valuable conversations on this topic. 
I also greatly benefited from useful discussions with and suggestions of 
Daniel Altschuler, Dan Freed, Eduard Looijenga, 
Alec Maassen van den Brink, Alosha Morozov, Peter van Driel, 
Alain Verberkmoes, Erik Verlinde, Herman Verlinde and especially Danny 
Birmingham and  Robbert Dijkgraaf. Special thanks are due to Peter van Driel 
for a collaboration in which we developed the concept of truncated braid 
groups~\cite{pema}.

\aanhangsel 
\sectiona{ Cohomological derivations}
\label{gc}

This appendix contains the derivation of some of the 
group cohomological results used in this paper. 
In passing, I stress that in contrast with the main text  
the cohomology and abelian groups will be presented in the 
additive rather than  multiplicative form.
In the additive presentation, a direct product 
of $k$ cyclic factors $\Z_N$, for example,
becomes the direct sum denoted by  $\Z_N^k := \oplus_{i=1}^k \Z_N$.

My first objective is to prove the isomorphism~(\ref{clasi}).
This will be done using  
the universal coefficients theorem (e.g.\ \cite{rotman})
\bea                         \label{uni}
H^n(X, {\mbox{\bf B}}) &\simeq& H^n(X, \Z) \otimes {\mbox{\bf B}}
 \oplus \mbox{Tor} (H^{n+1}(X,\Z),{\mbox{\bf B}}) \, ,
\eea
relating the cohomology of some topological
space $X$ with coefficients in some abelian group ${\mbox{\bf B}}$
and the cohomology of $X$ with integer coefficients $\Z$.
Here, $\ot$ stands for the symmetric tensor product and  
$\mbox{Tor}(\, . \, \, ,\, . \,)$ for the torsion product.
The symmetric tensor product ${\mbox{\bf A}} \otimes {\mbox{\bf B}}$ 
(over $\Z$) for abelian groups 
${\mbox{\bf A}}$ and ${\mbox{\bf B}}$ is the abelian group of all 
ordered pairs $a \otimes b$
($a \in {\mbox{\bf A}}$ and $b \in {\mbox{\bf B}}$) with 
relations~\cite{rotman} 
\beas
(a+a') \ot b &=& a \ot b + a' \ot b, \qquad 
a \ot (b+b') \; = \;  a \ot b + a \ot b'  \\
m(a \ot b )  &=& ma \ot b \; =\; a \ot mb \qquad \forall \, m \in \Z \,.
\eeas 
It is not difficult to check that these relations imply the 
following identifications
\bea      
\Z_N \otimes \Z_M & \simeq &  \Z_{{\gcd}(N,M)}  \label{begin} \\
\Z_N \otimes \Z &\simeq& \Z_N  \\
\Z_N \otimes U(1) &\simeq& 0  \label{nec0}   \\
\Z \otimes U(1) &\simeq& U(1)         \\
\Z \otimes \Z  &\simeq& \Z \label{nas} \, ,     
\eea
with ${\gcd}(N,M)$ the greatest common divisor of $N$ and $M$.
Finally, the symmetric tensor product $\otimes$ is obviously distributive 
\bea                                     \label{dizzy}
(\oplus_{i} {\mbox{\bf A}}_i)\otimes {\mbox{\bf B}} &\simeq& \oplus_{i} 
({\mbox{\bf A}}_{i} 
\otimes {\mbox{\bf B}}) \, .
\eea
The definition of  the torsion product $\mbox{Tor}(\, . \, \, ,\, . \,)$
can be found in any textbook on algebraic topology. 
For our purposes, the following properties suffice~\cite{rotman}.
Let ${\mbox{\bf A}}$ and ${\mbox{\bf B}}$ again be abelian groups, then
\beas
\mbox{Tor} ({\mbox{\bf A}},{\mbox{\bf B}}) 
&\simeq& \mbox{Tor}({\mbox{\bf B}},{\mbox{\bf A}})   \\
\mbox{Tor} (\Z_N, {\mbox{\bf B}}) &\simeq& {\mbox{\bf B}}[N] \; \simeq \; 
\{ b \in {\, \mbox{\bf B}}\,| \, N b =0 \,\} \, , 
\eeas
so in particular
\bea    
\mbox{Tor} (\Z_N, \Z_M) &\simeq& \Z_{{\gcd}(N,M)}  \label{difor} \\
\mbox{Tor} (\Z_N, U(1)) &\simeq& \Z_N   \label{nec}         \\
\mbox{Tor} ({\mbox{\bf A}}, \Z) &\simeq& 0    
\qquad \qquad \forall {\mbox{\bf A}} \, .  
\label{torp}
\eea
The last identity follows from  the fact
that the group of integers $\Z$
is torsion free, i.e.\ it does not contain elements of finite order.
Just as the symmetric tensor product, the torsion product is  distributive 
\bea                               \label{eind}
\mbox{Tor} (\oplus_{i} {\mbox{\bf A}}_{i}, {\mbox{\bf B}}) &\simeq& \oplus_{i} 
\mbox{Tor} ( {\mbox{\bf A}}_{i}, {\mbox{\bf B}}) \, .
\eea

The proof of the isomorphism~(\ref{clasi}) now goes as follows. 
First we note that  for finite
groups ${ H}$ all cohomology in fixed degree $n>0$ is finite.
With this knowledge, the universal coefficients theorem~(\ref{uni}) 
directly gives the desired result
\bea                                   
 H^n({ H}, U(1)) & \simeq &  H^n({ H}, \Z) \ot U(1) \oplus
               \mbox{Tor} (H^{n+1}({ H},\Z), U(1))    \nn      \\
             & \simeq &  H^{n+1}({ H}, \Z)  \qquad \mbox{for $n>0$} \, . 
\label{ole}
\eea
In the last step, we used the distributive property of the tensor 
product~(\ref{dizzy}) and 
the torsion product~(\ref{eind}) together with  the identities~(\ref{nec0}) 
and~(\ref{nec}).

We turn to the derivation of the identities~(\ref{conj1e})--(\ref{conj3e}). 
Our starting point will be the standard result  (e.g.\ \cite{spanier})
\bea
H^n(\Z_N,\Z) &\simeq&  \left\{ \ba{ll}
                 \Z_N & \mbox{if $n$ is even} \label{nec3} \\
                 0 & \mbox{if $n$ is odd} \\
                 \Z &  \mbox{if $n=0 \, ,$} \ea \right.
\eea 
which together with~(\ref{ole}) implies that
the identities in~(\ref{conj1e})--(\ref{conj3e})  are valid for $k=1$.
The extension to $k>1$  involves the so-called
K\"{u}nneth formula (e.g. \cite{rotman})
\bea                  \label{kun}
  H^n(X \times Y, \Z) \; \simeq 
 \sum_{i+j=n} H^i(X, \Z) \otimes H^j(Y, \Z) \oplus \!\!\!
 \sum_{p+q=n+1} \!\!\! \mbox{Tor} (H^{p}(X,\Z), H^{q}(Y,\Z)) \, ,   \;\;
\eea
which states that the cohomology of a direct product space is completely
determined in terms of the cohomology of its factors.
With the ingredients~(\ref{nec3}) and~(\ref{kun}),
the identities~(\ref{conj1e})--(\ref{conj3e}) can then be proven 
by induction. To lighten the notation a bit,
we will omit explicit mention of 
the coefficients of the cohomology groups
if the  integers $\Z$ are meant. So,  
$H^n(\Z_N^k) := H^n(\Z_N^k, \Z)$.  Let us start with  the 
trivial cohomology group $H^0(\Z_N^k)$.
Upon using the K\"{u}nneth formula~(\ref{kun}), 
the property~(\ref{torp}) of the torsion product
and the  result~(\ref{nec3}), we easily infer
\bea                   \label{eros}
H^0(\Z_N^k) \; \simeq \; H^0(\Z_N^{k-1}) \ot H^0(\Z_N^k)  
             \; \simeq \; H^0(\Z_N^{k-1}) \ot \Z  \; \simeq \; \Z \, ,  
\eea
where the last isomorphism follows by induction.
To be explicit, as indicated by~(\ref{nec3}) this isomorphism 
obviously holds for $k=1$. 
If we subsequently assume that this isomorphism is valid 
for some fixed $k$, we obtain with~(\ref{nas}) that it also holds for $k+1$.
To proceed, in a similar fashion, we arrive at 
\be                   
H^1(\Z_N^k) \; \simeq \; H^1(\Z_N^{k-1}) \ot H^0(\Z_N) 
         \; \simeq \; H^1(\Z_N^{k-1}) \; \simeq \; 0 \, .
\ee
These results enter the following derivation starting  from the K\"unneth
formula~(\ref{kun}) 
\bea                         \label{era}
H^2(\Z_N^k) &\simeq& H^0(\Z_N^{k-1}) \ot H^2(\Z_N) \oplus
                      H^2(\Z_N^{k-1}) \ot H^0(\Z_N)  \\
             &\simeq& \Z_N \oplus H^2(\Z_N^{k-1}) \; \simeq \; \Z_N^k 
\, .   \nn
\eea 
Here, we used the distributive property~(\ref{dizzy}) of the
tensor product and again induction 
to establish  the last isomorphism.
We continue with 
\bea                                  \label{era1}
H^3(\Z_N^k) &\simeq&  H^3(\Z_N^{k-1}) \ot H^0(\Z_N) \oplus 
                       \mbox{Tor}(H^2(\Z_N^{k-1}), H^2(\Z_N)) \\
             &\simeq&  H^3(\Z_N^{k-1}) \oplus \Z_N^{k-1} \; \simeq \;
                       \Z_N^{\frac{1}{2}k(k-1)} \, . \nn
\eea
Finally, using the previous results and  induction, we obtain
\bea                      \label{eureka}
H^4(\Z_N^k) &\simeq& H^0(\Z_N^{k-1})\ot H^4(\Z_N) \oplus 
                      H^2(\Z_N^{k-1}) \ot H^2(\Z_N)  \oplus     \\
            &  &    H^4(\Z_N^{k-1})\ot H^0(\Z_N) 
                  \oplus \mbox{Tor}(H^3(\Z_N^{k-1}), H^2(\Z_N))  \nn \\
            &\simeq& H^4(\Z_N) \oplus H^2(\Z_N^{k-1})\oplus  H^4(\Z_N^{k-1})
                    \oplus \mbox{Tor}(H^3(\Z_N^{k-1}), H^2(\Z_N))  \nn \\
            &\simeq &  \Z_N \oplus \Z_N^{k-1} \oplus H^4(\Z_N^{k-1})
                 \oplus \Z_N^{\frac{1}{2}(k-1)(k-2)}     \nn \\
            &\simeq & \Z_N^{k+\frac{1}{2}(k-1)(k-2)} \oplus H^4(\Z_N^{k-1}) 
                     \nn \\
     &\simeq &  \Z_N^{k+\frac{1}{2}k(k-1)+\frac{1}{3!}k(k-1)(k-2)} \, .   \nn
\eea
To conclude, the results~(\ref{era}),~(\ref{era1}) and~(\ref{eureka}) 
together with~(\ref{ole}) lead to the 
identities~(\ref{conj1e}),~(\ref{conj2e}) 
and~(\ref{conj3e}) respectively.

The foregoing  derivation also gives a nice insight into the 
structure of the terms building up the cohomology 
group $H^4(\Z_N^k)\simeq H^3(\Z_N^k, U(1))$.
We can, in fact, distinguish three types of terms that contribute here.
By induction, we find that there are  $k$ terms  
of the form  $H^4(\Z_N)$. 
These are the terms that label the 3-cocycles~(\ref{type1}) of type~I. 
By a similar argument, we infer that there are 
$\frac{1}{2} k(k-1)$ terms of the  form  
$H^2(\Z_N^{k-1})$. These terms 
label the type~II 3-cocycles~(\ref{type2}). 
Finally, the $\frac{1}{3!}k(k-1)(k-2)$ terms we are left with are 
entirely due to torsion products
and label the 3-cocycles~(\ref{type3}) of type~III.

The generalization of the above results 
to abelian groups ${ H}$ being 
direct products of cyclic groups possibly of different order is 
straightforward. The picture that the 3-cocycles divide 
into three different types remains unaltered. 
If the direct product ${H}$
consists of $k$ cyclic factors, then there are again $k$ different 
3-cocycles of type~I, $\frac{1}{2} 
k(k-1)$ different 3-cocycles of type~II and $\frac{1}{3!}k(k-1)(k-2)$ 
different 3-cocycles of type~III.
The only distinction is that through~(\ref{begin}) and~(\ref{difor})
the greatest common divisors of the orders of the different cyclic factors 
constituting the direct product group $H$ enter the scene 
for 3-cocycles of type~II and~III.
This is best illustrated by considering the direct product
group $H \simeq \Z_N \times \Z_M \times 
\Z_K$ being the simplest example where all three types of 3-cocycles
appear. The derivation~(\ref{eros})--(\ref{eureka}) for this 
case leads to the following content of the relevant cohomology groups 
\bea   \left\{    \ba{lcl}         \label{conj1do}         
H^1(\Z_N \times \Z_M \times \Z_K, U(1)) 
& \simeq & \Z_N \oplus \Z_M \oplus \Z_K \\
H^2(\Z_N \times \Z_M \times \Z_K,U(1)) & \simeq & 
\Z_{{\gcd}(N,M)} \oplus \Z_{{\gcd}(N,K)} \oplus \Z_{{\gcd}(M,K)}
\\
H^3(\Z_N \times \Z_M \times \Z_K,U(1)) & \simeq & 
\Z_N \oplus \Z_M \oplus \Z_K \oplus  \\ & & 
\Z_{{\gcd}(N,M)} \oplus \Z_{{\gcd}(N,K)} 
\oplus \Z_{{\gcd}(M,K)} \oplus \\
& & \Z_{{\gcd}(N,M,K)} \, .
\ea
\right.
\eea   
The 3-cocycles of type~I labeled by the terms $\Z_N$, $\Z_M$
and $\Z_K$ are of the form~(\ref{type1do}), whereas the explicit 
the 3-cocycles of type~II labeled by the terms
$\Z_{{\gcd}(N,M)}$, $\Z_{{\gcd}(N,K)}$ and $\Z_{{\gcd}(M,K)}$ 
take the form~(\ref{type2do}). The explicit realization of the 
3-cocycles of type~III corresponding to the term $\Z_{{\gcd}(N,M,K)}$
can be found in~(\ref{type3do}).

Let us close by establishing the isomorphism~(\ref{u1k}).
The standard result (e.g.\ \cite{diwi})
\bea
H^n(BU(1)) &\simeq&  \left\{ \ba{ll}
                 \Z & \mbox{if $n=0$ or $n$ even} \label{necu1} \\
                 0 & \mbox{otherwise,}
                 \ea \right.
\eea 
(generated by the first Chern class of degree 2)
indicates that~(\ref{u1k}) holds for $k=1$.
For $k > 1$, we may again appeal to the K\"unneth
formula, because the classifying space of the product group 
$U(1)^k$ is the same as the 
product of the classifying spaces of the factors. That is,
$B(U(1)^k) = B(U(1)^{k-1}) \times BU(1)$ 
(e.g.\ \cite{novikov}, page 132).   
The derivation of the result~(\ref{u1k}) then becomes 
similar to the one given for the finite abelian group $\Z_N^k$.
Since the group $\Z$ is torsion free, however, the terms 
due to  torsion products vanish in this case.
The terms that persist are the following.
First of all, there are $k$ terms of the form $H^4(BU(1))\simeq \Z$. These
label the different CS actions of type~I displayed in~(\ref{CSt1}).
In addition, there are $\frac{1}{2} k(k-1)$ terms 
of the form $H^2(BU(1)) \simeq \Z$ which label
the CS actions of type~II given in~(\ref{CSt2}).

\sectiona{Truncated braid groups}         \label{trubra}

A characteristic property of the braid operator~(\ref{braidaction}) 
is that it is of finite order. That is,  ${\cal R}^m = 
{\mbox{\bf 1}} \ot {\mbox{\bf 1}}$
with $\mbox{\bf 1}$ the identity operator and $m$ some integer 
depending on the particles on which the braid operator 
acts. Hence, we can assign a finite integral number  $m$ 
to any two particle 
internal Hilbert space $V_\alpha^A \ot V_\beta^B$ such that the effect
of $m$ braidings is trivial for all states in $V_\alpha^A \ot V_\beta^B$.
This result, which can be traced back directly to the 
finite order of ${ H}$, implies that the multi-particle 
configurations appearing in abelian discrete ${ H}$ 
CS theories actually realize representations of factor groups
of the ordinary braid groups.~\footnote{The same holds 
for multi-particle configurations in nonabelian discrete $H$ 
gauge theories with or without a CS action~\cite{banff,thesis}.}
This appendix is dedicated to the definition of these factor groups
and a subsequent identification of some of these factor groups with
well-known finite groups.

Let me first recall~\cite{wu} 
that the wave function of a system $n$ indistinguishable 
particles in the plane in general transforms as a nontrivial unitary 
representation of the braid group $B_n( \mbox{\bf R}^2)$.
For convenience, I suppose that the particles are 
numbered from $1$ to $n$. The braid group 
$B_n( \mbox{\bf R}^2)$ can then be presented by $n-1$ generators
$\tau_i$ (with $i \in 1, \ldots, n-1$) subject to the relations
\bea
\label{eqy}
\ba{rcll}
\tau_i\tau_{i+1}\tau_i &=& \tau_{i+1}\tau_i\tau_{i+1} &
\qquad i=1,\ldots,n-2  \\
\tau_i\tau_j &=& \tau_j\tau_i & \qquad |i-j|\geq 2 \, .
\ea
\eea
Here, the generator $\tau_i$ establishes a counterclockwise interchange 
of the particles $i$ and $i+1$. The braid relations~(\ref{eqy}) 
then express the fact that the particle trajectories corresponding to 
the composed interchange process at the l.h.s.\ of the equality
sign can be continuously deformed into the one at the r.h.s.\ 
of the equality sign.

Let us now focus on a system of $n$ indistinguishable particles in a
planar abelian discrete $H$ CS theory. So, all particles carry the same 
internal Hilbert space $V^A_{\alpha}$. Due to the finite order of the braid 
matrix~(\ref{braidaction}), the assignment~(\ref{brareco}) now 
furnishes a representation on the associated $n$-particle internal 
Hilbert space of the factor group of $B_n( \mbox{\bf R}^2)$
in which the generators $\tau_i$ satisfy the {\em extra} relation
\bea
\tau_i^m &=& e  \;\;\;\;\;\;\;\;\;\; i=1, \ldots , n-1 \, ,
\label{truncate}
\eea
with $e$ the unit element or trivial braid. Of course, the order $m$ of
the generators depends on the nature of the particles $(A, \alpha)$. 
For obvious reasons, we will call the foregoing factor group 
with defining relations~(\ref{eqy}) and the additional 
relations~(\ref{truncate}) the {\em truncated} braid group $B(n,m)$,  where
$n$ naturally stands for the number of particles and $m$ for the order of 
the generators $\tau_i$.

For a planar system consisting of $n$ distinguishable particles, in turn, 
only the monodromy operations on the particles are relevant. That is, 
the wave function of such a system generally transforms as a nontrivial 
unitary representation of the so-called pure or colored braid group 
$P_n(\mbox{\bf R}^2)$ being the subgroup of the braid group 
$B_n( \mbox{\bf R}^2)$ generated by the elements 
(see for instance~\cite{frohma})
\bea   \label{pure}                      
\gamma_{ij} &=& \tau_i \cdots \tau_{j-2}\; \tau_{j-1}^2 \; 
\tau_{j-2}^{-1}\cdots
\tau_i^{-1}   \qquad \qquad   1 \leq i<j \leq n \, ,
\eea
which establish a counterclockwise monodromy of the particles $i$ and $j$.  
The internal Hilbert space associated with a system of $n$ distinguishable 
particles in a discrete (CS) theory (i.e.\ the particles carry different 
colors or internal Hilbert spaces $V^{A_i}_{\alpha_i}$) then carries a 
representation of a truncated version or factor group $P(n,m)$ of the 
colored braid group $P_n(\mbox{\bf R}^2)$. To be specific, 
the truncated colored braid group $P(n,m)$ is the subgroup of $B(n,m)$ 
generated by the elements~(\ref{pure}) with the extra 
relation~(\ref{truncate}) implemented. So, the generators of $P(n,m)$ 
are of order $m/2$: 
\bea     \label{puretru}           
\gamma_{ij}^{m/2} &=& e \, ,  
\eea                                              
from which it is clear that the truncated 
colored braid group $P(n,m)$ is, in fact, just defined for even $m$.

Finally, a  `mixture' of the foregoing systems is of course also possible. 
That is, a system which contains a subsystem of $n_1$ particles 
with `color'  $V^{A_{1}}_{\alpha_{1}}$, a subsystem of $n_2$ particles 
carrying the different `color' $V^{A_{2}}_{\alpha_{2}}$ and so on.
Let $n=n_1+n_2+\ldots$ again be the total number of particles in the 
system. The $n$-particle internal Hilbert space associated to this system 
then carries a representation of the truncated partially colored braid group
being the subgroup of some truncated braid group $B(n,m)$ generated by 
the braid operations $\tau_i$ on  particles with the same `color' 
and the monodromy operations~(\ref{pure}) on differently `colored' 
particles.

The appearance of truncated rather than ordinary braid groups in 
discrete (CS) gauge theories facilitates the decomposition of a 
given multi-particle internal Hilbert space into 
irreducible subspaces under the braid/monodromy  operations. 
The point is that the representation theory of  
ordinary braid groups is quite complicated due to their infinite order.
The extra relation~(\ref{truncate}) for truncated braid groups $B(n,m)$, 
however, causes these to become finite for various values of the labels $n$ 
and $m$ leading to identifications  with well-known groups of finite 
order~\cite{pema}. The truncated braid group $B(2,m)$ for two 
indistinguishable particles, for instance, has only one generator $\tau$ 
satisfying  $\tau^m = e$. Thus, we obtain the isomorphism $B(2,m) \simeq \Z_m$.
For $m=2$, the relations~(\ref{eqy}) and~(\ref{truncate}) are the 
defining relations of the permutation group $S_n$ on $n$ strands:
$B(n,2) \simeq S_n$. A less trivial example is the nonabelian truncated 
braid group $B(3,3)$ for 3 indistinguishable particles. By explicit 
construction  from the defining relations~(\ref{eqy}) and~(\ref{truncate})
for $n=m=3$, we arrive at the identification $B(3,3) \simeq \bar{T}$ 
with $\bar{T}$ the lift of the tetrahedral group $T \subset SO(3)$ 
into $SU(2)$. We close this appendix with the structure of the truncated 
braid group $B(3,4)$ and the truncated colored braid group $P(3,4)$, 
representations of which are realized by certain three particle 
configurations in the planar type~III CS theory with finite gauge group 
$\Z_2 \times \Z_2 \times \Z_2$ discussed in section~\ref{typeIII}
(see subsection~\ref{z23ab}).

\begin{table}[htb]
\begin{tabular}{crrrrrrrrrrrrrrrr}
 \hline  \\[-4mm]
          & ${\ssc C_0^1}$ & ${\ssc C_0^2}$ & ${\ssc C_0^3}$ & 
            ${\ssc C_0^4}$ & ${\ssc C_1^1}$ & ${\ssc C_1^2}$ & 
            ${\ssc C_1^3}$ & ${\ssc C_1^4}$ & ${\ssc C_2^1}$ & 
            ${\ssc C^2_2}$ & ${\ssc C^3_2}$ & ${\ssc C^4_2}$ & 
            ${\ssc C_3^1}$ & ${\ssc C_3^2}$ & ${\ssc C_4^1}$ & 
            ${\ssc C_4^2}$  \\ \hline   \\[-4mm]
${\ssc \Lambda_0}$ & ${\ssc 1}$ & ${\ssc 1}$ & ${\ssc 1}$ & ${\ssc 1}$ & 
                     ${\ssc 1}$ & ${\ssc 1}$ & ${\ssc 1}$ & ${\ssc 1}$ &
                     ${\ssc 1}$ & ${\ssc 1}$ & ${\ssc 1}$ & ${\ssc 1}$ & 
                     ${\ssc 1}$ & ${\ssc 1}$ & ${\ssc 1}$ & ${\ssc 1}$   
                     \\ 
${\ssc \Lambda_1}$ & ${\ssc 1}$ & -${\ssc 1}$ & ${\ssc 1}$ & -${\ssc 1}$ & 
               ${\ssc \im}$ & -${\ssc \im}$ & ${\ssc \im}$ & -${\ssc \im}$ & 
                    -${\ssc 1}$ &  ${\ssc 1}$ & -${\ssc 1}$ & ${\ssc 1}$ & 
                    -${\ssc 1}$ & ${\ssc 1}$ & -${\ssc \im}$ & ${\ssc \im}$ 
                     \\
${\ssc \Lambda_2}$ & ${\ssc 1}$ & ${\ssc 1}$ & ${\ssc 1}$ & ${\ssc 1}$ & 
                    -${\ssc 1}$ & -${\ssc 1}$ & -${\ssc 1}$ & -${\ssc 1}$ & 
                     ${\ssc 1}$ & ${\ssc 1}$ & ${\ssc 1}$ & ${\ssc 1}$ & 
                     ${\ssc 1}$ & ${\ssc 1}$ & -${\ssc 1}$ & -${\ssc 1}$  
                     \\
${\ssc \Lambda_3}$ & ${\ssc 1}$ & -${\ssc 1}$ & ${\ssc 1}$ & -${\ssc 1}$ & 
               -${\ssc \im}$ & ${\ssc \im}$ & -${\ssc \im}$ & ${\ssc \im}$ & 
                    -${\ssc 1}$ & ${\ssc 1}$ & -${\ssc 1}$ & ${\ssc 1}$ & 
                    -${\ssc 1}$ & ${\ssc 1}$ & ${\ssc \im}$ & -${\ssc \im}$ 
                     \\ 
${\ssc \Lambda_4}$ & ${\ssc 2}$ & ${\ssc 2}$ & ${\ssc 2}$ & ${\ssc 2}$ & 
                     ${\ssc 0}$ & ${\ssc 0}$ & ${\ssc 0}$ & ${\ssc 0}$ & 
                    -${\ssc 1}$ & -${\ssc 1}$ & -${\ssc 1}$ & -${\ssc 1}$ & 
                     ${\ssc 2}$ & ${\ssc 2}$ & ${\ssc 0}$ & ${\ssc 0}$ \\ 
${\ssc \Lambda_5}$ & ${\ssc 2}$ & -${\ssc 2}$ & ${\ssc 2}$ & -${\ssc 2}$ & 
                     ${\ssc 0}$ & ${\ssc 0}$ & ${\ssc 0}$ & ${\ssc 0}$ & 
                     ${\ssc 1}$ & -${\ssc 1}$ & ${\ssc 1}$ & -${\ssc 1}$ & 
                    -${\ssc 2}$ & ${\ssc 2}$ & ${\ssc 0}$ & ${\ssc 0}$ \\ 
${\ssc \Lambda_6}$ & 
${\ssc 2}$ & ${\ssc 2\im}$ & -${\ssc 2}$ & -${\ssc 2\im}$ & 
                   ${\ssc \oq}$ & -${\ssc \oq^*}$ & -${\ssc \oq}$ & 
         ${\ssc \oq^*}$ & ${\ssc \im}$ & -${\ssc 1}$ & -${\ssc \im}$ & 
                   ${\ssc 1}$ & ${\ssc 0}$ & ${\ssc 0}$ & ${\ssc 0}$ & 
                   ${\ssc 0}$  \\ 
${\ssc \Lambda_7}$ & 
${\ssc 2}$ & ${\ssc 2\im}$ & -${\ssc 2}$ & -${\ssc 2\im}$ & 
                    -${\ssc \oq}$ & ${\ssc \oq^*}$ & ${\ssc \oq}$ & 
          -${\ssc \oq^*}$ & ${\ssc \im}$ & -${\ssc 1}$ & -${\ssc \im}$ & 
                     ${\ssc 1}$ & ${\ssc 0}$ & ${\ssc 0}$ & ${\ssc 0}$ & 
                     ${\ssc 0}$ \\ 
${\ssc \Lambda_8}$  & 
${\ssc 2}$ & -${\ssc 2\im}$ & -${\ssc 2}$ & ${\ssc 2\im}$ & 
                      -${\ssc \oq^*}$ & ${\ssc \oq}$ & ${\ssc \oq^*}$ & 
             -${\ssc \oq}$ & -${\ssc \im}$ & -${\ssc 1}$ & ${\ssc \im}$ & 
                      ${\ssc 1}$ & ${\ssc 0}$ & ${\ssc 0}$ & ${\ssc 0}$ & 
                      ${\ssc 0}$  \\ 
${\ssc \Lambda_9}$  & 
${\ssc 2}$ & -${\ssc 2\im}$ & -${\ssc 2}$ & ${\ssc 2\im}$ & 
                      ${\ssc \oq^*}$ & -${\ssc \oq}$ & -${\ssc \oq^*}$ & 
             ${\ssc \oq}$ & -${\ssc \im}$ & -${\ssc 1}$ & ${\ssc \im}$ & 
                      ${\ssc 1}$ & ${\ssc 0}$ & ${\ssc 0}$ & ${\ssc 0}$ & 
                      ${\ssc 0}$ \\ 
${\ssc \Lambda_{10}}$ & ${\ssc 3}$ & ${\ssc 3}$ & ${\ssc 3}$ & ${\ssc 3}$ & 
                        ${\ssc 1}$ & ${\ssc 1}$ & ${\ssc 1}$ & ${\ssc 1}$ & 
                        ${\ssc 0}$ & ${\ssc 0}$ & ${\ssc 0}$ & ${\ssc 0}$ & 
                       -${\ssc 1}$ & -${\ssc 1}$ & -${\ssc 1}$ & -${\ssc 1}$  
                        \\ 
${\ssc \Lambda_{11}}$ & ${\ssc 3}$ & -${\ssc 3}$ & ${\ssc 3}$ & -${\ssc 3}$ & 
          ${\ssc \im}$ & -${\ssc \im}$ & ${\ssc \im}$ & -${\ssc \im}$ & 
                        ${\ssc 0}$ &  ${\ssc 0}$ & ${\ssc 0}$ & ${\ssc 0}$ & 
                  ${\ssc 1}$ & -${\ssc 1}$ & ${\ssc \im}$ & -${\ssc \im}$ 
                        \\ 
${\ssc \Lambda_{12}}$ & ${\ssc 3}$ & ${\ssc 3}$ & ${\ssc 3}$ & ${\ssc 3}$ & 
                       -${\ssc 1}$ & -${\ssc 1}$ & -${\ssc 1}$ & -${\ssc 1}$ & 
                        ${\ssc 0}$ & ${\ssc 0}$ & ${\ssc 0}$ & ${\ssc 0}$ & 
                       -${\ssc 1}$ & -${\ssc 1}$ & ${\ssc 1}$ & ${\ssc 1}$ 
                       \\ 
${\ssc \Lambda_{13}}$ & ${\ssc 3}$ & -${\ssc 3}$ & ${\ssc 3}$ & -${\ssc 3}$ & 
               -${\ssc \im}$ & ${\ssc \im}$ & -${\ssc \im}$ & ${\ssc \im}$ & 
                         ${\ssc 0}$ & ${\ssc 0}$ & ${\ssc 0}$ & ${\ssc 0}$ & 
                         ${\ssc 1}$ & -${\ssc 1}$ & -${\ssc \im}$ & 
                         ${\ssc \im}$ \\ 
${\ssc \Lambda_{14}}$ & ${\ssc 4}$ & ${\ssc 4}$ & -${\ssc 4}$ & -${\ssc 4}$ & 
                          ${\ssc 0}$ & ${\ssc 0}$ & ${\ssc 0}$ & ${\ssc 0}$ & 
                          ${\ssc 1}$ & ${\ssc 1}$ & -${\ssc 1}$ & -${\ssc 1}$ &
                          ${\ssc 0}$ & ${\ssc 0}$ &  ${\ssc 0}$ & ${\ssc 0}$  
                          \\ 
${\ssc \Lambda_{15}}$ & ${\ssc 4}$ & -${\ssc 4}$ & -${\ssc 4}$ & ${\ssc 4}$ & 
                          ${\ssc 0}$ & ${\ssc 0}$ & ${\ssc 0}$ & ${\ssc 0}$ &
                         -${\ssc 1}$ & ${\ssc 1}$ & ${\ssc 1}$ & -${\ssc 1}$ & 
                          ${\ssc 0}$ & ${\ssc 0}$ & ${\ssc 0}$ & ${\ssc 0}$ 
                         \\[1mm] \hline
\end{tabular}  
\caption{\sl Character table of the truncated braid group 
$B(3,4)$. We used $\oq := \im+1$.}
\label{tab:tab4}
\end{table}

According to the general definition~(\ref{eqy})--(\ref{truncate}),
the truncated braid group $B(3,4)$ for three indistinguishable particles 
is generated
by two elements $\tau_1$ and $\tau_2$ subject to the relations
$\tau_1 \tau_2 \tau_1 = \tau_2   \tau_1 \tau_2$ and 
$\tau_1^4 = \tau_2^4  = e$.
By  explicit construction, which is a 
lengthy and not at all trivial job, it can be inferred
that $B(3,4)$ is a  group consisting 96 elements organized into the 
following 16 conjugacy classes 
\bea   \label{B34}
C_0^1 & = & \{ e \}    \\
C_0^2 & = & \{\tau_1\tau_2\tau_1\tau_2\tau_1\tau_2 \} \nn \\
C_0^3 & = & \{\tau_2^2\tau_1^2\tau_2^2\tau_1^2\} \nn \\
C_0^4 & = & \{\tau_2^2\tau_1^3\tau_2^2\tau_1^3\} \nn \\
C_1^1 & = & \{\tau_1\; , \; \tau_2\; , \; \tau_2\tau_1\tau_2^3\; , \; 
\tau_2^2\tau_1\tau_2^2\; , \; \tau_2^3\tau_1\tau_2\; , \; 
\tau_1^2\tau_2\tau_1^2\} \nn \\
C_1^2 & = & \{\tau_1^3\tau_2\tau_1^2\tau_2\; , \; 
\tau_2^3\tau_1\tau_2^2\tau_1\; , \; 
\tau_2\tau_1^3\tau_2\tau_1^2\; , \; 
\tau_2^2\tau_1^2\tau_2^2\tau_1\; , \; 
\tau_1\tau_2^3\tau_1\tau_2^2\; , \; 
\tau_1^2\tau_2^2\tau_1^2\tau_2\} \nn \\
C_1^3 & = & \{\tau_2\tau_1^3\tau_2\tau_1^3\tau_2\; , \; 
\tau_1^2\tau_2\tau_1^3\tau_2^2\tau_1\; , \; 
\tau_2^3\tau_1\tau_2^3\tau_1^2\; , \; 
\tau_1\tau_2^2\tau_1^3\tau_2^2\tau_1\; , \; 
\tau_2\tau_1\tau_2^3\tau_1^2\tau_2^2\; , \; 
\tau_2\tau_1^2\tau_2^3\tau_1^2\tau_2\} \nn \\
C_1^4 & = & \{\tau_2^2\tau_1^3\tau_2^2\; , \; 
\tau_1^2\tau_2^3\tau_1^2\; , \; 
\tau_2^3\tau_1^3\tau_2\; , \; 
\tau_1^3\; , \; \tau_2\tau_1^3\tau_2^3\; , \; 
\tau_2^3\}  \nn \\
C_2^1 & = & \{\tau_1\tau_2\; , \; \tau_2\tau_1\; , \; 
\tau_1^2\tau_2\tau_1^3\; , \; 
\tau_1^3\tau_2\tau_1^2\; , \; 
\tau_2\tau_1^2\tau_2^2\tau_1\; , \; 
\tau_2^2\tau_1\tau_2^3\; , \; 
\tau_2^3\tau_1\tau_2^2\; , \; 
\tau_1\tau_2^2\tau_1^2\tau_2\} \nn \\
C_2^2 & = & \{\tau_1^2\tau_2\tau_1^3\tau_2\tau_1\; , \; 
\tau_1\tau_2\tau_1^3\tau_2\tau_1^2\; , \; 
\tau_2\tau_1^2\tau_2^2\tau_1^3\; , \; 
\tau_1\tau_2\tau_1^2\tau_2^2\tau_1^2\; , \; \nn \\
      &   & \; \tau_2\tau_1^3\tau_2^2\tau_1^2\; , \; 
\tau_1\tau_2\tau_1^3\tau_2^2\tau_1\; , 
\; \tau_1^2\tau_2\tau_1^3\tau_2^2\; , 
\; \tau_1^2\tau_2^2\tau_1^3\tau_2\} \nn \\
C_2^3 & = & \{\tau_1^3\tau_2\tau_1^3\tau_2^2\tau_1\; , \; 
\tau_1\tau_2^2\tau_1^3\tau_2\tau_1^3\; , \; 
\tau_2\tau_1^3\tau_2^2\; , \; \tau_2^2\tau_1^3\tau_2\; , \; 
 \tau_2^3\tau_1^3\; , \; 
\tau_1\tau_2^3\tau_1^2\; , \; 
\tau_1^2\tau_2^3\tau_1\; , \; 
\tau_1^3\tau_2^3\}  \nn  \\
C_2^4 & = & \{\tau_1^3\tau_2^3\tau_1^2\; , \; 
\tau_1^2\tau_2^3\tau_1^3\; , \; 
\tau_2^3\tau_1\; , \; \tau_1\tau_2^3\; , \; 
\tau_1\tau_2\tau_1\tau_2\; , \; \tau_1^3\tau_2\; , \; 
\tau_2\tau_1^3\; , \; \tau_2\tau_1\tau_2\tau_1\} \nn \\
C_3^1 & = & \{\tau_1^2\; , \; \tau_2^2\; , \; 
\tau_1\tau_2^2\tau_1^3\; , \; 
\tau_2^2\tau_1^2\tau_2^2\; , \; 
\tau_1^2\tau_2^2\tau_1^2\; , \; 
\tau_1^3\tau_2^2\tau_1\} \nn \\
C_3^2 & = & \{\tau_2\tau_1^2\tau_2\; , \; 
\tau_1\tau_2^2\tau_1\; , \; \tau_1^2\tau_2^2\; , \; 
\tau_2^3\tau_1^2\tau_2^3\; , \; 
\tau_1^3\tau_2^2\tau_1^3\; , \; 
\tau_2^2\tau_1^2\} \nn \\
C_4^1 & = & \{\tau_1\tau_2\tau_1\; , \; \tau_1^2\tau_2\; , \; 
\tau_2^2\tau_1\; , \; \tau_2\tau_1^2\; , \; 
\tau_1\tau_2^2\; , \; \tau_1^3\tau_2\tau_1^3\; , \; 
\tau_1^3\tau_2\tau_1^3\tau_2^2\tau_1^2\; , \; \nn \\
      &   & \; \tau_2\tau_1^3\tau_2^2\tau_1\; , \; 
\tau_1^2\tau_2^2\tau_1^3\; , \; 
\tau_1\tau_2^2\tau_1^3\tau_2\; , \; 
\tau_1^3\tau_2^2\tau_1^2\; , \; 
\tau_1\tau_2\tau_1^3\tau_2^2\} \nn \\
C_4^2 & = & \{\tau_1\tau_2\tau_1\tau_2\tau_1\tau_2\tau_1\tau_2\tau_1\; , \; 
\tau_2\tau_1^2\tau_2^2\; , \; \tau_1\tau_2^2\tau_1^2\; , \; 
\tau_2^2\tau_1^2\tau_2\; , \; \tau_1^2\tau_2^2\tau_1\; , \; 
\tau_2\tau_1^3\tau_2\; , \; \nn \\
      &   & \; \tau_2^3\tau_1^3\tau_2^3\; , \; 
\tau_2^3\tau_1^2\; , \; \tau_1^3\tau_2^2\; , \; 
\tau_2^2\tau_1^3\; , \; \tau_1^2\tau_2^3\; , \; 
\tau_1\tau_2^3\tau_1\} \, .           \nn
\eea
Here, the conjugacy classes are presented such 
that $C_k^{i+1} = z C_k^i$ with 
$z = \tau_1\tau_2\tau_1\tau_2\tau_1\tau_2$ the 
generator for the centre of order 4 of $B(3,4)$. 
The character table of the truncated braid group
$B(3,4)$ is displayed in table~\ref{tab:tab4}.

\begin{table}[hbt]  
\begin{center}
\begin{tabular}{crrrrrrrrrrr} \hl
$P(3,4)$ & & $C_0$ & $C_1$ & $C_2$ & $C_3$ & $C_4$ & 
$C_5$ & $C_6$ & $C_7$ & $C_8$ & $C_9$    \\ \hl  \\[-4mm] 
$\Omega_0$ & & $1$ & $1$  & $1$ & $1$ & $1$ & $1$ & $1$ & $1$ & $1$ & $1$ \\ 
$\Omega_1$ & & $1$ & $1$ & $1$ & $1$ & $-1$ & $-1$ & $1$ & 
$-1$ & $-1$ & $1$ \\ 
$\Omega_2$ & & $1$ & $1$ & $1$ & $1$ & $-1$ & $1$ & $-1$ & 
$1$ & $-1$ & $-1$ \\ 
$\Omega_3$ & & $1$ & $1$ & $1$ & $1$ & $1$ & $-1$ & $-1$  & $-1$ & $1$ & $-1$ 
 \\   $\Omega_4$ & & $1$ & $-1$ & $1$ & $-1$ & $-1$ & $1$ & $1$ & $-1$ & 
$1$ & $-1$ \\ 
$\Omega_5$ & & $1$ & $-1$ & $1$ & $-1$ & $1$ & $-1$ &
$1$ & $1$ & $-1$ & $-1$ \\ 
$\Omega_6$ & & $1$ & $-1$ & $1$ & $-1$ & $1$ & $1$ & $-1$ & $-1$ &
$-1$ & $1$ \\ 
$\Omega_7$ & & $1$ & $-1$  & $1$  & $-1$  & $-1$ & $-1$ & $-1$ & $1$ & $1$ & 
             $1$ \\  
$\Omega_8$& & $2$ & $2\im$  & $-2$ & $-2\im$ & $0$ & $0$ & $0$ & $0$ & $0$
           & $0$ \\
$\Omega_{9}$& & $2$ & $-2\im$  & $-2$ & $2\im$ & $0$ & $0$ & $0$ & $0$ 
               & $0$ & $0$ \\[1mm]              \hl
\end{tabular}
\end{center}
\caption{\sl Character table of 
the truncated colored braid group $P(3,4)$.}
\label{char}
\end{table}

The truncated colored braid group $P(3,4)$ consisting of the monodromy  
operations on a configuration of three distinguishable particles
is the subgroup of $B(3,4)$ generated by 
\bea  \label{p34mongen}
\gamma_{12} \; = \; \tau_1^2 \,, \qquad
\gamma_{13} \; = \; \tau_1 \tau_2^2 \tau_1^{-1}= \tau_1 \tau_2^2\tau_1^3 
\,, \qquad
\gamma_{23} \;= \; \tau_2^2 \, ,     
\eea
which satisfy
$
\gamma_{12}^2 \; = \; \gamma_{13}^2 \;= \; \gamma_{23}^2 \;=\; e.
$
It can be verified  that $P(3,4)$ is a group
of order 16 splitting up in the following 10 conjugacy classes
\bea          \label{p34}
\ba{lcl lcl}
C_0 &=& \{e \}  &   
\qquad  
C_1 &=& \{\gamma_{13} \gamma_{12} \gamma_{23}\} \\
C_2 &=& \{ \gamma_{23}\gamma_{12}\gamma_{23} \gamma_{12} \}           &   
\qquad
C_3 &=& \{ \gamma_{23}\gamma_{12}\gamma_{13} \} \\
C_4 &=& \{\gamma_{12} \; , \; \gamma_{23} \gamma_{12}\gamma_{23} \}   &   
\qquad
C_5 &=& \{\gamma_{23} \; , \; \gamma_{12} \gamma_{23} \gamma_{12} \}  \\
C_6 &=& \{\gamma_{13} \; , \; \gamma_{12} \gamma_{13} \gamma_{12} \}  &      
\qquad
C_7 &=& \{\gamma_{13}\gamma_{12} \; , \; \gamma_{12} \gamma_{13}  \}   \\
C_8 &=& \{\gamma_{23}\gamma_{13} \; , \; \gamma_{13}\gamma_{23}   \}  &  
\qquad
C_9 &=& \{\gamma_{12} \gamma_{23} \; , \; \gamma_{23} \gamma_{12} \} \,  .
\ea
\eea   
The centre of $P(3,4)$, contained in the first four conjugacy classes,
coincides with that of $B(3,4)$. 
Further, the truncated colored braid group
$P(3,4)$ turns out to be isomorphic to the coxeter 
group denoted as $16/8$ in~\cite{thomas}.
Finally, the character table of $P(3,4)$ is given 
in table~\ref{char}.

\sectiona{$D_4$ gauge theory}     \label{dvier}

In this last appendix, I derive the modular matrices for a 2+1 dimensional 
gauge theory with finite gauge group the dihedral group $D_4$ used in the 
analysis of section~\ref{emduality}. The discussion will be
concise. For a thorough treatment of planar gauge theories
with a nonabelian finite gauge group $H$, the interested reader is 
referred to~\cite{banff,thesis}.

Let me start with some general remarks.
First of all, the spectrum of a planar nonabelian discrete $H$ gauge 
theory (without CS term) can be presented as~\cite{banff,spm} 
 \bea                                            \label{conencen}
(\, ^A  C, \alpha \, ) \, ,
\eea
where $A$ labels the different conjugacy classes of ${ H}$ and $\alpha$
the inequivalent UIR's of the centralizer associated to the 
conjugacy class $^A C$.  The particles only carrying magnetic 
flux are labeled by the nontrivial conjugacy classes 
paired with a trivial centralizer representation.  
The pure charges, on the other hand, correspond to the trivial 
conjugacy class (with centralizer the full group ${H}$) and are thus 
labeled by the different nontrivial UIR's of ${ H}$. 
The other particles are dyons. Let us now introduce  
an arbitrary but fixed ordering of the group elements in 
the different conjugacy classes of ${ H}$:
\bea \label{order}
^A C &=& \{\, ^A h_1, \, ^A h_2, \ldots,\,^A h_k \, \} \, .
\eea 
Next, let $^A N \subset { H}$ be the centralizer of the group element 
$^A h_1$ (i.e.\ $^A N$ consists of the elements of $H$ that commute 
with $^A h_1$) and $\{\, ^A x_1,\,^A x_2,\ldots,\,^A x_k \, \}$ an
arbitrary but fixed set of representatives for the equivalence 
classes of ${ H}/^A N$  
such that $^A h_i=\,^A x_i\,^A h_1\,^A x_i^{-1}$.
With these conventions, the modular matrices for a discrete $H$
gauge theory are given as
\bea                                \label{fusionz}   
S^{AB}_{\alpha\beta} &:=& 
\frac{1}{|{ H}|} \, \sum_{\stackrel{\,^A h_i\in\,^A C\,,^B h_j\in\,
^B C}{[\,^A h_i,\,^B h_j \,]\,=\,e}} 
\mbox{tr} \left(\alpha (\,^A x_i^{-1}\,^B h_j\,^A x_i) \right)^*
\; \mbox{tr} \left( \beta(\,^B x_j^{-1}\,^A h_i\, ^B x_j) \right)^*  \\
T^{AB}_{\alpha\beta} &:=& 
\delta_{\alpha,\beta} \, \delta^{A,B} \; \exp(2\pi \im s_{(A,\alpha)})
\; = \;
\delta_{\alpha,\beta} \, \delta^{A,B} 
\frac{1}{d_\alpha}   \, \mbox{tr} \left(  \alpha(^A h_1) \right) \, , 
\label{modutz}
\eea
with $[\,^A h_i,\,^B h_j \,] := 
\,^A h_i\, ^B h_j \,^A h_i^{-1}\, ^B h_j^{-1}$.
Here, the capitals label the conjugacy classes of $H$ and 
the greek letters the associated centralizer representations. 
Further, $|H|$ denotes the order of $H$, $e$ the unit element of $H$, 
$*$ complex conjugation, $\delta$ the Kronecker delta function,
$s_{(A,\alpha)}$ the spin assigned to the particle $(\, ^A  C, \alpha \, )$
and $d_\alpha$ the dimension of the centralizer representation $\alpha$.

Let us now focus on the case $H\simeq D_4$. 
The dihedral group $D_4$ of order 8 is 
the semi-direct product of the cyclic groups $\Z_2$ and $\Z_4$. 
Specifically, $D_4$ is defined by two generators $X$ and $R$
subject to the relations
\bea    \label{feyt}
X^{2} \; = \; e \, , \qquad R^4 \; = \; e \, , \qquad
  XR \; = \; R^{-1} X \, . 
\eea
For convenience, we will label  the elements of $D_4$ by the 2-tuples
\bea    \label{defelemd}
(K,k) & := & X^K R^k  \qquad \qquad \mbox{with $K \in 0,1$ and               
$k \in -1, 0, 1, 2 \,. $} \qquad
\eea
Hence, the capital $K$ represents an element of the $\Z_2$ subgroup 
of $D_{4}$ generated by $X$ and the lower-case letter $k$ an 
element of the $\Z_{4}$ subgroup generated by $R$. 
From~(\ref{feyt}) and~(\ref{defelemd}), we then 
infer that the multiplication law becomes 
\bea                  \label{grmul}
(K,k) \cdot (L,l) &=& ([K+L],[(-)^L k + l])\, ,
\eea
where I used the abbrevation $(-) := (-1)$. 
The rectangular brackets appearing in the first entry
of the 2-tuple indicate modulo $2$ calculus such that the sum lies 
in the range $0,1$ and those for the second entry
modulo $4$ calculus in the range $-1,0,1,2$.

\begin{table}[htb] 
\begin{center}
\begin{tabular}[t]{lcl} \hline     \\[-4mm]
$ \mbox{Conjugacy class}                      $ & & $\str 
\mbox{Centralizer}       $\\ \hl   \\[-4mm]
$ ^0  C=\{(0,0)\} \str                   $& &$ {D}_4$\\ 
$ ^1  C=\{(0,1),(0,-1)\}\str              
$& &$ \Z_{4} = \{(0,0),(0,1),(0,2),(0,-1)\}$\\ 
$ ^2  C=\{(0,2)\} \str                 $& &$ {D}_4$\\ 
$ ^X C=\{(1,0),(1,2) \}       
$& &$ \str \Z_2 \times \Z_2 = \{(0,0),(1,0),(0,2),(1,2)\} $\\ 
$ ^{\bar{X}}  C=\{(1,1),(1,-1)\}
$& &$ \str \Z_2 \times \Z_2 = \{(0,0),(1,1),(0,2),(1,-1)\} $\\[1mm] \hl
\end{tabular}
\end{center} 
\caption{\sl Conjugacy classes of the dihedral 
group $D_4$ and the associated centralizers.} 
\label{concend4}  
\end{table} 
\begin{table}[htb] 
\begin{center}
\begin{tabular}[t]{lccccc} \hline      \\[-4mm]
$\str {D}_4\qquad$&$ ^0 C $&$ ^1 C$&$ ^2 C  $&$ ^X C  $&$ ^{\bar{X}} C$
\\ \hl \\[-4mm]
$ \pi^{++} \str$&$ 1 $&$ 1      $&$ 1  $&$ 1  $&$ 1  $\\ 
$ \pi^{+-} \str$&$ 1 $&$ 1      $&$ 1  $&$-1  $&$-1  $\\ 
$ \pi^{-+} \str$&$ 1 $&$ -1 $&$ 1  $&$ 1  $&$ -1 $\\ 
$ \pi^{--} \str$&$ 1 $&$ -1 $&$ 1  $&$-1  $&$ 1 $\\ 
$ \pi^1    \str$&$ 2 $&$ 0  $&$ -2 $&$ 0  $&$ 0  $\\[1mm]   \hl
\end{tabular} \qquad \qquad
\end{center}
\caption{\sl Character table of ${D}_4$.} 
\label{charis}
\end{table}  

With~(\ref{grmul}), it is easily verified that 
the elements~(\ref{defelemd}) of $D_4$ are organized in 5 conjugacy 
classes as displayed in table~\ref{concend4} together 
with their centralizers. From the character table~\ref{charis}, we 
subsequently  infer that the spectrum of a $D_4$ gauge theory features 4 
nontrivial pure charges: three singlet charges corresponding to 
the 1-dimensional UIR's $\pi^{+-}$, $\pi^{-+}$, $\pi^{--}$ and one 
doublet charge associated with the 2-dimensional UIR $\pi^1$. 
The trivial $D_4$ representation $\pi^{++}$ denotes the vacuum.
Further, as indicated by table~\ref{concend4},
the conjugacy class $^2 C$ consists of the nontrivial centre 
element element $(0,2)$ with  centralizer the full group $D_4$. 
Hence, there are 5 particles with the singlet flux $(0,2)$, namely 
the pure flux $(0,2)$ itself and 4 dyons carrying this flux and a 
nontrivial $D_4$ charge. 
In addition, the spectrum consists of 3 pure doublet fluxes corresponding 
to the conjugacy classes $^1 C$, $^X C$ and $^{\bar X} C$, which all 
contain two commuting elements. The 3 dyons associated with the doublet 
flux $^1 C$ carry a nontrivial $\Z_4$ charge $\Gamma^n$ with 
$n \in 1,2,3$ defined as 
\bea    \label{charz4}
\Gamma^n((0,k)) &=& \exp \left( \frac{\pi \im}{2} \, nk \right) \, .
\eea    
To proceed, there are 3 distinct dyons carrying doublet flux 
$^X C$ and a nontrivial $\Z_2 \times \Z_2$ charge 
$\Gamma^{rs}$ with $r,s \in +,-$. In our conventions, the element 
$(1,0)$ is the generator of the first and 
$(0,2)$ the generator of the second $\Z_2$ factor of the 
$\Z_2 \times \Z_2$ centralizer related to the conjugacy class $^X C$. The   
$\Z_2 \times \Z_2$ representation $\Gamma^{rs}$ is then given by 
\bea \label{charz2z2}
\Gamma^{rs}((1,0)) &=& r1\, , \qquad \Gamma^{rs}((0,2)) \; = \; s1 \, .
\eea  
Thus $r$ determines the sign assigned to the first $\Z_2$ generator 
in the UIR $\Gamma^{rs}$ and $s$ the sign of the second. 
Finally, there are 3 different dyons with doublet flux $^{\bar{X}} C$ 
and nontrivial $\Z_2 \times \Z_2$ charge 
$\Gamma^{rs}$ with $r,s \in +,-$. Here, $\Gamma^{rs}$ is defined as before
with the only distinction that the generator of the first  
$\Z_2$ factor for the  $\Z_2 \times \Z_2$ centralizer 
related to $^{\bar X} C$ is the element $(1,1)$.
To conclude, including the vacuum, the spectrum of a 
$D_4$ gauge theory features 22 particles which will be denoted as 
\be   \label{specd4}       \ba{rclrclrcl}
         (0,rs)   &:=& ( \,  ^0 C, \, \pi^{rs} \, )     \, ,
& \;   (0,1)    &:=& ( \, ^0 C, \, \pi^{1} \, )       \, ,
& \;   (1,n)    &:=& (\,^1 C, \, \Gamma^{n}\,)        \, ,  \\
         (2,rs)   &:=& (\, ^2 C, \,  \pi^{rs} \, )      \, ,   
& \;   (2,1)    &:=& ( \, ^2 C, \, \pi^{1} \, )       \, , 
&                 &  &                                        \\
( \, {\bar{X}}, \, rs \,) 
                  &:=& ( \, ^{\bar{X}} C , \, {\Gamma^{rs}} \, ) \, ,  
&\;    (X,rs)   &:=& ( \, ^X C , \, {\Gamma^{rs}}\, )          \, , 
&                 &  &
\ea
\ee
where $r,s\in +, -$ label both the four ${D}_4$ singlet charges and 
the four $\Z_2 \times \Z_2$ dyon charges and $n \in 0,1,2,3$ 
denote the four $\Z_4$ dyon charges.

\begin{table}[htb]
\begin{center}
\begin{tabular}{cccccccc}      
\hline \\[-4mm]
$S$ & $(0,rs)$ & $(0,1)$ & $(1,n)$ & $(2,rs)$ & $(2,1)$ & $(X,rs)$ &
$(\bar{X},rs)$  
\\ \hl
\\[-4mm]  
$(0,r's')$ & $1$   & $2$  & $r'2$ & $1$    & $2$  & $ s'2 $ & $r's'2$   \\ 
$(0,1)$    & $2$   & $4$  &  $0$  & $-2$   & $-4$ & $0$     & $0$       \\
$(1,n')$   & $r2$  & $0$  & $4\cos\left(\frac{\pi}{2}(n'+n)\right)$ 
                                    &$ r(-)^{n'} 2$ 
                                            & $0$  & $0$     & $0$       \\
$(2,r's')$ & $1$   & $-2$ & $r'(-)^n 2$
                                    & $1$    & $-2$ & $s's2 $ & $r's's 2$ \\
$(2,1)$    & $2$   & $-4$ &  $0$  & $-2$   & $4$  & $0$     & $0$       \\
$(X,r's')$ & $s2$  & $0$  &  $0$  & $s's2$ & $0$  & $r'r4 \delta_{s',s}$
                                                             & $0$       \\ 
$(\bar{X},r's')$
           & $rs2$ & $0$  &  $0$  & $rss'2$
                                             & $0$  & $0$    
&$r'r4 \delta_{s',s}$
\\[1mm]   \hline
\end{tabular}
\end{center}
\caption{\sl Modular $S$ matrix for a $D_4$ gauge theory up to an 
overall factor $\frac{1}{8}$. Here, $\delta_{s',s}$ denotes the Kronecker
delta function. We also used the algebra of signs. For instance,
the matrix element $S^{\;\; 2 \;\;\; \bar{X}}_{r's' \; rs} = r's's 2$ 
takes the value $(-)\cdot(-)\cdot(-)2=-2$ for $r'=s'=s=-$.} 
\label{modsd4}
\end{table}

Let us finally turn to the modular matrices for a  $D_4$ gauge theory.
We will work with the ordering of the elements of $D_4$ indicated 
in table~\ref{concend4} and the following representatives 
(see the discussion concerning relation~(\ref{order})) 
\bea   \label{convrep}
^0 x_1 &=& ^1 x_1  \;=\; ^2 x_1  \;=\; ^X x_1  \;=\; ^{\bar{X}} x_1  \;=\; 
(0,0) \\
^1 x_2 &=& (1,0) \, , \;\; ^X x_2  \;=\; ^{\bar{X}} x_2  \;=\; (0,1) \,.
\eea    
The modular $T$ matrix~(\ref{modutz}) contains the spin factors 
$\exp(2 \pi \im s_{(A,\alpha)}) = 
\mbox{tr} \left( \alpha(^A h_1) \right)/ d_\alpha$ assigned to 
the different particles~(\ref{conencen}) in the spectrum of 
a discrete $H$ gauge theory.
From table~\ref{charis} and the relations~(\ref{charz4}) 
and~(\ref{charz2z2}), we easily infer the following spin factors
for the particles~(\ref{specd4}) in our $D_4$ gauge theory:  
\bea             \label{spd4}
\ba{cc}
\mbox{particle}                &\qquad   \exp (2\pi \im  s)    \\
(0,rs), \, (0,1)         & \qquad       1                      \\
(1, n)                   & \qquad       \im^n                  \\ 
(2,rs)                   & \qquad       1                      \\
(2,1)                    & \qquad      -1                      \\
(X,rs), \,(\bar{X},rs)   & \qquad       r1 \, .                 
\ea          
\eea  
To conclude, a lengthy but straightforward calculation involving
table~\ref{concend4}, the character table~\ref{charis} and 
the relations~(\ref{charz4}), (\ref{charz2z2}) and~(\ref{convrep})
shows that the modular $S$ matrix~(\ref{fusionz}) for this $D_4$ gauge theory
is of the form displayed in table~\ref{modsd4}.

\sectie

\end{document}